\documentclass[apj,onecolumn,numberedappendix]{emulateapj}

\usepackage{apjfonts}

\newcommand{\M}{\mathrm}
\newcommand{\E}{\,\mathrm}

\def\hmpc{\ifmmode{h_{100}^{-1}\,\hbox{Mpc}}\else{$h^{-1}$\thinspace Mpc}\fi}

\newcommand{\iss}{{s}}

\newcommand{\mc}{\multicolumn}

\newcommand{\darray}{\renewcommand{\arraystretch}{2.0}}

\newcommand{\ang}{\phi}

\newcommand{\Dol}{{D_\mathrm{ol}}}

\newcommand{\Dos}{{D_\mathrm{os}}}
\newcommand{\RE}{{R_\mathrm{E}}}

\newcommand{\tE}{{t_\mathrm{E}}}

\newcommand{\const}{{\mathrm{const}}}
\newcommand{\Min}{{\mathrm{min}}}
\newcommand{\Max}{{\mathrm{max}}}
\newcommand{\Int}{\int\limits}
\newcommand{\Intinf}{\int\limits_{-\infty}^{+\infty}}
\newcommand{\Intninf}{\int\limits_0^{\infty}}
\newcommand{\Sum}{\sum\limits}
\newcommand{\tfwhm}{{t_\mathrm{FWHM}}}
\newcommand{\tfwhmmin}{{t_\mathrm{FWHM}^\mathrm{min}}}
\newcommand{\tfwhmmax}{{t_\mathrm{FWHM}^\mathrm{max}}}
\newcommand{\vt}{{v_\mathrm{t}}}

\newcommand{\hs}{\hspace{1cm}}

\newcommand{\bT}{{b_\mathrm{T}}}

\newcommand{\pphi}{{p_\ang}}
\newcommand{\pv}{{p_\vt}}

\newcommand{\ptE}{{p_{\tE}}}
\newcommand{\Dt}{{\Delta t}}
\newcommand{\Dvt}{{\vt\Dt}}

\newcommand{\pcmd}{{p_\M{cmd}}}

\newcommand{\uT}{{u_\M{T}}}
\newcommand{\AT}{{A_\M{T}}}
\newcommand{\sigl}{{\sigma_\M{l}}}
\newcommand{\sigs}{{\sigma_\M{s}}}
\newcommand{\sigls}{{\sigma_\M{ls}}}
\newcommand{\sigst}{{\tilde{\sigma}_\M{s}}}
\newcommand{\vl}{{v_{l}}}
\newcommand{\vli}{{v_{l,i}}}
\newcommand{\vlx}{{v_{l,x}}}
\newcommand{\vly}{{v_{l,y}}}

\newcommand{\vsx}{{v_{s,x}}}
\newcommand{\vsy}{{v_{s,y}}}
\newcommand{\vsz}{{v_{s,z}}}
\newcommand{\vlsx}{{v_{ls,x}}}
\newcommand{\vlsy}{{v_{ls,y}}}
\newcommand{\vlvec}{{\vec{v}_{l}}}
\newcommand{\vsvec}{{\vec{v}_{s}}}

\newcommand{\vx}{{v_{0,x}}}
\newcommand{\vy}{{v_{0,y}}}

\newcommand{\ms}{{M}}

\newcommand{\Feff}{{F_\mathrm{eff}}}
\newcommand{\teff}{{t_\mathrm{eff}}}

\newcommand{\LF}{\tilde{\Phi}}

\newcommand{\Ns}{{N_\iss}}

\newcommand{\A}{{A_0}}
\newcommand{\uub}{{u_0}}

\newcommand{\XX}{{X}}
\newcommand{\CX}{{X-X'}}

\newcommand{\F}{{F_0}}
\newcommand{\Flum}{{\mathcal{F}}}

\newcommand{\DF}{{\Delta_F}}
\newcommand{\DFmin}{{\Delta_F^\mathrm{min}}}
\newcommand{\DFmax}{{\Delta_{F,\mathrm{max}}}}

\newcommand{\Mlum}{{\mathcal{M}}}
\newcommand{\FVega}{{F_\M{Vega}}}
\newcommand{\SB}{{\mu}}

\newcommand{\ml}{{M}}

\newcommand{\Mlens}{{\ml_0}}

\newcommand{\Col}{{\mathcal{C}}}
\newcommand{\Colmin}{{\mathcal{C}^\mathrm{min}}}
\newcommand{\Colmax}{{\mathcal{C}^\mathrm{max}}}

\newcommand{\GAM}{\frac{d^4 \Gamma}{d\Dol \,d\ml \,d\vt \,d\uub }}
\newcommand{\GAMb}{\frac{d^4 \Gamma}{d\Dol \,d\ml \,d\vt \,db }}

\newcommand{\ML}{{\left(\frac{M}{L}\right)}}

\newcommand{\ext}{\mathcal{A}}

\newcommand{\dfak}{\left(\frac{10\E{pc}}{\Dos}\right)^2}

\newcommand{\texp}{t_\M{exp}}
\newcommand{\ZP}{\M{ZP}}
\newcommand{\pixelsize}{l_\M{pixel}}
\newcommand{\skymag}{\M{sky}}

\newcommand{\SN}{{\left(\frac{S}{N}\right)}}

\newcommand{\GT}{\Gamma_\M{T}}
\newcommand{\G}{\Gamma}

\newcommand{\tEmeas}{t_\M{E}^\M{meas}}
\newcommand{\uubmeas}{u_0^\M{meas}}
\newcommand{\Ameas}{A_0^\M{meas}}
\newcommand{\xmeas}{x^\mathrm{meas}}
\newcommand{\ymeas}{y^\mathrm{meas}}
\newcommand{\DFmeas}{{\Delta_F^\mathrm{meas}}}
\newcommand{\Colmeas}{{\mathcal{C}^\mathrm{meas}}}
\newcommand{\tfmeas}{{t_\mathrm{FWHM}^\mathrm{meas}}}

\newcommand{\dudA}{\left|\frac{d\uub}{d\A}\right|}

\newcommand{\Opsf}{\Omega_\M{psf}}

\newcommand{\eff}{{\epsilon}}

\newcommand{\Q}{{Q}}
\newcommand{\Qgou}{{Q_\M{event}}}

\newcommand{\ufs}{{u_0^\ast}}
\newcommand{\ANfs}{{A_{0}^\ast}}
\newcommand{\Afs}{{A^\ast}}
\newcommand{\uFStfwhm}{u{\left(\frac{\ANfs+1}{2}\right)}}
\newcommand{\tfsFWHM}{{t^\ast_\M{FWHM}}}
\newcommand{\tfsFWHMmin}{{t^\ast_\M{FWHM,min}}}
\newcommand{\tfsFWHMmax}{{t^\ast_\M{FWHM,max}}}
\newcommand{\Upsfs}{{\Upsilon^\ast}}
\newcommand{\Dolfs}{{D_\mathrm{ol}^\ast}}
\newcommand{\Cfs}{{C_\Dolfs}}
\newcommand{\Omegafs}{{\Omega^\ast}}

\newcommand{\lgt}{\log}
 
\newcommand{\Rstar}{R_\ast}

\newcommand{\Msun}{{\ml_\odot}}

\newcommand{\FsunX}{{\Flum_{\odot}}}
\newcommand{\FsunR}{{\Flum_{\odot,R}}}

\newcommand{\lsim}{{\,\stackrel{\scriptstyle <}{\scriptstyle\sim}\,}}

\newcommand{\tV}{\times 10^{-5}}
\newcommand{\tVI}{\times 10^{-6}}
\newcommand{\tVII}{\times 10^{-7}}

\journalinfo{The Astrophysical Journal Supplement Series, 163:225-269, 2006 April}
\submitted{Received 2005 May 18; accepted 2005 November 12} 

\shorttitle{\scshape Microlensing toward crowded fields}
\shortauthors{\scshape Riffeser et al.}

\begin{document}

\title{Microlensing toward crowded fields: Theory and applications to M31}

  \author{A.~Riffeser\altaffilmark{1}, J.~Fliri, S.~Seitz, and R.~Bender\altaffilmark{1}}

  \affil{Universit\"ats-Sternwarte M\"unchen, Scheinerstrasse 1, 81679
    M\"unchen, Germany; arri@usm.lmu.de}

	 \altaffiltext{1}{Max-Planck-Institut f\"ur Extraterrestrische
     Physik, Giessenbachstrasse, 85748 Garching bei M\"unchen,
     Germany.}

   \submitted{Received 2005 May 18; accepted 2005 November 12}

\begin{abstract}
  We present a comprehensive treatment of the pixel-lensing theory and
  apply it to lensing experiments and their results toward M31.  Using
  distribution functions for the distances, velocities, masses, and
  luminosities of stars, we derive lensing event rates as a function
  of the event observables.  In contrast to the microlensing regime,
  in the pixel-lensing regime (crowded or unresolved sources) the
  observables are the maximum excess flux of the source above a
  background and the full width at half-maximum (FWHM) time of the
  event.  To calculate lensing event distribution functions depending
  on these observables for the specific case of M31, we use data from
  the literature to construct a model of M31, reproducing consistently
  photometry, kinematics and stellar population.  We predict the halo-
  and self-lensing event rates for bulge and disk stars in M31 and
  treat events with and without finite source signatures separately.
  We use the M31 photon noise profile and obtain the event rates as a
  function of position, field of view, and S/N threshold at maximum
  magnification.  We calculate the expected rates for WeCAPP and for a
  potential Advanced Camera for Surveys (ACS) lensing campaign.  The
  detection of two events with a peak signal-to-noise ratio larger
  than 10 and a timescale larger than 1 day in the WeCAPP 2000/2001
  data is in good agreement with our theoretical calculations.  We
  investigate the luminosity function of lensed stars for noise
  characteristics of WeCAPP and ACS.  For the pixel-lensing regime, we
  derive the probability distribution for the lens masses in M31 as a
  function of the FWHM timescale, flux excess and color, including the
  errors of these observables.
\end{abstract}


\hspace{10cm}

\section{Introduction}

Searches for compact dark matter toward the Large and Small Magellanic
Clouds (LMC and SMC) and the Galactic bulge identified numerous
microlensing events in the past decade (MACHO,
\citeauthor{1997ApJ...486..697A} \citeyear{1997ApJ...486..697A}; EROS,
\citeauthor{1993Natur.365..623A} \citeyear{1993Natur.365..623A}; OGLE,
\citeauthor{2000AcA....50....1U} \citeyear{2000AcA....50....1U}; DUO,
\citeauthor{1997A&A...326....1A} \citeyear{1997A&A...326....1A}).  In
parallel to these observations, a lot of effort has been spent on the
prediction of the number, the spatial distribution, the amplitude, and
the duration of lensing events toward these targets. The underlying
models require knowledge of density and velocity distribution, as well
as of the luminosity and mass function of lensing and lensed stars.
The halo MACHO mass fraction and lens mass are free parameters. From
that, the contributions of self-lensing and halo-lensing is obtained.
The self-lensing predictions (minimum lensing that has to occur due to
star-star lensing) serve as a sanity check for observations and
models. An excess of lensing relative to self-lensing can then be
attributed to halo lensing, from which the MACHO parameters are
finally inferred.

\cite{1986ApJ...304....1P} was the first to present such a lensing
model for the Galaxy halo and to estimate the probability of lensing
(i.e., a magnification larger than 1.34) taking place at any time. This
probability is also called the microlensing optical depth. On the
basis of this work \citet{1991ApJ...366..412G} evaluated the optical
depth with more realistic assumptions on halo density and velocity
structure. He also obtained the event rate and distributions for
lensing timescales and amplifications.  \cite{1995ApJ...449...28A}
related the Einstein timescale distribution of the events to the
microlensing rate and optical depth. They evaluated these
distributions for several axisymmetric disk-halo models in the
framework of the MACHO project.

Any microlensing light curve can be characterized by the maximum
magnification, the time to cross the Einstein radius (Einstein time)
and the time of the event.  The first two observables depend on the
line-of-sight distance of the source and lens, the minimum projected
transverse lens-source distance (impact parameter), transverse
lens-source velocity, and lens mass.  These quantities therefore
cannot be extracted separately from an individual lensing event;
instead, one can only derive probability distributions for them (see
\citeauthor{1991MNRAS.250..348D} \citeyear{1991MNRAS.250..348D} and
\citeauthor{1998A&A...330..963D} \citeyear{1998A&A...330..963D}).
Most interesting are of course the object masses responsible for the
measured lensing light curves: \cite{1994PhLB..323..347J} have derived
the lens mass probability function for an event with given Einstein
time and amplification.  \cite{1996ApJ...467..540H} have determined
the MACHO mass spectrum from 51 MACHO candidates using their observed
Einstein times.

Blending has proven to be a severe limitation in the analysis of
microlensing events. It can be overcome partly by using low-noise,
high spatial resolution {\it Hubble Space Telescope} ({\it HST})
images for measurements of the unlensed source fluxes (see
\citeauthor{2001ApJ...552..582A} \citeyear{2001ApJ...552..582A}).  For
extragalactic objects, however, this can provide a precise source flux
for a fraction of lensed stars only.

One can also use an advanced technique called difference imaging
analysis, which is insensitive to crowding and allows to measure pixel
flux differences in highly crowded fields at the Poisson noise level.
Therefore, lensing searches could be extended to more distant targets
like M31 (AGAPE, \citeauthor{1999A&A...344L..49A}
\citeyear{1999A&A...344L..49A}; Columbia-VATT,
\citeauthor{1996ApJ...473L..87C} \citeyear{1996ApJ...473L..87C};
WeCAPP, \citeauthor{2003ApJ...599L..17R}
\citeyear{2001A&A...379..362R,2003ApJ...599L..17R}; POINT-AGAPE,
\citeauthor{2003A&A...405...15P} \citeyear{2003A&A...405...15P},
\citeauthor{2005A&A...443..911C} \citeyear{2005A&A...443..911C}; MEGA,
\citeauthor{2004A&A...417..461D} \citeyear{2004A&A...417..461D};
SLOTT-AGAPE, \citeauthor{2000MmSAI..71.1113B}
\citeyear{2000MmSAI..71.1113B}, \citeauthor{2003A&A...405..851C}
\citeyear{2003A&A...405..851C}; NMS, \citeauthor{2001BASI...29..531J}
\citeyear{2001BASI...29..531J}), or M87 \citep{2004ApJ...610..691B}.

\cite{1996ApJ...470..201G} called microlensing of unresolved sources
``pixel-lensing''.  This definition encompasses surveys at the
crowding limit as well as extragalactic microlensing experiments
(e.g., toward M31 or M87) where hundreds of stars contribute to the
flux within 1 pixel.  Gould (and also \cite{1997A&A...324..843A})
showed that the comparison of pixel fluxes at different epochs can
extend the search for microlensing events up to distances of a few
megaparsecs.  Applying his equations \cite{1996ApJ...470..201G},
\cite{1996ApJ...472..108H} and \cite{1996ApJ...473..230H} obtained the
optical depth and distributions of timescales and event rates for a
pixel-lensing survey toward M31.  If one does not know the flux of
the unlensed source accurately (i.e., if one is not in the classical
microlensing regime anymore), the information that can be extracted
from light curves is reduced.

\cite{1997ApJ...487...55W} were the first to note that the light curve
maximum does not provide the maximum magnification of the source
anymore, and, second, one cannot obtain the Einstein time from the
FWHM time of an event (since the latter is a product of the Einstein
time and a function of the magnification at maximum).  This initiated
efforts to deal with the lacking knowledge of the Einstein timescales
in the pixel-lensing regime (see
\cite{1999ApJ...510L..29G,2001MNRAS.320..341A}) and the suggestion to
extract the Einstein time using the width of the ``tails'' of the
lensing light curves by \cite{2000ApJ...530..578B} and
\cite{1996ApJ...470..201G}.

However, it is more straightforward to compare quantities that one can
easily measure in an experiment with model predictions for the same
quantity.  The two independent and most precisely measurable
observables are the flux excess of the light curve at its maximum and
its FWHM timescale.  \cite{2000ApJ...530..578B} followed that strategy
and derived the event rate as a function of the FWHM timescale of the
events.  We proceed in that direction and calculate the contributions
to the event rate as a function of the event's FWHM time and maximum
excess flux, because both the excess flux and timescale determine the
event's detectability.

The definition of Gould for pixel-lensing may imply that a
pixel-lensing event should be called a microlensing event, if its
source has been resolved (e.g., with {\it HST} images) after the event
has been identified from ground. Analogously, one could feel forced to
call a microlensing event a pixel-lensing event, once it has turned
out that ``the source star'' is a blend of several stars, and
therefore the source flux is not known. Therefore, the classification
of an event as a pixel-lensing event or a microlensing event is not
unique.

One can take the following viewpoint: the physical processes are the same, and
therefore classical microlensing is a special case of pixel-lensing, in which
the source flux probability distribution is much more narrow than the stellar
luminosity function, i.e., the distribution function used in the pixel-lensing
regime.  The two methods only differ in how to analyze a light curve and how
to derive the probability distribution for the source flux: One can make use
of a noisy and potentially biased baseline value of the light curve (hence,
stay in the classical microlensing regime), or ignore the baseline value and
obtain a source flux estimate from the wings of the light curve (analyze the
difference light curve).  Other possibilities are to obtain the source flux
from an additional, direct measurement or to constrain its distribution by
theory.  After having determined the source flux probability distribution by
one of these methods, one can use the formalism described in this paper to
derive, e.g., the lens mass probability function.

Our paper is organized as follows: We introduce our notation for the
microlensing and pixel-lensing regime in \S~\ref{sec.definitions}. We also
describe the treatment of finite source effects and how to extract the
observables from the light curves. In \S~\ref{sec.prob_lens} we combine the
probability distributions for location, mass, source-lens velocity and impact
parameter distribution to obtain the lensing event rate distribution as a
function of these parameters.  Section~\ref{sec.source_dis} summarizes the
statistical properties of the source populations, i.e., luminosity function,
number density, color-magnitude and luminosity-radius relation.  In
\S~\ref{sec.app} we calculate the optical depth and the observables in the
microlensing regime: single-star event rate, amplification distribution of the
events, Einstein timescale distribution, and FWHM distribution of the events.
Section~\ref{sec.app_source} deals with the pixel-lensing regime.  We calculate
the event rate as a function of the maximum excess flux and FWHM time (and
color) of the event in the point-source approximation.  We also show how the
event rate changes, if source sizes (shifting events to larger timescales and
smaller flux excesses) are taken into account.  In \S~\ref{sec.exp} we obtain
the event rate for pixel-lensing surveys with spatially varying photon noise
(related to the surface brightness contours of M31) but fixed signal-to-noise
threshold for the excess flux at maximum magnification.  We predict the number
of halo- and self-lensing events in the WeCAPP survey (without taking into
account the sampling efficiency of the survey) for the M31 model presented in
\S~\ref{app.model}. We demonstrate that accounting for the minimum FWHM
of the events is extremely important to correctly predict the number of events
and the luminosity distribution of the lensed sources. We also compare the
characteristics of self-lensing events with halo-lensing events.  Finally
\S~\ref{sec.massprob} derives the lens mass probability distribution from the
observables and errors as obtained from light curve fits.  The paper is
summarized in \S~\ref{sec.outlook}.  In Appendix~\ref{standard_vs_newdef} we
motivate an alternative event definition.  In Appendix~\ref{app.model} we
describe and construct ingredients of the M31 lens model, which we use
throughout the paper to calculate examples and applications.

\section{Basics of Lensing by a Point Mass}
\label{sec.definitions}

In this section we summarize the basics of microlensing theory and introduce
our notation. The change in flux $\DF(t)$ caused by a microlensing event
depends on the unlensed flux $\F$ and the magnification $A(t)$:
\begin{equation}
  \DF(t) := \F \,\left[ A(t) - 1 \right]    .
 \label{eq.lc_diff}
\end{equation}
For a pointlike deflector and a pointlike source moving with constant
relative transversal velocity $\vt$, the amplification is symmetric
around its time of maximum $t_0$ and is connected to the Einstein
radius $\RE$ and the impact parameter $b$ as follows
\citep{1986ApJ...304....1P}:
\begin{equation}
  A(u(t)) = \frac{u^2+2}{u \sqrt{u^2+4}}  
  \stackrel{u \ll 1}{\approx} \frac{1}{u}   , 
  \label{eq.pac_LC}
\end{equation}
\begin{equation}
  u(r(t)) := \frac{r(t)}{\RE} := \sqrt{\frac{\vt^2 (t-t_0)^2 + b^2}{\RE^2}}     ,
\end{equation}
\begin{equation}
  \RE :=\frac{\sqrt{4G\ml}}{c} \sqrt{\frac{\Dol (\Dos-\Dol)}{\Dos}}   ,
\label{eq.pac_re}
\end{equation}
where $\ml$ is the mass of the lens, $\Dol$ and $\Dos$ are
the distances to the lens, and $r(t)$ is the distance between source and 
lens in the lens plane.

With the Einstein timescale\footnote{\citet{1991ApJ...366..412G}
  defines the event duration $t_e$ as the time span where the lens is
  closer than a relative impact parameter $\uT$ to the source. This
  can be converted to the Einstein timescale using $t_e \equiv
  2\tE\sqrt{\uT^2-\uub^2}$, $\cos\theta \equiv
  {\left(1-\uub^2/\uT^2\right)}^{1/2}$.  \citet{2000ApJ...530..578B}
  use a different definition for the Einstein timescale $\tE$; for
  comparison use $\tE \equiv 2 \tE$ in their formulas.}  $\tE
:=\frac{\RE}{\vt}$ and the normalized impact parameter $\uub :=
\frac{b}{\RE}$ we obtain
\begin{equation}
  u(t) = \sqrt{\frac{(t-t_0)^2}{\tE^2} + {\uub}^2} .
  \label{eq.pac_u}
\end{equation}
The maximum amplification (at $t=t_0$) becomes
\begin{equation}
  \A := \frac{\uub^2+2}{\uub \sqrt{\uub^2+4}} \stackrel{\uub \ll 1}{\approx} \frac{1}{\uub}
  .
  \label{eq.A0_u0}
\end{equation}
Equation~(\ref{eq.pac_LC}) can be inverted to 
\begin{equation}
 u(A) = \left[2A(A^2-1)^{-1/2}-2\right]^{1/2} \stackrel{A \gg 1}{\approx}  \frac{1}{A}
  . 
 \label{eq.u0_A0}
\end{equation}
Inserting $\A$ in equation (\ref{eq.u0_A0}) its derivative can be written as
\begin{equation}
    \frac{d\uub}{d\A} 
    = - \frac{2\left[(\A^2-1)^{-1/2}-1/2 \A(\A^2-1)^{-3/2}2\A\right]}
              {2\left[2\A(\A^2-1)^{-1/2}-2\right]^{1/2}} 
     = - \left\{2 \left[\frac{\A}{(\A^2-1)^{1/2}}-1\right] (\A^2-1)^3\right\}^{-1/2} 
     = \frac{-\sqrt{2}}{2}\frac{\left[\A+(\A^2-1)^{1/2}\right]^{1/2}}
        {(\A^2-1)^{5/4}}.
 \label{eq.du0dA0}
\end{equation}
The FWHM timescale $\tfwhm$ of a light curve
is defined by $A{\left(\frac{\tfwhm}{2}\right)}-1 := \frac{\A-1}{2}$. 
It is related to the Einstein timescale $\tE$ by
\begin{equation}
  \tfwhm  = \tE \, w(\uub) =  \tE \Upsilon(\A) 
  ,
\label{eq.t12_tE}
\end{equation}
where $w(\uub)$ was first obtained by
\cite{1999ApJ...510L..29G}\footnote{With $\beta \equiv \uub$ and
  $\delta(\beta) \equiv \A-1 .$}:
\begin{equation}
  \darray
\begin{array}{ll}
   w(\uub) & 
:= 2 \sqrt{ u{\left(\frac{A(\uub)+1}{2}\right)^2 - {\uub}^2} }
     =  2 \sqrt{\frac{2[A(\uub)+1]}{\sqrt{[A(\uub)-1][A(\uub)+3]}}-2 - {\uub}^2} 
     \stackrel{\uub \ll 1}{\approx} \sqrt{12}\, \uub      ,
\end{array}
  \label{eq.w_u0}
\end{equation}
and $\Upsilon(\A):= w(u(\A))$ is
\begin{equation}
  \darray
\begin{array}{ll}
 \Upsilon(\A) & 
= 2\sqrt{ u{\left(\frac{\A+1}{2}\right)^2- {u(\A)}^2}} 
=  \sqrt 8 \,\frac{\left[(\A+1)^{3/2}-\A(\A+3)^{1/2}\right]^{1/2}}{[(\A-1)(\A+1)(\A+3)]^{1/4}} \stackrel{\A \gg 1}{\approx} \frac{\sqrt{12}}{\A}
  .
\end{array}
  \label{eq.Upsilon_A0}
\end{equation}
Hence, the easy measurable timescale $\tfwhm$ is a product of the
quantity $\tE$, which contains the physical information about the lens, and 
the magnification of the source at maximum light $\A$.

\subsection{Finite Source Effects}
\label{sec.def_fs}

If the impact parameter of a source-lens system becomes comparable to the
source radius projected on the lens plane $\Rstar\frac{\Dol}{\Dos}$, the
point-source approximation is not valid anymore. The amplification then
saturates at a level below the maximum magnification in equation
(\ref{eq.A0_u0}).

The finite source light curve for extended sources can be derived for a
disk-like homogeneously radiating source,

\begin{equation}
  \darray
  \begin{array}{rl}
  \Afs(u) 
& =
\Int_0^{\frac{\Rstar\Dol}{\Dos}} \,
\Int_0^{2\pi} 
A{\left(\left(u^2+\frac{\tilde{r}^2}{\RE^2}-2 u \frac{\tilde{r}}{\RE} \cos \theta \right)^{1/2}\right)} 
\frac{\tilde{r}\,d\theta \;d\tilde{r}}{\pi \left(\Rstar\frac{\Dol}{\Dos}\right)^2} 
\\
& = \frac{1}{\pi} \left(\frac{\RE \,\Dos}{\Rstar \,\Dol}\right)^2 
\Int_0^{2\pi} \,
\Int_0^{\frac{\Rstar\Dol}{\RE\Dos}} 
A{\left(\left(u^2+\tilde{u}^2-2\tilde{u}\cos \theta \right)^{1/2}\right)} \, \tilde{u}\,d\tilde{u} \,d\theta
\\
& =: 
\frac{z^2}{\pi}  
\Int_0^{2\pi} \,
\Int_0^{1/z} 
A{\left(u \left( 1+q^2-2 q \,\cos \theta \right)^{1/2}\right)} 
\, q \, dq \,d\theta
\\
& =
\frac{z^2}{\pi}  
\Int_0^{2\pi} \,
\Int_0^{1/z} 
\frac{\left( 1+q^2-2 q \,\cos \theta \right)+2/u^2}{\sqrt{ \left( 1+q^2-2 q \,\cos \theta \right)^2+4/u^2 \left(1+q^2-2 q \,\cos \theta\right)}} \, q \, dq \,d\theta
\end{array}
\quad   ,
\label{eq.A_fin}
\end{equation}
with source-lens separation $r(t)$ and where the definitions 
\begin{displaymath}
  z(t):=u(t) \frac{\RE
    \,\Dos}{\Rstar \,\Dol} =\frac{r(t)}{\Rstar} \frac{\Dos}{\Dol} 
\end{displaymath}
and
$q:=\frac{\tilde{u}}{u}$ have been inserted. For high magnifications, where
$A(u)\approx u^{-1}$ is a valid approximation, equation (\ref{eq.A_fin})
becomes equivalent to \citet[eq.~(2.5)]{1994ApJ...421L..71G}.

The maximum amplification in the finite source regime then becomes
\begin{equation}
  \darray
  \begin{array}{rl}
  \ANfs & = \Int_0^{\Rstar\frac{\Dol}{\Dos}} A(r/\RE) \frac{2\pi r \;dr}{\pi \left(\Rstar\frac{\Dol}{\Dos}\right)^2} 
  = 2 \left(\frac{\RE \,\Dos}{\Rstar \,\Dol}\right)^2 \Int_0^{\frac{\Rstar\Dol}{\RE \Dos}} \frac{\tilde{u}^2+2}{\sqrt{\tilde{u}^2+4}} \,d\tilde{u} 
    = \sqrt{1+\left(\frac{2 \,\RE \,\Dos}{\Rstar \,\Dol}\right)^2}
  \stackrel{\ANfs\gg 1}{\approx}  \frac{2 \,\RE \,\Dos}{\Rstar \,\Dol}
  ,
\end{array}
  \label{eq.A0_fin}
\end{equation}
which equals the approximation of \citet[eq.~(19)]{2000ApJ...530..578B} 
for high amplifications.

\begin{figure}  
  \epsscale{0.5}\plotone{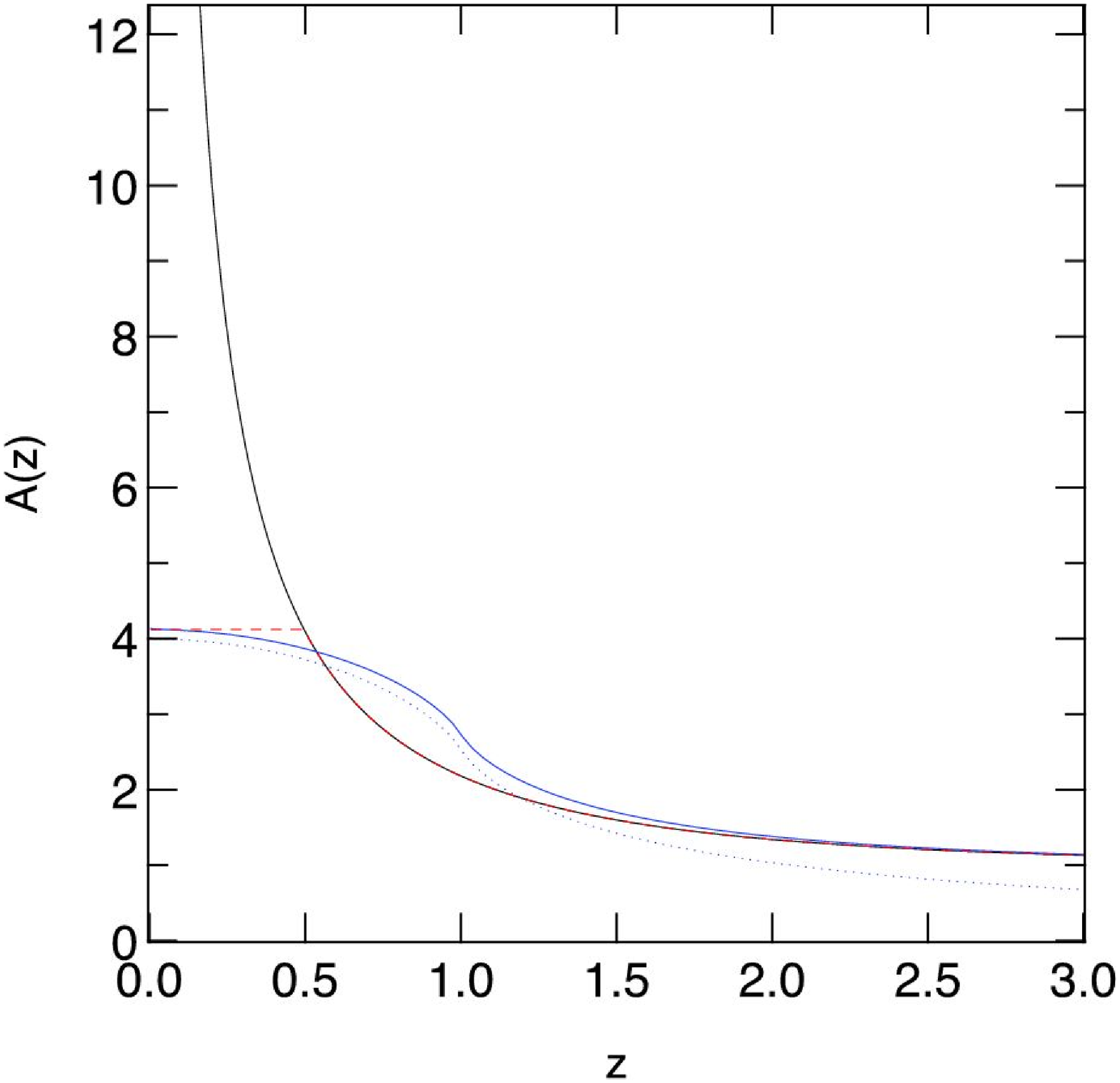}
  \caption{Amplification $A(z)$ versus
    $z(t):=\left[r(t)/\Rstar\right]\left(\Dos/\Dol\right)$ plotted for
    $\Rstar/\Dos = 0.5 \left(\RE/\Dol\right)$. 
    {\it Black curve:} point-source approximation, see equation (\ref{eq.pac_LC}). 
    {\it Blue curve:} 
		 finite source magnification  $\Afs(z)$ for a 
		 homogeneously radiating disk of size $\Rstar$, exact solution, 
     see equation (\ref{eq.A_fin}).
    {\it Red dashed curve:} 
		 simple approximation $\Afs(z)$ for finite source effects 
     according to equation (\ref{eq.Afin}). 
    {\it Blue dots:} 
		 finite source size approximation in the high-magnification regime, 
     introduced by \cite[eq. (2.5)]{1994ApJ...421L..71G}.
  }
  \label{fig.fs_Az}
\end{figure}

For small source-lens distances with $\Dol\approx\Dos$ (e.g., for
bulge-bulge self-lensing) the above relation becomes 
\begin{displaymath}
\ANfs \approx
\sqrt{1+1.5\times10^6\,\frac{\ml}{\ml_\odot}
\left(\frac{\Rstar}{R_\odot}\right)^{-2} \frac{\Dos-\Dol}{1\E{kpc}}}.
\end{displaymath}
For a source radius of supergiants of $\Rstar \approx 200 R_\odot$ a
source-lens distance of $1\E{kpc}$, and a lens with $\ml=1\ml_\odot$
finite source effects already arise above a magnification of $\ANfs
\approx 6.2$.  For smaller masses $\ml=0.1\ml_\odot$ finite source
effects become important even at a low magnification $\ANfs \approx
2$.  Although typical source radii are smaller, this example shows
that finite-source effects cannot be neglected.  We will show in
\S~\ref{ssec.finite_source2} and Table~\ref{tab.Gammatot_WeCAPP} that
indeed a large fraction of the M31 bulge-bulge lensing events will
show finite source effects. 

Figure~\ref{fig.fs_Az} shows that for $u<\ufs$ (or $z\lsim\frac{1}{2}$) with 
\begin{equation}
  \ufs :=  u(\ANfs) = 
  \,\left\{2\left[1+\left(\frac{\Rstar \,\Dol}{2 \,\RE \,\Dos}\right)^2\right]^{1/2}-2\right\}^{1/2} 
 \stackrel{\ANfs\gg 1}{\approx} \frac{\Rstar\Dol}{2\RE\Dos}   , 
  \label{eq.fs}
\end{equation}
the amplification is no longer directly connected
to the source-lens separation \citep{1995ApJ...455...44G}, 
but all $u<\ufs$
have nearly the same amplification equal to the point-source approximation $A(u)$ at 
$\ufs$.
Therefore we generalize equation (\ref{eq.pac_LC}) to approximately account for 
finite-source effects

\begin{equation}
  \Afs (u) \approx \left\{
  \begin{array}{ll}
    \sqrt{1+\left(\frac{2 \,\RE \,\Dos}{\Rstar \,\Dol}\right)^2}, &    u <   \ufs \\
    \frac{u^2+2}{u \sqrt{u^2+4}},                                 &    u \ge \ufs  .\\
  \end{array}
  \right.
 \label{eq.Afin}
\end{equation}
For light curves with finite source signatures ($u_0<\ufs$) at an impact parameter \\
\begin{displaymath}
  u{\left(1+(\ANfs-1)/2\right)} \stackrel{\ANfs\gg 1}{\approx}
  (\Rstar \; \Dol)/(\RE \; \Dos)
\end{displaymath}
(or $z\approx 1$) the amplification of our approximation is half of
the maximum and can be used to define the $\tfsFWHM$:
\begin{equation}
  \darray
  \begin{array}{rcl}
    \tfsFWHM & := & \tE \Upsfs :=
    \frac{2\RE}{\vt}\sqrt{\uFStfwhm^2-\uub^2} \stackrel{\ANfs\gg
    1}{\approx} 2 \tE \; \sqrt { \left(\frac{2}{\ANfs}\right)^2 -
    \uub^2 } \\ & \approx & 2 \tE \; \sqrt { \left( \frac{\Rstar \;
    \Dol} {\RE \; \Dos} \right)^2 - \uub^2 } \approx \tfwhm \;
    \frac{1}{\sqrt{3}} \sqrt { \left( \frac{\Rstar \; \Dol} { b \;
    \Dos} \right)^2 -1 } \ge \tfwhm \quad \uub < \ufs .
  \end{array}
 \label{eq.t12_fin}
\end{equation}
with 
\begin{displaymath}
\Upsfs(\uub,\Rstar,\Dol,\Dos,\ml) := 2\sqrt{\uFStfwhm^2-\uub^2}
= 2 \sqrt{ \frac{2 (\ANfs+1)}{\sqrt{ (\ANfs-1)(\ANfs+3)}} - 2 - \uub^2}.
\end{displaymath}
In equation (\ref{eq.t12_fin}) the FWHM timescales for light curves
that show finite source signatures are related to the values $\tfwhm $
for the point-source approximation using equations (\ref{eq.t12_tE}) and
(\ref{eq.w_u0}).  This demonstrates that the source does affect the
timescale of an event severely: a source with an impact parameter of
one-tenth the projected source radius will have an event timescale
almost 6 times as long as that in the point-source approximation.

The shortest and longest FWHM timescales for an event with finite
source signature ($\uub\le\ufs$) are equal (insert $\uub=\ufs$ and
$\uub=0$ into eq.~[\ref{eq.t12_fin}]),
\begin{equation}
  \darray
  \begin{array}{l}
  \tfsFWHMmin = \tE \Upsilon(\ANfs)  \stackrel{\ANfs\gg 1}{\approx}  \sqrt{3} \; \frac{\Rstar\,\Dol}{\vt \; \Dos}  , \\
  \tfsFWHMmax  = 2 \tE \,\uFStfwhm  \stackrel{\ANfs\gg 1}{\approx}  2 \; \frac{\Rstar\,\Dol}{\vt \; \Dos}  .
  \end{array}
\label{min_max_tfwhm_fs}
\end{equation}
For a given transversal velocity the minimum timescale 
becomes the larger, the larger the source sizes 
are. 

The largest flux excess of a lensed, extended star becomes
\begin{equation}
  \DFmax = F_0 (\ANfs -1) = F_0 \left[\sqrt{1+\frac{16 G \ml\,\Dos
  (\Dos-\Dol)}{c^2\,\Rstar^2 \,\Dol}} -1\right] \stackrel{\ANfs\gg
  1}{\approx} \frac{4\sqrt{G}}{c}
  \,\sqrt{\frac{\Dos(\Dos-\Dol)}{\Dol}} \;
  {\sqrt{\ml}}\;\frac{F_0}{\Rstar} ,
\label{eq.largest_deltaF}
\end{equation}
irrespective of whether the light curve shows finite 
source signatures.

\subsection{Extracting Observables from Light Curves}
\label{sec.lightcurves}
 
\subsubsection{Measuring $\DF$ and $\tfwhm$}

In this section we present three methods for measuring the excess flux
$\DF$ at maximum and the FWHM time $\tfwhm$.  One can see in equations
(\ref{eq.t12_tE}) and (\ref{eq.t12_fin}) that $\tE$ and $u_0$ (or
$A_0$) enter the value of $\tfwhm$ as a product, giving rise to the
``Einstein time magnification'' degeneracy, which may lead to poor
error estimates for $\tE$ (and $u_0$) even for well-determined values
of $\tfwhm$ and $\DF$.

Accounting for this degeneracy,
\citet{1996ApJ...470..201G}\footnote{
\citeauthor{1996ApJ...470..201G}'s (\citeyear{1996ApJ...470..201G})
 eq. (2.4) with
  $\beta \equiv \uub$, $\omega \equiv \tE^{-1} .$}
approximated the Paczynski light curve with one fewer parameter for
the special case of high amplification:
\begin{equation}
  \Delta_F^\M{Gould}(t) \approx \Feff
  \left[\frac{(t-t_0)^2}{\teff^2} + 1\right]^{-1/2}.
  \label{eq.gouLC}
\end{equation}
The three free parameters are $\Feff := \frac{\F}{\uub}$, $\teff :=
\uub\,\tE$, and $t_0$.  This approximation has turned out to be a very
useful filter for detecting lensing events; however, it fails to describe
light curves when the magnification is not very large.
We suggest using

\begin{equation}
  \DF(t) \approx \DF \left[\frac{12 (t-t_0)^2}{\tfwhm^2} + 1\right]^{-1/2}
  \label{eq.modgouLC}
\end{equation}
instead. 
This approximation provides a good description also for lower magnifications. 
The three free parameters of this approximation are the time of maximum $t_0$, 
the excess flux $\DF$, and the FWHM timescale $\tfwhm$.

\begin{figure}  
  \epsscale{0.7}\plotone{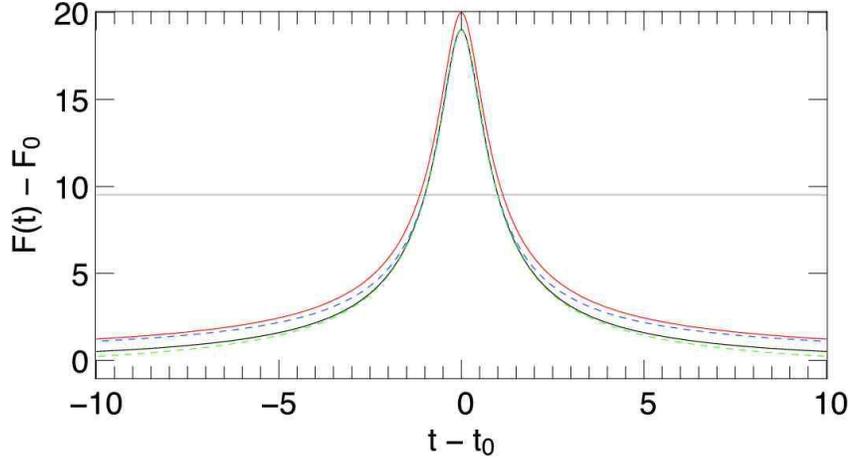} 
  \caption{Different light curve approximations using the
following parameters: $\tfwhm=2$, $\A=20$, $F_0=1$, 
    $\uub=0.05005$, and $\tE=12.28$.    
   {\it Black curve:} Paczynski (eqs.~[\ref{eq.lc_diff}] and [\ref{eq.pac_LC}]).
   {\it Red curve:} Gould (eq. [\ref{eq.gouLC}]). 
   {\it Blue dashed curve:} Eq. [\ref{eq.modgouLC}].
   {\it Green dashed curve:} Gould fit with additional free constant (eq. [\ref{eq.newLC0}]).
   {\it Gray line:} $\DF/2$ marks the flux level where $\tfwhm$ is defined
   for the Paczynski curve.
  }
  \label{fig.LC_theoretical}  
\end{figure}

Figure~\ref{fig.LC_theoretical} shows that equation (\ref{eq.modgouLC}) 
better approximates the Paczynski light curve than the Gould approximation
in the core and in the inner part of the
wings, and also provides the correct value for $\tfwhm$ and $\DF$.

There are two situations that can require a fourth, additive, free
parameter in the light curve fit.  The first one is the transition
regime from pixel-lensing to microlensing (i.e., where the errors are
small enough to sample the wings of the light curve).  We suggest
using
\begin{equation}
  \DF(t) \approx \Feff \left[\frac{(t-t_0)^2}{\teff^2}  + 1\right]^{-1/2} - \F   ,
  \label{eq.newLC0}
\end{equation}
which provides an excellent fit to the Paczynski light curve (see
Fig.~\ref{fig.LC_theoretical}, {\it green curve}).  

The second situation is the following: imagine that the photon noise
of the background becoming larger and finally exceeding the unlensed
flux of the star $\F$. Then the star cannot be resolved anymore and
the rms error of the baseline of the light curve becomes proportional
$N_\M{data\ points}^{-1/2} \,\sigma$.  The (minimum) systematic error is
given by the fact that the subtracted reference image (with error
$\sigma_\M{ref}$) is a sum of (high-quality) images, potentially
including some of the amplified phases of the sources.\footnote{ The
  measured light curve varies around the theoretical light curve
  $L(t)$ due to noise of amplitude $\sigma$. We can therefore write
  the light curve measured at times $t$ as
\begin{displaymath}
L(t) + \sigma := [\DF(t) + B + \sigma] 
                     - [\DF(t_\M{ref}) + B + \sigma_\M{ref}] 
     = \DF(t) - \DF(t_\M{ref}) - \sigma_\M{ref} + \sigma 
     = \DF(t) + \sigma + C',
\end{displaymath}
with the background $B$, the epoch for the reference measurement
$t_\M{ref}$, and a constant $C':= - \DF(t_\M{ref}) - \sigma_\M{ref}$.
} 
This implies that there are fundamental limits to the accuracy of the
baseline, and we thus require an additive parameter to account for
that.  The approximation of any pixel-lensing light curve then becomes
\begin{equation}
  \DF(t) \approx \Feff \left[\frac{(t-t_0)^2}{\teff^2}  + 1\right]^{-1/2} + C   .
  \label{eq.newLC}
\end{equation}
In fact, numerical simulations showed that much more accurate values
are derived for $\Feff$ and $\teff$ if this additional constant $C$ is
allowed for.

\subsubsection{Constraining $\F$}

In this section we address the important question, how to extract the
source flux $\F$ from a lensing light curve.  There are four potential
ways to constrain the flux of the lensed star:
\begin{enumerate}
\item The lensed star is resolved and isolated, and therefore a bias in
the flux measurement (by crowding) can be excluded (assuming no
systematic effects in the baseline). One would of course call such an
event a classical microlensing event.  A microlensing fit (using
$\chi^2$ analysis methods) to the light curve then directly provides
$\F$ and its probability distribution, ideally given by an Gaussian
error $\sigma_\F$.  In this case the flux measurement error is
directly correlated to $Q$, the signal-to-noise ratio at maximum
magnification.\footnote{If the noise is dominated by background sky,
  one can write $\sigma_\F = \DF \Q^{-1} N_\M{data\ points}^{-1/2}$,
  where $N_\M{data\ points}$ is the number of the light curve data
  points.}
\item The flux $\F$ is obtained through the information that is in the
shape of the wings of the difference light curve $\F(A(t)-1) + C
\equiv \F A(t) + B$.  The $\chi^2$ analysis leads to a probability
distribution for $\F$.  This flux estimate method is used if no
alternative unbiased flux measurement is available, i.e., cases in
which the source star is resolved but blended (see
\citep{1999A&A...343...10A} for applications in the microlensing
regime), and cases in which the source star is not resolved (usually
called a pixel-lensing event).  Note that other methods using the
shape of the wings \citep{2000ApJ...530..578B} provide similar
results.
\item  The flux $\F$ is obtained from an additional, direct measurement,
e.g.,  low-noise, high spatial resolution photometry from space.
\item  The flux $\F$ is constrained by theory through plausible
distribution functions, e.g., the luminosity function $\Phi$, the
color-magnitude relation of stars, and the distance distribution of
stars, which together yield the source flux distribution function (see
\S~\ref{sec.masstfwhmDF}).  Another constraining example is an upper
source flux limit that can be obtained from the fact that the source
star is not resolved in the absence of lensing.
\end{enumerate}
Since the physical processes are the same in pixel-lensing and
microlensing, microlensing is a special case of pixel-lensing, where
the source flux probability distribution is much more narrow than the
stellar luminosity function, i.e., the distribution function used in
the pixel-lensing regime.  The methods only differ in how to analyze a
light curve and how to derive the probability distribution for the
source flux.

\subsubsection{Evaluating $\tE$}

In this section we use the distribution of $\F$ (from measurement or theory;
see previous section) to estimate the probability distribution for a value of
$\tE$.  Note that transforming the distribution of $\F$ to a distribution of
$\tE$ can lead to a different value compared to a $\tE$ obtained directly from
the best estimate for $\F$.

As the fitting process in the light curve analysis yields the non degenerate
observables $\tfwhm$ and $\DF$, we can combine their (Gaussian) measurement
errors with the probability distribution for the source flux $\F$ and obtain
the probablity distribution for $\tE$:
     
\begin{equation}
  \darray
  \begin{array}{ll}
    \ptE(\tE) & =
    \int \int \int 
    \,p_\tfwhm(\tfwhm)
    \,p_\DF(\DF)
    \,p_\F(\F)                      ) 
    \,\delta{\left(\tE -  \frac{\tfwhm}{\Upsilon}\right)}       
    \,d\tfwhm \,d\DF \,d\F 
    \\
    & = 
    \int \int \int 
    \,p_\tfwhm(\tfwhm)
    \,p_\DF(\DF)
    \,p_\F(\F)
    \,\frac{\delta{\left(\tfwhm -  \tE \Upsilon \right)}}{\left|\Upsilon^{-1}\right|}
    \,d\tfwhm \,d\DF \,d\F
    \\
    & = 
    \int \int 
    \,p_\tfwhm(\tE \Upsilon )
    \,p_\DF(\DF)
    \,p_\F(\F)
    \Upsilon{\left(\frac{\DF}{\F}+1\right)} 
    \,d\DF \,d\F
     .                      
  \end{array}
  \label{eq.micropixel-lensing}
\end{equation}
This also allows to include non-Gaussian distributions for the source
flux.

By transforming the measurements of $\DF$ and $\tfwhm$ together with a
probability distribution of $\F$, we derive a general formalism that
is applicable to all microlensing and pixel-lensing problems.  In
\S~\ref{sec.massprob} we further develop this idea using plausible
distribution functions as physical constraints, which narrows the
width of the distribution of the lens mass $\ml$ (connected to $\tE$).

\section{Distribution Function for Lens Parameters}
\label{sec.prob_lens}

For a source of fixed intrinsic flux $\F$, position
$\vec{r}_\iss=(x,y,\Dos)$ and velocity vector
$\vsvec=(\vsx,\vsy,\vsz)$, the number and characteristics of lensing
events are determined by the probability function $
p(\vec{r}_l,\vlvec,\ml)$ for a lens with mass $\ml$ and velocity
$\vlvec$ being at position $\vec{r}_l$. For the change of
magnification of the background source, only the transversal velocity
components of source and lens are relevant (we assume velocities to be
constant).  For parallax microlensing events
\citep{1994ApJ...421L..75G,1994ApJ...421L..71G} the nonuniform
velocity of the observer changes the observed light curves, since the
observer's reference frame is not fixed.  However, this effect is
unimportant for extragalactic microlensing events.

Therefore, in addition to $\ml$ and $\Dol$ only the projected relative
transversal positions $r:=r_{\M{t},l}-\frac{\Dol}{\Dos} r_{\M{t},s}$
and velocities $\vt:=v_{\M{t},l}-\frac{\Dol}{\Dos} v_{\M{t},s}$ and
the angle $\ang$ enclosed by relative position and velocity vector
enter the lensing properties.  The distributions in $r$ and $\ang$ can
be reduced to the distribution of one parameter, the impact parameter
$b$ of the lens-source trajectory.  This is obvious, since in a
symmetric potential the trajectory of a particle is fully described by
its minimum distance.

So, the relevant lens parameters are $\Dol$, $\vt$, $\ml$, and $b$.
We introduce the lens density and the distributions of $\Dol$, $\vt$,
and $\ml$ in the next two subsections and then come up with a new
lensing event definition in \S~3.3. For those lenses that satisfy
the event definition, i.e., those which cause events, we will then
derive the distribution of the impact parameters $dN/db$.  We will
show that our event definition gives the familiar relation for the
event rate but is more easy to implement in numerical simulations.

\subsection{Distance and Mass Distribution}

The probability distributions for a lens with mass $\ml$ being at
distance $\Dol$ are given by
\begin{equation}
  p_\Dol = \rho(\Dol) \,\left[ \Int_0^\Dos{\rho(\Dol) \,d\Dol} \right]^{-1} ,
\end{equation}
\begin{equation}
  p_\ml = \xi(\ml)  \,\left[  \Int_0^\infty {\xi(\ml) \,d\ml} \right]^{-1}
 ,
\end{equation}
where $\rho(\Dol)$ is the lens mass density and $\xi(\ml)$ is the lens
mass function (which itself is normalized to $\int \xi(\ml) \ml d\ml
=1$; see \citet[p. 747]{1987gady.book.....B}).  The number density per
lens mass interval finally is defined by
\begin{equation}
  n(\Dol,\ml) := \rho(\Dol) \,\xi(\ml) ,
\label{eq.numberdens}
\end{equation}
where $n(\Dol,\ml)$ has units of $\M{length}^{-3}\E{mass}^{-1}$.

\subsection{Velocity Distribution for Lenses}

We assume that the velocity distribution of the lenses around their
mean streaming velocity is Gaussian:
\begin{eqnarray*}
  p(\vli) & = & C_l \,e^{-\frac{\vl_{i}^2}{2\sigl^2}},\quad i=x,y,z,\quad C_l=\frac{1}{\sqrt{2\pi}\,\sigl} .
\label{isov}
\end{eqnarray*}
where $\sigl$ is the dispersion and depends on the position $(x,y,z)$.
We furthermore assume that the combined transverse motion of observer
and source relative to the mean transverse streaming velocity of the
lenses is known and occurs in the $x$-direction with amplitude
$v_0(x,y,z)$ {as projected onto the lens plane}. This means that the
velocity $v_s$ of the source turns into a projected velocity $v_p =
\Dol/\Dos \, v_s$ (lensing timescales are determined by relative
proper motions not absolute motions of lens and source).

We now define the relative projected velocity $\vlsx:=\vlx+v_0$
(analogously $\vlsy:=\vly+0$) and obtain the transverse lens-source
velocity distribution as\footnote{\label{calc} We extract the desired
  distribution functions using
\begin{displaymath}
   p(s) =
      \Int_{y_0}^{y_1} \Int_{x_0}^{x_1}
     p(x,y) \,\delta(s-s(x,y)) \,dx \,dy 
      =   \Int_{\tilde{y}_0}^{\tilde{y}_1} \Int_{x_0}^{x_1}
      p(x,y) \,\frac{\delta(x-x(s,y))}{\left|\frac{ds(x,y)}{dx}\right|} \,dx \,dy 
      =   \Int_{\tilde{y}_0}^{\tilde{y}_1}  
     p(x(s,y),y) \,\left|\left.\frac{ds(x,y)}{dx}\right|_{x=x(s,y)}\right|^{-1} \,dy.
\end{displaymath}
Note that if $\left.ds(x,y)/dx\right|_{x=x(s,y)}$ has a different
domain for $y$ than $f(x,y)$, the limits for $y$ have to change to
$\tilde{y}_0$ and $\tilde{y}_1.$ }
\begin{equation}
  \darray
\begin{array}{rcl}
  \pv(\vt,v_0) 
  &= & \frac{1}{2 \pi \sigl^2}  \Intinf\Intinf 
  \delta{\left(\vt-\sqrt{\vlsx^2+\vlsy^2}\right)}  
  \,\exp\left(-\frac{(\vlsx - v_0)^2}{2\sigl^2}\right)  
  \,\exp\left(-\frac{\vlsy^2}{2\sigl^2}\right) \,d\vlsy \,d\vlsx \\
  &= & \frac{1}{2 \pi \sigl^2} \Int_{-\vt}^{\vt}\Intinf   
  \frac{\delta{\left(\vlsy - \sqrt{\vt^2-\vlsx^2}\right)} + 
    \delta{\left(\vlsy + \sqrt{\vt^2-\vlsx^2}\right)}}  
  {\left|\vlsy/\sqrt{\vlsx^2+\vlsy^2}\right|}
  \,\exp\left(-\frac{(\vlsx - v_0)^2}{2\sigl^2}\right)   
  \,\exp\left(-\frac{\vlsy^2}{2\sigl^2}\right)  
  \,d\vlsy \,d\vlsx  \\  
  &= & \Int_{-\vt}^{\vt} \frac{1}{\pi \sigl^2} \vt   
  \,\exp\left(-\frac{\vt^2+v_0^2}{2\sigl^2}\right)   
  \,\frac{\exp\left(-\frac{-2v_0\vlsx}{2\sigl^2}\right)}  
  {\sqrt{\vt^2-\vlsx^2}} \,d\vlsx \\  
  &= &  \frac{1}{\sigl^2} \,\vt 
  \,\exp\left(-\frac{\vt^2+v_0^2}{2\sigl^2}\right) 
  I_{0}\left(\frac{v_0\vt}{\sigl^2}\right) .
\end{array}
\label{eq.vt_l} 
\end{equation}
Here the Bessel function $I_{0}$ stretches the distribution depending on $v_0$.

\subsection{Impact Parameter Distribution for Events}
\label{pb}

In \cite{1986ApJ...304....1P} definition for lensing events (hereafter
called standard definition) lens-source configurations become lensing
events if the magnification of a source rises above a given threshold
within the survey time interval $\Delta t$.  This means that for each
lens mass one can define a ``microlensing-tube'' along the
line-of-sight to the source, which separates the high-magnification
region from the low-magnification region, and a lens causes an event
if it enters the tube.

\begin{figure}             
  \epsscale{0.5}\plotone{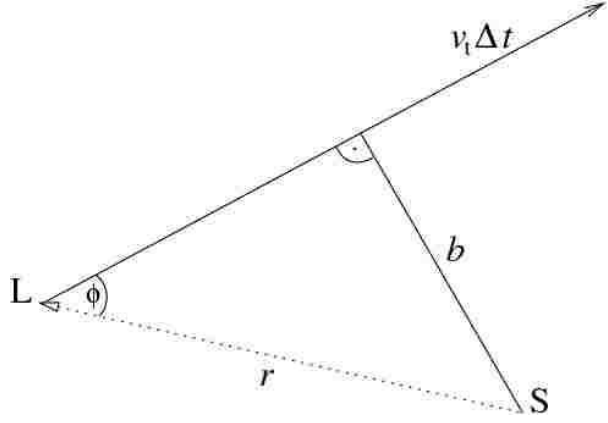}  
  \caption{For a projected lens-source separation $r$ and an angle
    $\phi$ between the projected distance vector and the projected
    relative velocity vector the impact parameter of the source-lens
    configuration is $b=r \sin (\phi) $. The lens approaches the
    source only for angles between $-\pi/2$ and $\pi/2$. }
  \label{b_max}                                           
\end{figure}                                            

We use (for the motivation, see \S~\ref{standard_vs_newdef}) an
alternative event definition: a lens-source configuration becomes an
event, if the lensing light curve reaches the maximum within the
survey time $\Delta t$. This definition does not specify any specific
magnification threshold at the time of maximum magnification because
this magnification threshold will in reality depend on the
observational setup and the brightness of the source.  We show that
the impact parameter distribution for the maximum light curve event
definition agrees with the standard definition, if the same
magnification threshold is used.

For simplicity we consider lenses with one mass, distance, and
velocity, for the moment only.  The lenses are homogeneously
distributed points (in two dimensions) with density $n$ and velocities
of $\vt$ (the velocities can have arbitrary directions, but the
angular distribution of the velocities must be the same for all the
points).  The number of {lenses} per radius interval around the
line-of-sight to the source is
\begin{equation} 
  \frac{dN}{dr}(r) =  n \,2 \pi r  .
  \label{eq.N0}
\end{equation}
If $r$ is the source-lens distance at the beginning of the survey and
$\phi$ is the angle that the lens's velocity vector encloses with the
lens-source vector at that time, then the configuration will become an
{event} with impact parameter $b$ if $ b \le r \le \sqrt{b^2 + (\vt
  \Delta t )^2 }$ and $ b = r |\sin (\phi)| $ with $ \phi \in
[-\pi/2,\pi/2]$ holds (see Figure~\ref{b_max} and
\S~\ref{standard_vs_newdef}). Therefore, $\frac{dN}{db}$ can be
derived from the spatial distribution of the lenses relative to the
source, $\frac{dN}{dr}$, and the distribution of the angles between
velocity vector and distance to the source. For the special case in
which all lenses have {isotropic} velocities of $\vt$, the probability
for the angle between radius vector and velocity vector is independent
of the location of the lens and equals

\begin{equation}
  \pphi(\ang) := \frac{1}{2 \pi}, \quad 0 \le \ang \le 2 \pi,  
\end{equation}
and
\begin{equation}
 \darray
\begin{array}{ll}  
  \frac{dN}{db} (b) &= 
      \Int_{b}^{\sqrt{ (\vt\Delta t)^2 + b^2 }}  \, 2  \Int_{0}^{\pi/2}  \, \frac{dN}{dr} \,\frac{1}{2\pi}
             \, \delta(b-r\sin\ang) \,d\ang \,dr  
        =  n \,   \Int_{b}^{\sqrt{ (\vt \Delta t )^2 + b^2 }} \, 2 
 \Int_{0}^{\pi/2} r \, \frac{\delta{\left(\ang-\arcsin\frac{b}{r}\right)}}{\sqrt{r^2-(r\sin\ang)^2}}    
         \,d\ang \,dr \\
& =  n \,   \Int_{b}^{\sqrt{ (\vt \Delta t )^2 + b^2 }} \, 
 \frac{2r }{\sqrt{r^2-b^2}}     \,dr = 2 n \, \vt \Delta t 
 .
\end{array}
 \label{eq.pb_b_max}
\end{equation}
In this equation, the radial integration limits correspond to the
minimum and maximum source-lens separation for an event with impact
parameter $b$ within $\Delta t$, and the $\delta$-function then allows
only for those trajectories through $r$ that have the correct angle
$\phi$ for the impact parameter $b=r \sin (\phi)$ of interest.  The
factor of 2 accounts for integrating from $0$ to $\pi/2$ instead of
$-\pi/2$ to $\pi/2$ in the angle.  In the second line of this equation
we have changed the variable in the $\delta$-function from
$r\sin(\phi)$ to $\phi$ and then have carried out the angle
integration and finally the r-integration. The quantity
$\frac{dN}{db}$ has units of $\M{length}^{-1}$.

Note that $\frac{dN}{db}$ is independent of $b$; i.e., the impact
parameters of the events are uniformly distributed. Of course, in
reality, an upper limit $b_{\rm max} $ will be present, depending on
the source brightness, background light and the observing conditions.
The integral $\int_{0}^{b_{\rm max}} \frac{dN}{db} db = 2 n \,\vt \,
\Delta t \times b_{\rm max}$ is dimensionless and equals (for the
considered line-of-sight) the number of lenses that cause an event
above a minimum magnification (corresponding to $b_{\rm max}$) within
$\Delta t$.

Equation~(\ref{eq.pb_b_max}) can also be obtained from geometrical
arguments: a circle with radius $b$ embedded into a two dimensional
plane defines a cross section of $2b$ to streaming particles in that
plane, independent of the streaming direction. Therefore, the number
of particles passing through that aperture with diameter $2b$ in a
time $\Delta t$ is $n \vt \Delta t 2 b$.  Hence, equation
(\ref{eq.pb_b_max}) also holds for a coherent particle stream, with
any velocity direction. Therefore equation (\ref{eq.pb_b_max}) is also
valid for any probability distribution of the velocity angles.

The number of events per line-of-sight distance $\Dol$, lens mass $\ml$, 
transversal velocity $\vt$, and impact parameter $b$ follows from 
equation (\ref{eq.pb_b_max}) by replacing $n$ with 
$n(\Dol,\ml,\vt)=n(\Dol,\ml)\,\pv(\vt,\Dol)$:
\begin{equation}
    \frac{d^4N}{d\Dol \,d\ml \,d\vt \,db} = 
    2 \, n(\Dol,\ml) \,\pv(\vt,\Dol)  \,\vt \Delta t  
    = 2 \,\rho(\Dol) \,\xi(\ml) \,\pv(\vt,\Dol) \,\vt \Dt.
  \label{eq.dNdb}
\end{equation}
We now transfer the number $N$ of the events per line-of-sight to the
event rate (per line-of-sight), $\Gamma := \frac{N}{\Dt}$, and write
equation (\ref{eq.dNdb}) as
\begin{equation}
  \GAMb
  =  2 \,\rho(\Dol) \,\xi(\ml) \,\pv(\vt,\Dol) \,\vt .
  \label{eq.GAMb}
\end{equation}
With the relative impact parameter $\uub$ defined as
$\uub=\frac{b}{\RE(\Dol,\ml)}$ this distribution can be rewritten as
\begin{equation}
    \GAM =  2 \,\rho(\Dol) \,\xi(\ml) \,\pv(\vt,\Dol) \,\vt \,\RE(\Dol,\ml)  ,
  \label{eq.dNdu0}
\end{equation}
which corresponds to the event rate for the standard definition (see
\citet{1991MNRAS.250..348D}).\footnote{\citeauthor{1991MNRAS.250..348D}'s
(\citeyear{1991MNRAS.250..348D}) eq.~(10) with
  $d\Gamma \equiv d^4 \Gamma$, $D \equiv \Dos$, $x \equiv \Dol/\Dos$,
  $r_\M{E} [\mu x(1-x)]^{1/2} \propto \RE$, $\rho_0 H(x) \equiv
  \rho(\Dol)$, $dn_0/d\mu \equiv \xi(\ml)$, $u_\Min \equiv \uub$
  yields
	$d^4 \Gamma = 2 \Dos \,\vt \,\pv(\vt) \,\RE 
  \,\rho(\Dol) \,\xi(\ml) \,d\ml \,d\uub \,d\vt \,(d\Dol/\Dos)$.
}

\section{The Source Distributions}
\label{sec.source_dis}

In the case of pixel-lensing the parameters of the source cannot be
determined. Therefore, we now introduce probability distributions for
the source distance $\Dos$, velocity $\vsvec$, unlensed flux $\F$,
color $\Col$, and radius $\Rstar$ (for finite source effects).

\subsection{The Transverse Lens-source Velocity Distribution}

We again assume that the velocity distributions of lenses $l$ and
sources $s$ are approximately isotropic around their mean respective
streaming velocities (cf. equation (\ref{isov})). The {projected}
velocity dispersion of the source population we call $\sigst =
\Dol/\Dos \,\sigs$.  We define $\vec{v}_0=(\vx,\vy)$ as the difference
between the {projected} streaming velocities of the source and lens
populations. Then, the transverse velocity differences in $x$ and $y$
between a lens and a source, each drawn from their respective
distributions, are: $\vlsx := \vlx-\vsx+\vx$ and $\vlsy :=
\vly-\vsy+\vy$.

Similar to equation (\ref{eq.vt_l}), we obtain for the distribution of
the transverse velocities
\begin{equation}
    \pv(\vt,v_0)  =  \Int \delta\left(\vt-\sqrt{\vlsx^2+\vlsy^2}\right)
    p(\vlsx) \,p(\vlsy) \,d\vlsx \,d\vlsy ,
\end{equation}
where $p(\vlsx)$ [and $p(\vlsy)$ analogously] is given by
\begin{equation}
  \darray
\begin{array}{ll}
  p(\vlsx)  
& =  C_l C_\iss \Intinf \Intinf \exp\left(-\frac{\sigst^2 \vlx^2+\sigl^2 \vsx^2}{2\sigl^2\sigst^2}\right)  \delta(\vlsx- (\vlx-\vsx+\vx)) \,d\vlx \,d\vsx  \\  
    & = C_l C_\iss \Intinf \Intinf \exp\left(-\frac{\sigst^2 \vlx^2+\sigl^2 \vsx^2}{2\sigl^2\sigst^2}\right)  \delta(\vlx-(\vlsx+ \vsx-\vx)) \,d\vlx \,d\vsx \\  
    & = C_l C_\iss  \Intinf \exp\left(-\frac{(\vlsx-\vx)^2}{2(\sigl^2+\sigst^2)}\right) \sqrt{\frac{2\pi\sigl^2\sigst^2}{\sigl^2+\sigst^2}} \,d\vlsx \\
    & = \frac{1}{\sqrt{\sigl^2+\sigst^2} \,\sqrt{2\pi}} \,\exp\left(-\frac{(\vlsx-\vx)^2}{2(\sigl^2+\sigst^2)}\right)  \\  
    & = \frac{1}{\sigls \,\sqrt{2\pi}}
   \,\exp\left(-\frac{(\vlsx-\vx)^2}{2 \sigls^2}\right)
     . \\ 
\end{array} 
\end{equation} 
In the last step we have defined 
\begin{equation}
   \sigls := \sqrt{\sigl^2+\left(\frac{\Dol}{\Dos}\right)^2\sigs^2}
 ,
\end{equation}
which is the combined width of the velocity distribution of the lenses
and that of the sources, projected onto the lens plane.

Finally, analogously to equation (\ref{eq.vt_l}), we obtain
\begin{equation}
  \pv(\vt,v_0) =  \frac{1}{\sigls^2}
  \,\vt 
  \,\exp{\left(- \frac{\vt^2+v_0^2}{2\sigls^2}  \right)}
\, I_{0}{\left(\frac{v_0\,\vt}{\sigls^2}\right)}   ,
\end{equation}
with $v_0(x,y,\Dol,\Dos)$, $\sigl(x,y,\Dol)$, and $\sigs(x,y,\Dos)$.

\subsection{The Luminosity Function}
\label{sec.LF}

The luminosity function (LF) $\phi \left(\M{flux}^{-1}\right)$
or $\Phi \left(\M{mag}^{-1}\right)$ is usually defined as the number of
stars per luminosity bin\footnote[100]{
Note that we neglect the correct indizes refering to the 
band $\XX$ and define
$\F \equiv F_{0,\XX}$,
$\FVega\equiv{F_{\M{Vega},\XX}}$,
$\Flum\equiv{\mathcal{F}_\XX}$,
$\FsunX\equiv\Flum_{\odot,\XX}$,
$\DF\equiv{\Delta_{F_\XX}}$,
$\Mlum\equiv{\mathcal{M}_\XX}$,
$\Col\equiv{\mathcal{C}_\CX}$,
$\SB\equiv\mu_\XX$,
$(M/L) \equiv (M/L)_\XX$,
$\ext\equiv A_\XX$.}.

The mean, or so-called characteristic flux of a stellar population is
\begin{equation}
  <\Flum>:=\frac{\int\Flum\,\phi(\Flum) \,d\Flum}{\int \phi(\Flum) \,d\Flum}
           ,
\end{equation}
or, if one instead uses the luminosity function $\Phi$ in
magnitudes,\footnote{\label{foot.f2mag} With $d\Mlum=-(2.5/\ln
  10)\,d\Flum/\Flum$ the conversion of the luminosity function from
  flux to magnitudes becomes $\Phi(\Mlum) = -0.4 \ln10 \,\FVega
  10^{-0.4 \Mlum} \,\phi{\left(\FVega \, 10^{-0.4 \Mlum}\right)}$.}

\begin{equation}
  <\Flum>:=
  \frac{\int \FVega 10^{-0.4 \Mlum}\,\Phi(\Mlum) \,d\Mlum}{\int \Phi(\Mlum) \,d\Mlum}   ,
\end{equation}
where $\FVega$ is the flux of Vega.

We use a luminosity function normalized equal to 1,
\begin{equation}
  \Intninf \tilde{\phi}(\Flum) \,d\Flum := \Intinf \LF(\Mlum) \,d\Mlum :=  1   ,
\end{equation}
as we obtain the amplitude of the LF from the matter density and the
mass-to-light ratio of the matter components (bulge, disk) later on. 

The luminosity functions in the literature are usually given for stars
at a distance of 10 pc. The relations for the source flux $\F$ at a
distance $\Dos$ and its flux $\Flum$ at 10 pc, or its absolute
magnitude $\Mlum$ are given in the following two equations, allowing for
extinction along the line-of-sight:
\begin{equation}
  \F(\Flum,x,y,\Dos) := \Flum \times \left(\frac{10\E{pc}}{\Dos}\right)^2  \,10^{-0.4\,\ext(x,y,\Dos)}  ,
  \label{eq.F_Flum}
\end{equation}
\begin{equation}
  \darray
\begin{array}{l}
  \F(\Mlum,x,y,\Dos) := 
\FVega 10^{-0.4 \Mlum} \times \left(\frac{10\E{pc}}{\Dos}\right)^2  \,10^{-0.4\,\ext(x,y,\Dos)}
   .
  \end{array}
 \label{eq.F_Mlum}
\end{equation}

\subsection{The Number Density of Sources}

We characterize different source components (bulge and disk) by an index $\iss$
with corresponding indices in the density, luminosity, and mass-function of
that component.  $(M/L)_\iss$ is the mass-to-light ratio of that component in
solar units.

The number density of sources is a function of the  mass density, 
the mass-to-light ratio, and the characteristic flux of each component: 
\begin{equation}
  n_\iss(x,y,\Dos) := \frac{d^3 \Ns}{dx\,dy\,d\Dos}  = 
  \frac{\,\rho_\iss(x,y,\Dos)}{\ML_\iss  
  \,\frac{\ms_\odot}{\Flum_\odot}
  \,<\Flum>_\iss}    .
  \label{eq.nsourcedis}
\end{equation}
Note that $(M/L)$ is the mass-to-light ratio of the total disk or bulge
component, and has to include the mass in stellar remnants or in gas.
Therefore, the value of $(M/L)$ is not necessarily equal to the stellar
mass-to-light-ratio in the bulge and the disk.

The normalized probability distribution for sources $p_\iss(\Dos)$ at distance
$\Dos$ is
\begin{equation}
   p_\iss(\Dos) := \frac{\rho_\iss(\Dos)}{\Int_0^\infty{\rho_\iss(\Dos) \,d\Dos}}
   \label{eq.sourcedis}   .
\end{equation}

\subsection{Including the Color and Radius Information}
\label{sec.color}

To use the color information, $\Col:=\Mlum-\Mlum`$, we construct a
normalized color-flux distribution $\pcmd(\Mlum,\Col)$ from the
color-magnitude diagram of stars,
\begin{equation}
 \int \int \pcmd(\Mlum,\Col) \,d\Mlum  \,d\Col \stackrel{!}{=} 1 
  , 
\end{equation} 
which is related to the luminosity function as  
\begin{equation}
 \LF(\Mlum) = \int \pcmd(\Mlum,\Col) \,d\Col    .
\end{equation}
The radius is related to the luminosity and color as $\Rstar(\Mlum,\Col)$
(see \S~\ref{sec.radius_flux}).

\clearpage

\section{Applications for the Microlensing Regime}
\label{sec.app}

In this section we derive the basic microlensing quantities and
distributions using the four-dimensional event rate differential
derived in \S~\ref{sec.prob_lens}. We apply the equations to M31 using
the M31 model in \S~\ref{app.model}.

\subsection{Optical Depth $\tau$}

The optical depth $\tau$ is defined as the number of lenses that are
closer than their own Einstein radius $\RE$ to a line-of-sight.  The
optical depth $\tau$ is therefore the instantaneous probability of
lensing taking place, given a line-of-sight and a density distribution
of the lenses.  For a given source star at distance $\Dos$, the
optical depth equals the number of lenses within the microlensing tube
defined by the Einstein radius $\RE(\ml,\Dol,\Dos)$
(eq.~[\ref{eq.pac_re}]) along the line-of-sight:
\begin{equation}
  \darray
\begin{array}{ll}
  \tau(\Dos) & = 
     \Int_0^\Dos \Int_0^\infty \Int_0^\infty  \Int_{0}^{\RE} n(\Dol,\ml,\vt) \times  2 \pi r \,dr  \,d\vt \,d\ml \,d\Dol \\
    & = \Int_0^\Dos \Int_0^\infty \Int_0^\infty  \rho(\Dol) \,\xi(\ml) \, \pv(\vt,\Dol)  \,\Int_{0}^{\RE} 2 \pi r \,dr  \,d\vt \,d\ml \,d\Dol \\
    & =\Int_0^\Dos \Int_0^\infty \rho(\Dol) \,\xi(\ml) \,\pi\RE^2\,d\ml \,d\Dol \\    
    & =  \frac{4 \pi G}{c^2} \Int_0^\Dos \rho(\Dol)\, D(\Dol) \,d\Dol   , 
\end{array}
\label{eq.tau}
\end{equation}
with $D(\Dol):=\Dol (\Dos-\Dol)/\Dos$, equal to
\citet[eq. (9)]{1986ApJ...304....1P}.  Equation~(\ref{eq.tau})
demonstrates that the optical depth depends on the mass density, but
not on the mass function $\xi(\ml)$ of the lenses.

In the past, the optical depth along a line-of-sight to M31 was often
calculated by setting $\Dos$ equal to the distance to the plane of the
disk of M31 \citep{2000ApJ...535..621G,2000ApJ...530..578B}.  This is
like treating the sources for lensing as a two dimensional
distribution. It yields fairly adequate results for the optical depth
of disk stars but cannot be justified for the bulge stars in M31.  We
use the source distance probability distribution (equation
(\ref{eq.sourcedis}) to obtain the line-of-sight distance--averaged
optical depth:
\begin{equation}
  <\tau>_\iss \;:= \int p_\iss(\Dos) \,\tau(\Dos) \,d\Dos .
  \label{eq.tau_mean}
\end{equation}
\begin{figure}
  \begin{center}
    \epsscale{1.0}\plotone{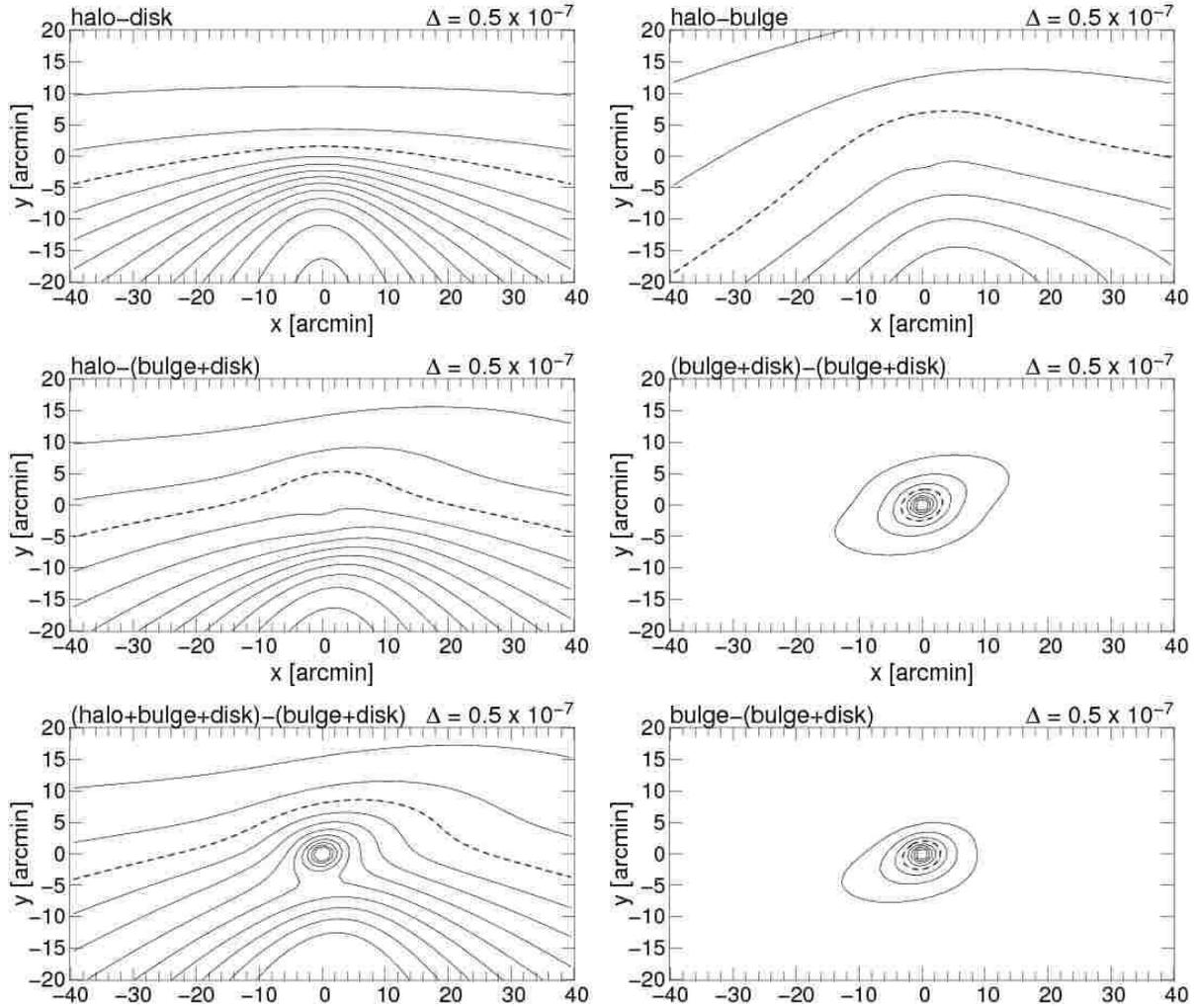}
  \end{center}
  \caption{Contours of the line-of-sight--averaged optical depth
    $\left<\tau\right>_\iss$ (eq. [\ref{eq.tau_mean}]). $x$ and $y$ are
    given in the intrinsic M31 coordinate system, which is centered on
    the nucleus of M31 and where the M31 disk major axis is orientated
    horizontally ($\M{P}.\M{A}.=38^\circ$).  Halo lensing of disk
    sources ({\it first row left}, a), halo-bulge lensing ({\it first
      row right,}), halo-lensing of bulge \& disk sources ({\it second
      row left}). The average optical depth for self-lensing of
    sources in M31 is shown in the second row on the right.  In the
    third row, left panel, we show the resulting total optical depth
    with the contributions of all lenses.  The third row, right panel,
    displays the optical depth due to bulge lenses.  The optical depth
    caused by the MW (not shown), is nearly constant $\tau_\M{MW} =
    0.78\times 10^{-6}$.  To obtain the values of $<\tau>_\iss$ we used
    the model of the luminous and dark matter of M31 presented in
    \S~\ref{app.model}. Here and in all following calculations a
    MACHO fraction in the dark halo of M31 of unity was assumed. The
    spacing between adjacent contours are shown as inserts in each
    diagram. The contour lines $<\tau>_\iss=2 \times 10^{-6}$ are shown
    as dashed curve.  }
  \label{fig.tau_wecapp}
\end{figure}
Figure~\ref{fig.tau_wecapp} shows the average optical depth for the
central part of M31 for lenses in the halo of M31 (``halo-lensing''),
and for stellar lenses in the bulge and disk of M31
(``self-lensing''). The self-lensing optical depth is symmetric (with
respect to the near and far side of M31) and dominates the optical
depth in the central arcminute of M31.  The halo-lensing optical depth
is asymmetric and rises toward the far side of the M31 disk, since
there are more halo lenses in front of the disk.

Figure~\ref{fig.tau_wecapp} ({\it first row, left}) shows the
halo-disk optical depth. The results do not depend so much on the
three-dimensional structure of the disk but much more on the halo core
radius assumed. We use $r_\M{c}=2\E{kpc}$ (see
\S~\ref{app.model}).  \cite{2000ApJ...535..621G} used core radii
of $r_\M{c}=1\E{kpc}$ and $r_\M{c}=5\E{kpc}$ for their Figures~1{\it
  c} and 1{\it d}, and our result is in between their results, as
expected.  \cite{2000ApJ...530..578B} have obtained qualitatively
similar results using $r_\M{c}=5\E{kpc}$, but assuming an M31 distance
of 725 kpc and a slightly less massive halo than we do. The optical
depth caused by all M31 components is shown in
Figure~\ref{fig.tau_wecapp} ({\it third row, left}).  The result of
\cite[see his Fig.~1]{1996ApJ...472..108H} using a halo core radius of
$r_\M{c}=6.5\E{kpc}$ looks strikingly different.  Comparison to
Figure~\ref{fig.tau_wecapp} ({\it third row, left}) demonstrates that
the total optical depth is dominated by bulge lenses in the central
part of M31.  The last panel of this figure shows the optical depth
for bulge-lensing toward M31 sources.  The bulge-lensing optical depth
had been obtained by \cite[see their Fig.~5]{2000ApJ...535..621G}, but
the values that they obtained are up to a factor of 5 larger than ours
(which probably is due to their different M31 model).

\subsection{Single-star Event Rate}

The optical depth is the probability of stars to be magnified above a
threshold of 1.34 at any time. Observations, however, usually measure
only a temporal change of magnification. Therefore, the event rate,
which is the number of events per time interval, is the relevant
quantity for observations. The event rate is the integral of equation
(\ref{eq.GAMb}) over lens masses, lens distances, relative velocities,
and impact parameters $b$ smaller than a threshold $\uT\RE$:
\begin{equation}
  \darray
  \begin{array}{ll}
    \GT(\Dos) & := 
     \Int_0^\infty \Int_0^\Dos \Int_0^\infty \,\Int_0^{\uT\RE}  \GAMb \,db \,d\vt \,d\ml \,d\Dol \\
     & = 2 \uT \Int_0^\infty \,\xi(\ml) \Int_0^\Dos \,\rho(\Dol)\,\RE(\Dol,\ml)  \Int_0^\infty \vt  \,\pv(\vt,\Dol) \,d\vt  \,d\Dol \,d\ml\\
     & = 2 \uT \frac{\sqrt{4G}}{c} \Int_0^\Dos \rho(\Dol) \,\sqrt{D(\Dol)} \,\Int_0^\infty \sqrt{\ml}\,\xi(\ml)   \Int_0^\infty \vt\,\pv(\vt,\Dol) 
    \,d\vt  \,d\Dol \,d\ml \\
    & =:  \uT \,\G_1(\Dos)   .
\end{array}
\label{eq.gamma_ml}
\end{equation}
This had been first evaluated [using a single mass instead of
$\xi(\ml)$] by \citet{1991ApJ...366..412G}.\footnote{Eq. (11):
  changing his notation with $\Gamma \equiv \GT$, $v_c \equiv
  \sqrt{2}\sigl$, $L \equiv \Dos$, $\sqrt{m} \equiv \sqrt{\ml_0}$,
  $\frac{\rho_0 A'}{A'+B x'+x`^2} \equiv \rho(\Dol)$, $x' \equiv
  \Dol/\Dos$, $\eta \equiv v_0/(\sqrt{2}\sigl)$, $u' \equiv
  \vt/(\sqrt{2}\sigl)$:
\begin{displaymath}
  \GT  =  4 \, \sqrt{\frac{G}{c^2}}
  \frac{\uT}{\sqrt{\ml_0}} \int_0^\Dos  d\Dol \,\rho(\Dol) \,\sqrt{D(\Dol)} 
  \,e^{-v_0^2/(2 \sigl^2)} 
  \int_0^\infty d\vt \,\frac{\vt^2}{\sigl^2} 
  \,e^{-\vt^2/(2\sigl^2)} 
  \,I_{0}{\left(\frac{2 v_0 \vt}{2 \sigl^2}\right)}
\end{displaymath}
corresponds to our formula setting $\xi(\ml)=\delta(\ml-\ml_0)/\ml_0$,
$\sigs=0$.}

The impact parameter threshold $\uT$ is equivalent to a magnification
threshold $\AT$. Therefore, the number of events with amplifications
larger than $\AT(\uT)$ is proportional to the threshold parameter
$\uT$.

$\Gamma_1(\Dos)$ is the event rate along a chosen line-of-sight to a
distance of $\Dos$. Analogously to the optical depth, we also define
the line-of-sight distance--averaged single-star event rate
\begin{figure}
   \epsscale{0.9}\plotone{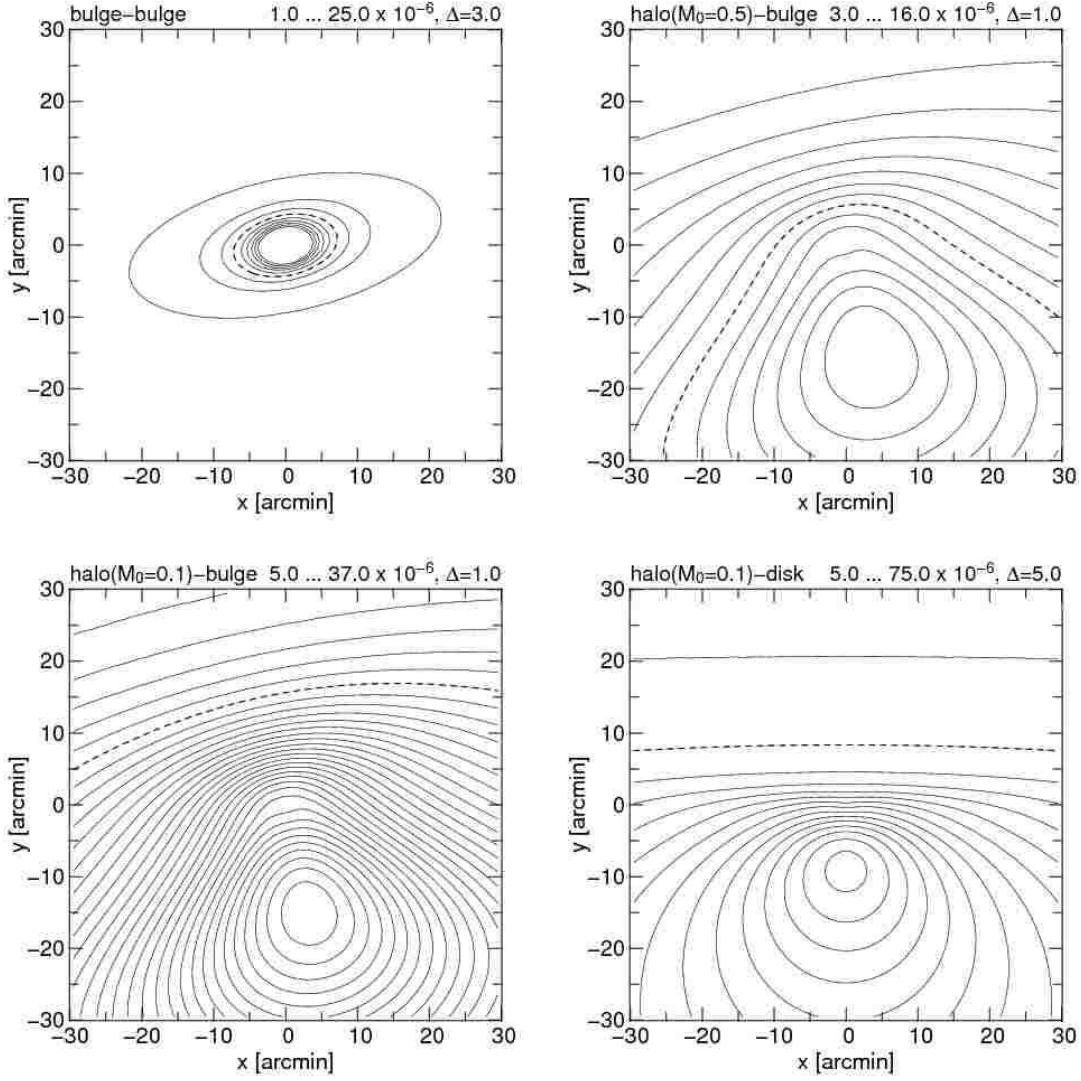}
   \caption{Averaged event rate $<\Gamma_1>_\iss$ ($\M{y}^{-1}$) toward
     M31 for bulge-bulge lensing {\it first row, left}), halo-bulge
     lensing with $\Mlens= 0.5\,\Msun$ (i.e., a mass function
     $\xi(\ml)=\delta(\ml_0 - 0.5\,\Msun)/0.5\,\Msun$) ({\it first
       row, right}), halo-bulge lensing with $\Mlens= 0.1\,\Msun$
     ({second row, left}), and halo-disk lensing with $\Mlens=
     0.1\,\Msun$ ({second row, right}). Contour levels and the spacing
     between adjacent contours are given on top of each diagram. The
     dashed line marks the $10^{-5}$ events $\E{yr}^{-1}$
     level. Whereas self-lensing is symmetric, halo lensing shows a
     clear asymmetry.  The event rate shows a maximum at the far side
     of the M31 disk (negative $y$-values).  These contours cannot be
     compared with an experiment since, first, one could certainly not
     identify all objects with a threshold of $\uT=1$ or a
     magnification of $1.34$ and second, one has to convolve the
     single-star event rate with the density of sources. The proper
     event rate maps in the pixel-lensing regime can be seen in
     \S~\ref{sec.exp}.}
  \label{fig.dis_u0}
\end{figure}
\begin{equation}
  <\Gamma_1>_\iss \;=\int p_\iss(\Dos) \,\Gamma_1(\Dos) \,d\Dos
\end{equation}
toward M31. 

We show these line-of-sight distance--averaged event rates for the
halo of M31 and the stellar lenses in the bulge and disk of M31
(self-lensing) in Figure \ref{fig.dis_u0}; the single-star
halo-lensing event rate is evidently asymmetric, whereas the
single-star self-lensing event rate is symmetric.  The levels of the
event rates (for each line-of-sight) are of the order $\sim 10^{-5}$
events $\E{yr}^{-1}$ ({\it dashed}), which implies that at least a few
times $10^4$ source stars are needed to identify one lensing event
(even if all lensing events below the threshold $\uT=1$ could be
observed).  It can also be seen in Figure \ref{fig.dis_u0} that only
in the innermost part ($r\le 5\arcmin$) the self-lensing event rate
exceeds the halo-lensing event rate (for a 100\% MACHO halo).  As
mentioned earlier, the optical depth does not depend on the lens-mass
distribution (for the same matter density) because the decrease of
number of lenses with lens-mass is balanced by the increased area of
the Einstein disks around them.  However, the events take longer,
since larger Einstein radii have to be crossed. For the same optical
depth, this then must imply a decrease in event rate: $\G \sim
\ml_0^{-1/2}$, setting $\xi(\ml)=\delta(\ml-\ml_0)/\ml_0$ in equation
(\ref{eq.gamma_ml}). The decrease of the event rate with increasing mass
of the lenses can be seen in Figure \ref{fig.dis_u0} ({\it first row,
  right panel}) and Figure \ref{fig.dis_u0} ({\it second row, left
  panel}).

The relations above give the event rate per line-of-sight or per
star. To compare this with measurements of the lensing rate for
resolved stars, one has to account for the source density.

\subsection{Distribution for the Einstein Timescale}
Not only the number of lensing events per time and their spatial
distribution but also their duration (Einstein time) is a key
observable in microlensing surveys.  The distribution of the Einstein
timescales of the events is
\begin{equation}
  \darray
\begin{array}{ll}
  \frac{d^2 \Gamma}{d\tE \,d\uub} & = 
    \Int_0^\infty \Int_0^\Dos \Int_0^\infty  
       \GAM  \,\delta{\left(\tE-\frac{\RE}{\vt}\right)} \,d\vt  \,d\Dol\,d\ml 
\\ 
  & =  \Int_0^\infty \Int_0^\Dos \Int_0^\infty       
       \rho(\Dol) \,\xi(\ml) \,\pv(\vt)   2      
       \,\vt \,\RE      
       \,\delta{\left(\tE-\frac{\RE}{\vt}\right)}      
       \,d\vt \,d\Dol\,d\ml \\     
   & =  2 \Int_0^\infty \Int_0^\Dos \Int_0^\infty  \rho(\Dol) \,\xi(\ml) \,\pv(\vt) \,\vt\,\RE    
    \,\frac{\delta{\left(\vt-\frac{\RE}{\tE}\right)}}{\left|\RE/\vt^2\right|} \,d\vt \,d\Dol\,d\ml \\    
   & =  \frac{2 }{\tE^3} \Int_0^\infty \Int_0^\Dos 
\rho(\Dol) \,\xi(\ml)  \,\pv{\left(\frac{\RE}{\tE}\right)} \,\RE^3  
   \,d\Dol\,d\ml  .\,
\end{array}
\label{eq.dG_du0_dtE}
\end{equation}

The second line of equation (\ref{eq.dG_du0_dtE}) is proportional to
the equation presented in \citet{1996ApJ...473..230H}.\footnote{Their
  eqs.~(2.2.6) and (2.2.7) with $f(\tE) \propto p(\tE)$, $d_\Min = 0$,
  $d_\Max \equiv \Dos$, and $f_M(M) \propto \xi(\ml) ,$
  \begin{displaymath}
    p(\tE) \propto \int
    \xi(\ml) \Int_0^{\Dos} \rho(\Dol) \,\RE(\ml,\Dol,\Dos) \Intninf \vt \,\pv(\vt)
    \, \,\Intninf \delta{\left(\tE'-\frac{\RE(\Msun,\Dol,\Dos)}{\vt}\right)}
    \,\delta{\left(\tE- \left(\frac{\ml}{\Msun}\right)^\frac{1}{2} \tE'\right)}
    \,d\tE' \,d\vt \,d\Dol \,d\ml .
  \end{displaymath}
}

The result is of course independent of the relative impact parameter
$\uub$.  If one carries out an (microlensing) experiment with a
threshold $\uT$, one obtains with equation (\ref{eq.dG_du0_dtE}) the
Einstein timescale distribution of events as
\begin{equation}
  \frac{d\GT}{d\tE} = \frac{2 \uT}{\tE^3} \Int_0^\infty \Int_0^\Dos
  \rho(\Dol) \,\xi(\ml) \,\pv{\left(\frac{\RE}{\tE}\right)} \,\RE^3
  \,d\Dol\,d\ml .
\end{equation}
This result corresponds to that of
\citet{1997PhR...279...67R}\footnote{Their eq.~(31) corresponds to our
  formula converting their notation to ours $\Gamma \equiv
  \Gamma/\uT$, $dn/dm \equiv \rho(\Dol) \,\xi(\ml)$, $T \equiv \tE$,
  $v^\bot \equiv \vt$, $\int_0^{2\pi} d\gamma \,v^\bot
  \,G(v_\M{dis}^l) \,G(v_\M{dis}^b) \equiv \pv(\vt)$.}  
and
\citet{2000ApJ...530..578B}\footnote{Their eq.~(9) corresponds to our
  formula converting their notation to ours $L \equiv \Dos$, $\tE
  \equiv 2\tE$, $\beta_\M{T} \equiv \uT$, $v_\M{c} \equiv \sqrt 2
  \sigl$, $x \equiv \Dol/\Dos$, $\eta \equiv v_0/\sqrt 2 \sigl)$, $v
  \equiv \vt/(\sqrt 2 \sigl)$ and setting
  $\xi(\ml)=\delta(\ml-\ml_0)/\ml$, $\sigs=0$.}.
The (normalized) probability distribution for the Einstein timescales becomes
\begin{equation}
  p(\tE) := \frac{1}{\GT} \frac{d\GT}{\,d\tE}  .
  \label{eq.prob_tE}
\end{equation}
With this probability distribution the average timescale
$\overline{\tE}$ of an event with line-of-sight distance $\Dos$ can be
obtained:
\begin{equation}
  \darray
  \begin{array}{ll}
    \overline{\tE}(\Dos)  & :=  \Int_0^\infty \tE \,p(\tE) \,d\tE  
    =  \frac{2\uT}{\GT} \Int_0^\infty \Int_0^\infty \Int_0^\Dos      
    \rho(\Dol) \,\xi(\ml) \,\pv{\left(\frac{\RE}{\tE}\right)} \,\tE \,\frac{\RE^3}{\tE^3}  \,d\Dol\,d\ml \,d\tE \\     
    &  =  \frac{2\uT}{\uT\G_1} \Int_0^\infty \Int_0^\infty \Int_0^\Dos      
    \rho(\Dol) \,\xi(\ml)  \,\pv(\vt) \,\RE^2  \,d\Dol\,d\ml \,d\vt     
    =   \frac{2}{\pi \G_1} \Int_0^\infty \Int_0^\Dos      
    \rho(\Dol) \,\xi(\ml)  \,\pi\RE^2  \,d\Dol\,d\ml       
    =  \frac{2}{\pi}  \frac{\tau(\Dos)}{\G_1(\Dos)}  , 
  \end{array}
  \label{eq.tE_bar}
\end{equation}
which equals the result of \citet{1995ApJ...449...28A}.\footnote{Their
  eq.~(2) with $\hat{t} \equiv 2 \tE$ .}

We instead aim for the line-of-sight distance--averaged mean Einstein
timescale [at an arbitrary position $(x,y)$]. We start from the
line-of-sight distance--averaged event rate per Einstein time $\tE$,
\begin{equation}
  \left<\frac{d\GT}{d\tE} \right>_\iss = \int \,p_\iss(\Dos) \frac{d\GT}{d\tE} \,d\Dos  .
  \label{eq.tE_3d}
\end{equation}
\begin{figure}
  \epsscale{0.6}\plotone{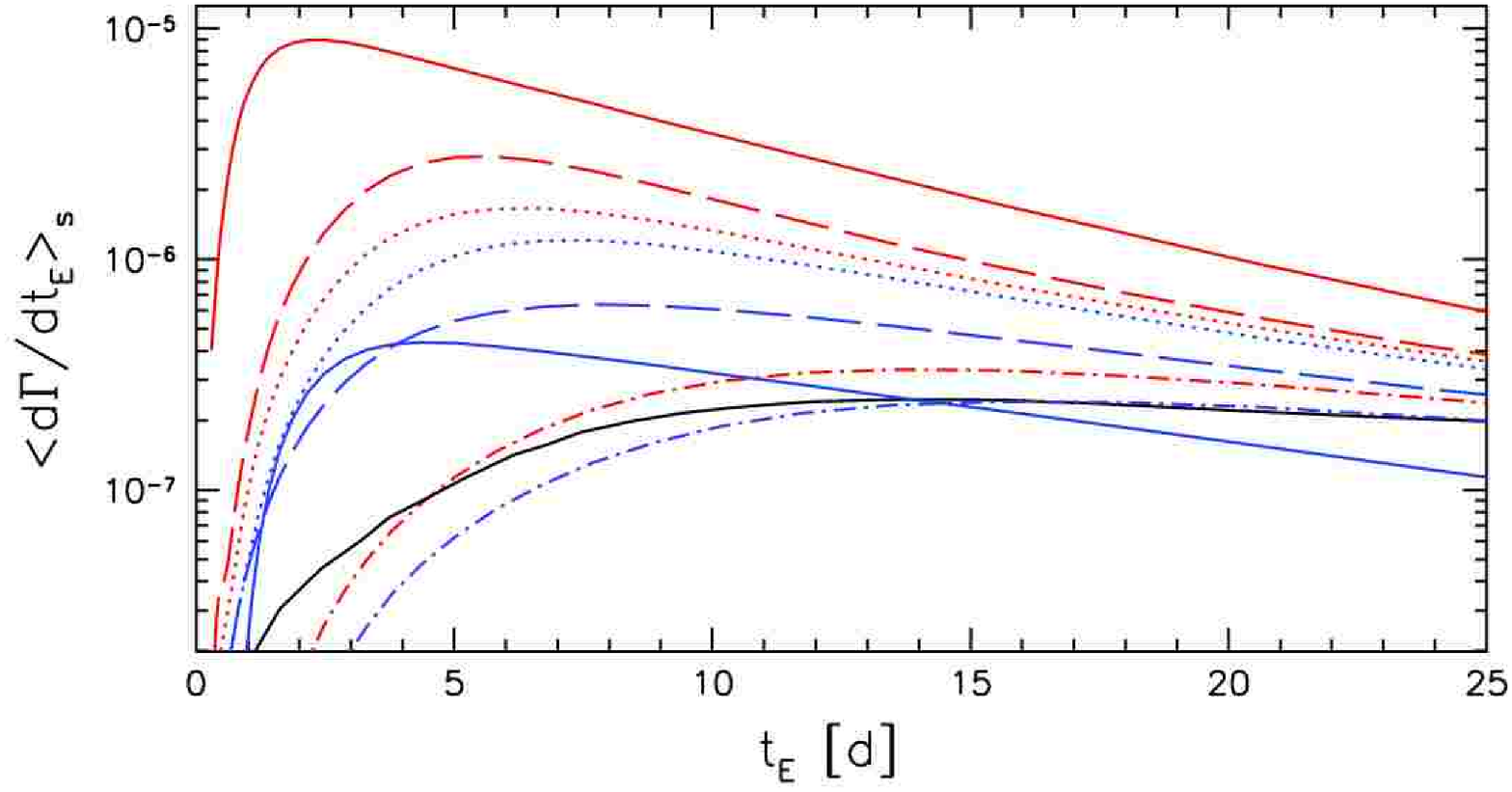}
  \caption{Line-of-sight distance--averaged distribution of the event
    rate with Einstein timescale $\left<d\G/d\tE\right>_\iss$
    ($\M{yr}^{-1}$) using the model of M31 presented in
    \S~\ref{app.model} and assuming a 100\% MACHO halo. Results
    are shown for two positions in the intrinsic M31 coordinate system
    (see Figure~\ref{fig.tau_wecapp}), at $(x,y)=(1\arcmin,0\arcmin)$
    ({\it red}), and at $(x,y)=(4.46\arcmin,4.46\arcmin)$
    (corresponding to $(x,y)=(1\E{kpc},1\E{kpc})$, {\it blue}). The
    bulge-bulge Einstein time distribution is shown as solid line. The
    halo-bulge distributions have been evaluated for a MACHO mass of
    $\ml_0 = 0.1\,\Msun$ ({\it dotted line}) and $\ml_0 = 0.5\,\Msun$
    ({\it dot-dashed line}).  The halo-disk lensing case is shown for
    a MACHO mass of $\ml_0 = 0.1\,\Msun$ ({\it dashed line}).  The
    Einstein time distributions of the event rate differ considerably
    for halo-disk, halo-bulge and bulge-bulge lensing and also vary
    significantly with the line-of-sight position.  For comparison we
    also plot the Einstein timescale distribution for M31 halo-lensing
    derived by \cite{1996ApJ...473..230H} up to a pre-factor (that we
    chose equal to $3 \times 10^{-6}$) as a {black solid
      curve}. \cite{1996ApJ...473..230H} considered the distributions
    for the halo-disk and halo-bulge lensing to be similar and not
    distinguish between them further. They used a MACHO mass of
    $\ml_0=0.1\,\Msun$ for their curve. However, it looks more similar
    to our halo-bulge curve for $\ml_0=0.5\,\Msun$, and cannot be moved
    on the halo-bulge or halo-disk for $\ml_0=0.1\,\Msun$ curve with
    another choice of the prefactor. }
\label{fig.Gamma_tE} 
\end{figure}
Figure~\ref{fig.Gamma_tE} shows examples for this line-of-sight
distance--averaged distribution $\left<d\GT/d\tE\right>_\iss$ for two
different positions in the intrinsic M31 coordinate system (see
Fig.~\ref{fig.tau_wecapp}), at $(x,y)=(1\arcmin,0\arcmin)$ and
$(x,y)=(4.46\arcmin,4.46\arcmin)=(1\E{kpc},1\E{kpc})$. The
distributions show a strong dependence on the line-of-sight
position. The halo-bulge and halo-disk lensing timescales are longer
than those of bulge-bulge lensing. An increase in MACHO mass decreases
the event rate (see Fig.~\ref{fig.dis_u0}), and the timescale of the
events becomes longer (see the examples for $\ml_0=0.1\,\Msun$ and
$\ml_0=0.5\,\Msun$ in Fig.~\ref{fig.Gamma_tE}).

\begin{figure}
  \epsscale{0.9} \plotone{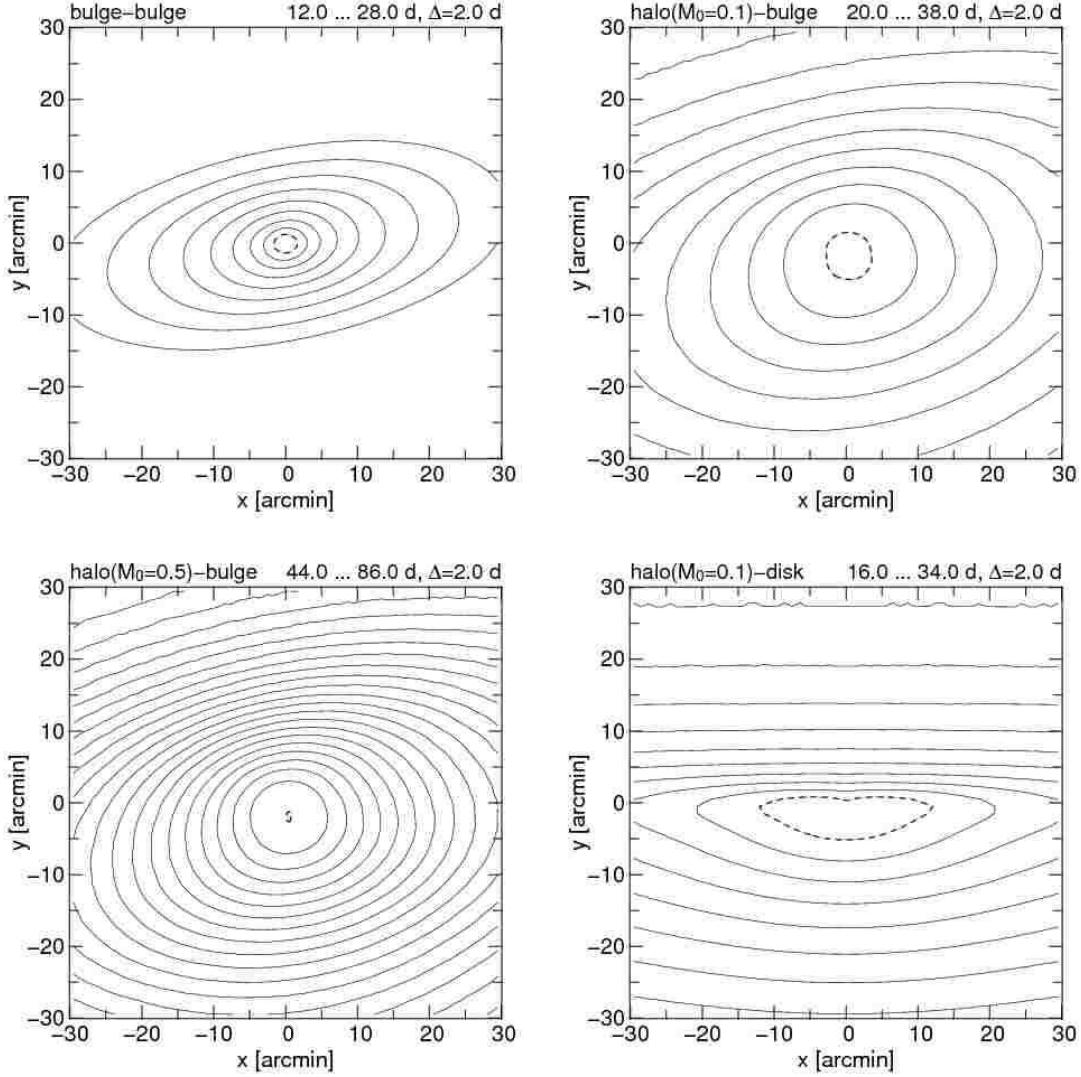}
  \caption{Einstein timescale averaged over all sources
    $\overline{\left<\tE\right>_\iss}$ calculated for lines of sights
    toward the center of M31 using the model of M31 presented in
    \S~\ref{app.model}. We show the distributions for
    bulge-bulge self-lensing ({\it first row, left}), halo-bulge
    lensing for lenses of $\ml_0 = 0.1\,\Msun$ [i.e., a mass function
    $\xi(\ml)=\delta(\ml_0 - 0.1\,\Msun)/0.1\,\Msun$] ({\it first
      row, right}), halo-bulge lensing for lenses of $\ml_0 =
    0.5\,\Msun$ ({\it second row, left}), and halo-disk lensing with
    $\Mlens= 0.1\,\Msun$ ({\it second row, right}).  Contour levels and
    spacing are shown on top of each diagram.  We assume the MACHO
    fraction in the dark halo to be unity.  The dashed line marks the
    lowest contour level.}
\label{fig.Gamma_tEavg} 
\end{figure}

Weighting $\tE$ with this function and integrating over all timescales
finally yields the desired mean line-of-sight distance--averaged
Einstein timescale of an event:
\begin{equation}
  \overline{\left<\tE\right>_\iss} \;:=  
  \frac{\Intninf \tE \left<\frac{d\Gamma}{d\tE}\right>_\iss \,d\tE}
       {\Intninf \left<\frac{d\Gamma}{d\tE}\right>_\iss \,d\tE} =
  \frac{2}{\pi} \frac{\left<\tau\right>_\iss}{\left<\G_1\right>_\iss} \ne  \left<\overline{\tE}\right>_\iss .
\end{equation}
Mean Einstein timescales $\overline{\left<\tE\right>_\iss}$ are shown
for lensing and self-lensing in Figure~\ref{fig.Gamma_tEavg}.
Generally, the minimum of $\overline{\left<\tE\right>_\iss}$ is near
the M31 center, irrespective of the lens-source configuration.  The
mean Einstein timescale is smaller for lower MACHO masses, since the
Einstein radii become smaller and are faster to cross (compare the two
middle panels in Fig.~\ref{fig.Gamma_tEavg}). The bulge-bulge
lensing events ({\it first panel}) are the shortest. This is caused by
the small lens-source distances, which reduce the sizes of the
Einstein radii.

\subsection{The Amplification Distribution}

The magnification distribution of the event rate is 
\begin{equation}
     \frac{d\Gamma}{d\A}  = \left|\frac{d\uub}{d\A}\right| 
    \frac{d\Gamma}{d\uub} =  \left|\frac{d\uub}{d\A}\right|  \frac{\Gamma}{\uT} .
    \label{eq.dNdA}
\end{equation}
Inserting equation (\ref{eq.du0dA0}) makes the result equal to that of
\citet{1991ApJ...366..412G}.\footnote{Eq.~(22): with $A \equiv \A$,
  $\Gamma' \equiv \frac{\Gamma}{\uT}$ ,
  $\frac{d\Gamma}{d\A}=-\frac{\Gamma/\uT}{\sqrt{2}} \,[\A
  (\A^2-1)^{-1/2}-1]^{-1/2} (\A^2-1)^{-3/2}$.}

Transforming equation (\ref{eq.dG_du0_dtE}) we can write
\begin{equation}
  \frac{d^2 \Gamma}{d\tE \,d\A} =  \frac{2 }{\tE^3} \left|\frac{d\uub}{d\A}\right| \Int_0^\infty \Int_0^\Dos 
  \rho(\Dol) \,\xi(\ml)  \,\pv{\left(\frac{\RE}{\tE}\right)} \,\RE^3  
  \,d\Dol\,d\ml  .\,
  \label{eq.dG_dA_dtE}
\end{equation}
using $d\uub/d\A$ from equation (\ref{eq.du0dA0}).

\subsection{The Distribution for the FWHM Timescale}
\label{sec.dis_tfwhm}

Although the Einstein timescale $\tE$ contains all the relevant
physical properties (mass, position, and velocity) of the lens, it is
of limited practical use in the case of an ill-determined source flux
(``Einstein time - magnification degeneracy'', see
\S~\ref{sec.lightcurves}).  In this case $\tfwhm$ is the only properly
measurable timescale of a light curve. We obtain the distribution
function for $\tfwhm$ {(neglecting finite-source effects)}, starting
from \S~\ref{sec.prob_lens}, using $\tfwhm(\vt,\ml,\Dol,\uub) =
\frac{\RE(\ml,\Dol)}{\vt} w(\uub)$ (see eq. [\ref{eq.t12_tE}]):
\begin{equation}
  \darray
  \begin{array}{ll}
    \frac{d^2 \Gamma}{d\tfwhm \,d\uub} & =
    \Int_0^\infty \Int_0^\Dos \Int_0^\infty         
    \,\rho(\Dol) \,\xi(\ml) \,\pv(\vt)  \,2\vt\,\RE        
    \,\delta{\left(\tfwhm -  \frac{\RE}{\vt} \,w(\uub)\right)}         
    \,d\vt\,d\Dol\,d\ml \\         
    & =  2 \Int_0^\infty \Int_0^\Dos \Int_0^\infty         
    \,\rho(\Dol) \,\xi(\ml) \,\pv(\vt)  \, \vt \RE        
    \,\frac{\delta{\left(\vt-\frac{\RE}{\tfwhm}\,        
          w(\uub)\right)}}        
    {\left|- \frac{\RE}{\vt^2} \,w(\uub)\right|}        
    \,d\vt\,d\Dol\,d\ml \\         
    & =   \frac{2}{\tfwhm^3} \Int_0^\infty \Int_0^\infty         
    \,\rho(\Dol) \,\xi(\ml)  \,w^2(\uub)   \,\RE^{3}       
    \,\pv{\left(\frac{\RE}{\tfwhm}\,w(\uub)\right)}        
    \,d\Dol\,d\ml \\        
    & =  \frac{2 \,w^2(\uub)}{\tfwhm^3} \Int_0^\infty \xi(\ml)  \Int_0^\Dos 
    \rho(\Dol) \,\pv{\left(\frac{\RE}{\tfwhm} \,w(\uub)\right)} 
    \,\RE^{3} \,d\Dol\,d\ml  .                      
  \end{array}
  \label{eq.dN_dtfwhm_du}
\end{equation}
\citet{2000ApJ...530..578B} expressed the same relation in an
alternative way\footnote{We can derive their expression in eq.~(10)
  with $\eta \equiv \frac{v_0}{\sqrt 2 \sigl}$, $v \equiv
  \frac{\vt}{\sqrt 2 \sigl}$, $M \equiv \ml_0$:
\begin{displaymath}
  \begin{array}{ll}
    \frac{d\Gamma}{d\tfwhm} & = 
    \Int_0^\infty \Int_0^\Dos \Int_0^\infty        
    \Int_0^{\uT\RE}  \GAMb         
    \,\delta{\left(\tfwhm -  2\frac{\RE}{\vt} \,w(\uub)\right)}  
    \,db \,d\vt\,d\Dol\,d\ml \\         
    & = 
    \Int_0^\infty \Int_0^\Dos  \Int_0^\infty         
    \,\rho(\Dol) \,\xi(\ml)          
    \,\vt^2 \,\pv(\vt) \,\left\{w'{\left[w^{-1}{\left(\frac{\tfwhm \vt}{2\RE}\right)}\right]}\right\}^{-1}        
    \,\theta{\left(2\frac{\RE}{\tfwhm} \,w(\uT) - \vt \right)} 
    \,d\vt \,d\Dol\,d\ml
  \end{array}
\end{displaymath}
using $\xi(\ml)=\delta(\ml-\ml_0)/\ml_0$.
\label{fn.baltz}
} 
and already motivated the same change of variables from $\vt$ to
$\tfwhm$.  Our relation for the FWHM time distribution of the event
rate in equation (\ref{eq.dN_dtfwhm_du}) does not include any
derivative or inversion of $w(\uub)$ and thus is very easy to evaluate
numerically. Note that one can use $w(u_0) \approx \sqrt{12} \,u_0$ as
high-magnification approximation.

Replacing the relative impact parameter $\uub$ by the maximum
amplification $\A$ (using eqs.~[\ref{eq.u0_A0}] and [\ref{eq.du0dA0}])
yields an equivalent description of this result:

\begin{equation}
  \frac{d^2\Gamma}{d\tfwhm \,d\A}  =
  \frac{2 \Psi(\A)}{\tfwhm^3} \Int_0^\infty \xi(\ml)  \Int_0^\Dos 
  \rho(\Dol) \,\pv{\left(\frac{\RE}{\tfwhm} \,\Upsilon(\A)\right)} 
  \,\RE^{3} \,d\Dol\,d\ml   ,
  \label{eq.dgam_dt12_dA}
\end{equation}
with $\RE(\ml,\Dol,\Dos)$ and $\Psi(\A)$ as
\begin{equation}
    \Psi(\A) :=  \dudA \, \Upsilon^2(\A)   
    =  4 \sqrt{2} \,\frac{\left[\A+(\A^2-1)^{1/2}\right]^{1/2} \,
      \left[(\A+1)^{3/2}-\A(\A+3)^{1/2}\right]}{(\A^2-1)^{7/4} (\A+3)^{1/2}}
    \stackrel{\A \gg 1}{\approx}  \frac{12}{\A^{4}},
  \label{eq.psi}
\end{equation}
where $\dudA$ was defined in equation (\ref{eq.du0dA0}).

\begin{figure} 
  \epsscale{1.0}\plotone{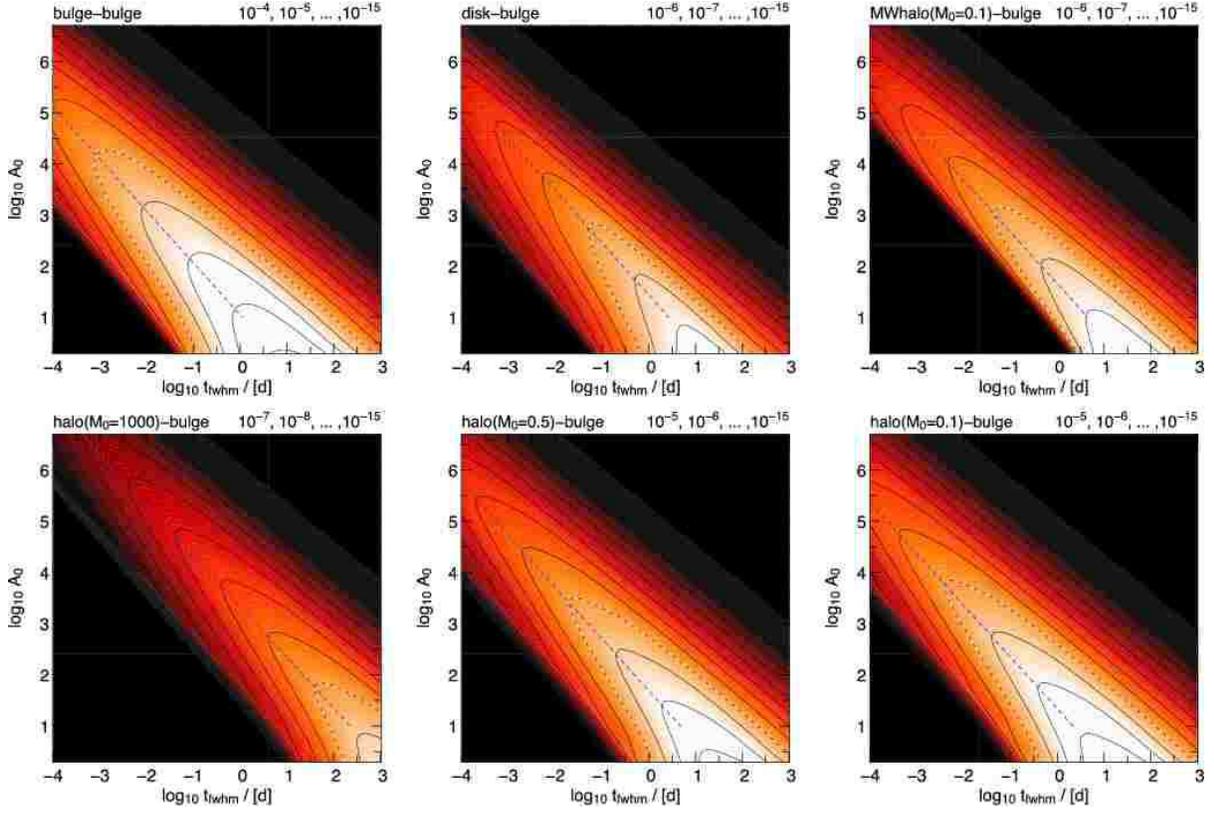}
	\caption{Distribution $\left<d^2
    \Gamma/\left(d\lgt \tfwhm \,d\lgt\A\right)\right>_\iss$ (y$^{-1}$)		
    (i.e., rate per year and line-of-sight) for different lens-source
    configurations, calculated at the position
    $(x,y)=(1\arcmin,0\arcmin)$ in the intrinsic M31 coordinate system
    (see Fig.~\ref{fig.tau_wecapp}).  The values of timescale and
    magnification are largely confined to a linear region within the
    logarithmic timescale-magnification plane.  {\it First row, left
      panel:} Bulge-bulge lensing; {\it middle panel:} disk-bulge
    lensing; {\it right panel:} MW-halo-bulge lensing for MW lenses
    with $\Mlens=0.1\,\Msun$.  {\it Second row, left panel:} Halo-bulge
    lensing with $\Mlens=1000\,\Msun$; {\it middle panel:} halo-bulge
    lensing with $\Mlens=0.5\,\Msun$; {\it right panel:} halo-bulge
    lensing with $\Mlens=0.1\,\Msun$.  The {dashed} contour level line
    marks the $10^{-8}$ (y$^{-1}$) level in each diagram.  The number
    of events rises toward smaller amplifications and larger
    timescales. Small amplifications are strongly favored for all
    lens-source configurations.  The number of high-amplification
    events rises for smaller $\tfwhm$ timescale. Detecting events
    toward M31 with maximum amplifications of $\A>10^3$ therefore
    requires an experiment that is particularly sensitive to short
    timescale events.  The {dashed blue} line shows for each selected
    timescale the amplification where the distribution is maximal
    [$\partial \lgt(\A)=0$].}
  \label{fig.dGamma_dA0_dtfwhm}
\end{figure}

Figure~\ref{fig.dGamma_dA0_dtfwhm} shows the distribution of events
$\left<\frac{d^2 \Gamma}{d\lgt(\tfwhm/[d])\,d\lgt \A}\right>_\iss$, at
the position $(x,y)=(1\arcmin,0\arcmin)$ in the intrinsic M31 coordinate
system (see Fig.~\ref{fig.tau_wecapp}), i.e., on the disk major
axis.\footnote{We have now changed to logarithmic units for timescale
  and magnification, and also converted the probability density
  according to that.}

Small amplifications are favored, which implies a strong dependency of
the total number of events on the experimental limit of $\A$ (e.g.,
$A_\M{T}$).  Figure~\ref{fig.dGamma_dA0_dtfwhm} can be compared with
sensitivity regions of current experimental setups for microlensing
experiments toward M31. As these are usually only sensitive to
$\tfwhm$ of larger than 1~day, it is extremely unlikely to detect
maximum magnifications larger than $10^3$.  These high-magnification
events can only be routinely detected with combined observations from
several sights located on different longitudes, with large telescopes
allowing short integration times, or from space. Note that recently
after an alert detection and intensive follow-up monitoring,
\cite{2006ApJ...642..842D} could measure a lensing time-scale of
$\tfwhm\approx 0.05\E{d}$ and a magnification of the order 3000.

\section{Applications for the Pixel-Lensing Regime}
\label{sec.app_source}

The microlensing parameters ($\F$, $\tE$, and $\uub$) are not directly
observable anymore in crowded or unresolved stellar fields. In that
case, the two measurable quantities are the full-width timescale
$\tfwhm$ and the difference flux $\DF$ of an event.

We now make use of the luminosity function $\LF(\Mlum)$, the source
number density $n_\iss(x,y,\Dos)$, and the color distribution
$\pcmd(\Mlum,\Col)$ of the source stars introduced in
\S~\ref{sec.source_dis} and derive the event rate distribution
function $\frac{d^2\Gamma}{d\tfwhm \,d\DF}$. This quantity can then be
linked to the measured distributions most
straightforwardly.\footnote{Note that this distribution function is
  different from $d\Gamma/dw_F=d\Gamma / d\left(\DF\,\tfwhm\right)$
  derived by \citet{2000ApJ...530..578B} for the flux-weighted
  timescale $w_F:=\DF\,\tfwhm$.}

In the first two subsections (\S\S~\ref{sec.Gamma_tfwhm_DF} and
\ref{ssec.Gamma_color}) we derive the required distributions
neglecting finite source effects.  However, the high magnifications
needed to boost MS stars to large flux excesses go in parallel with
finite source effects that make these large flux excesses hardly
possible.  We show this in detail in \S~\ref{ssec.finite_source2},
where we incorporate finite source effects in the calculations.

\subsection{Changing Variables of $\Gamma$ to  $\tfwhm$ and $\DF$}
\label{sec.Gamma_tfwhm_DF}

\subsubsection{Event Rate per Star with Absolute Magnitude $\Mlum$}

We now use the relations $\RE(\ml,\Dol,\Dos)$, $\A(\F,\DF)$,
$\tfwhm(\vt,\RE,b)$, $\Psi[\A(\F,\DF)]$, and $\Upsilon[\A(\F,\DF)]$
from \S~\ref{sec.definitions} and the equations
$\tfwhm(\vt,\RE,\F,\DF) = \frac{\RE}{\vt} \,\Upsilon(\A)$,
$\DF(\F,b,\RE) =\F \left[ A(b/\RE) -1 \right]$ and
$\frac{d\DF}{db}=-8\frac{\F}{\RE} \,\uub^{-2}
\,\left[\uub^2+4\right]^{-3/2}\,,$ and we obtain the event rate per
FWHM time, per flux excess, per lens mass and per source star with an
absolute magnitude $\Mlum$:
\begin{equation}
  \darray
  \begin{array}{rl}
    \frac{d^4 \Gamma}{d\tfwhm\,d\DF\,d\ml \,d\Mlum}   
    & = \LF(\Mlum)  \Int_0^\Dos \Intninf
    \Intninf  \GAMb
    \,\delta{\left\{\DF -\F \left[ A\left(\frac{b}{\RE}\right) -1 \right] \right\}} 
    \,\delta{\left\{\tfwhm -  \frac{\RE}{\vt} \,w\right\}} 
    \,db \,d\vt\,d\Dol\\ 
    &=    \LF             
    \Int_0^\Dos \Intninf            
    \Intninf  \rho \,\xi \,\pv \,2 \vt             
    \,\frac{\delta{\left\{b - \RE \left[2 \A (\A^2-1)^{-1/2}-2\right]^{1/2} \right\}}}            
    {\left|\frac{d\DF}{db} \right|}             
    \,\delta{\left(\tfwhm -  \frac{\RE}{\vt} \,\omega\right)}            
    \,db \,d\vt\,d\Dol\\               
    &=    2\LF               
    \frac{1}{\F} \frac{d\uub}{d\A} 
    \Int_0^\Dos \Intninf              
    \rho \,\xi \,\pv  \,\vt \,\RE              
    \,\delta{\left(\tfwhm -  \frac{\RE}{\vt} \,\Upsilon\right)}               
    \,d\vt\,d\Dol\\               
    &=   2\LF                 
    \frac{1}{\F} \frac{d\uub}{d\A}    
    \Int_0^\Dos \Intninf              
    \rho \,\xi \,\pv \,\vt \,\RE              
    \,\frac{\delta{\left(\vt -  \frac{\RE}{\tfwhm} \,\Upsilon\right)}}{\left|-\frac{\RE}{\vt^2} \Upsilon\right|}     
    \,d\vt\,d\Dol\\               
    &=   2 \LF             
    \frac{1}{\F}  \frac{d\uub}{d\A}  \,\Upsilon^2               
    \,\frac{1}{\tfwhm^3}               
    \Int_0^\Dos               
    \rho \,\xi \,\pv{\left(\frac{\RE}{\tfwhm}\,\Upsilon\right)} \,\RE^{3}     
    \,d\Dol\\               
    &=     \frac{2}{\tfwhm^3} \,\LF(\Mlum) 
    \,\xi(\ml) \,\frac{\Psi}{\F} \Int_0^\Dos \rho(\Dol)  \,\RE^3
    \,\pv{\left(\frac{\RE}{\tfwhm}\,\Upsilon\right)} 
    \,d\Dol .
\end{array}
\label{eq.dgam_dt12_dMlum}
\end{equation}
using the luminosity function in magnitudes $\LF(\Mlum)$ and the
conversion from absolute magnitudes to intrinsic source fluxes
$\F(\Mlum,\Dos)$ (eq.~[\ref{eq.F_Mlum}]).  Equation
(\ref{eq.dgam_dt12_dMlum}) is the transformation of equation
(\ref{eq.GAMb}) to the observables relevant in the pixel-lensing
regime. It gives the event rate per star with absolute magnitude
$\Mlum$ and will be converted to the event rate per area using the
density of stars below. For {the special case of highly amplified
  events}, ($\A\gg1$), the approximations $\Psi \approx 12(\F/\DF)^4$
and $\Upsilon \approx \sqrt{12} (\F/\DF)$ can be inserted into
equation (\ref{eq.dgam_dt12_dMlum}).

\subsubsection{Event Rate per Area}

All previously derived event rates are per star, or per star with a
given absolute magnitude $\Mlum$. Observed, however, are event rates
per area. These are obtained from the source density distribution
along the line-of-sight $n_\iss(x,y,\Dos)$ and equation
(\ref{eq.dgam_dt12_dMlum}):
\begin{equation}
    \frac{d^6 \Gamma_{s,l}}{dx\,dy \,d\tfwhm\,d\DF \,d\ml \,d\Mlum} = 
    \frac{2\,\LF_\iss(\Mlum)\,\xi_l(\ml)}{\tfwhm^3}
    \Intninf n_\iss(x,y,\Dos) \,\frac{\Psi}{\F}
    \Int_0^\Dos \rho_l(\Dol)  \,\RE^{3} 
    \,\pv{\left(\frac{\RE\Upsilon}{\tfwhm},v_0\right)} 
    \,d\Dol \,d\Dos, 
  \label{eq.n_dgam_dt12_dMlum}
\end{equation}
where the quantities in the integral have the following functional
dependences $\F(\Mlum,\Dos)$, $\rho_l(x,y,\Dol)$,
$\RE(\Dol,\ml,\Dos)$, $\Psi(\A(\F,\DF))$, $\Upsilon(\A(\F,\DF))$,
$v_0(x,y,\Dol,\Dos)$.  Equation (\ref{eq.n_dgam_dt12_dMlum}) is the
event rate per interval of lens plane area, FWHM time flux excess,
lens mass and absolute magnitude of the lensed star.  For highly
amplified events one can replace $\Psi $ and $\Upsilon $ in the
integral by $12(\F/\DF)^4$ and $\sqrt{12}(\F/\DF)$, respectively.

Different lens (disk, bulge, or halo) and source (disk or bulge)
populations are characterized by an index $l$ and $s$ in equation
(\ref{eq.n_dgam_dt12_dMlum}).  For the total event rate $\Gamma_{\rm
  tot}$ one has to sum up the contributions of all lens-source
configurations:
\begin{equation}
  \frac{d^6 \Gamma_\M{tot}}{\,dx\,dy \,d\tfwhm\,d\DF \,d\ml \,d\Mlum} :=
  \Sum_{s} \Sum_{l}
  \frac{d^6 \Gamma_{s,l}}{dx\,dy \,d\tfwhm\,d\DF \,d\ml \,d\Mlum}
  .
  \label{eq.n_dgamtot_dt12_dF0}
\end{equation}
The event rate per area is then obtained by multiplying equation
(\ref{eq.n_dgamtot_dt12_dF0}) with the efficiency
$\eff(x,y,\DF,\tfwhm)$ of the experiment and integrating over all lens
masses and source magnitudes, and the timescale and flux excess.  The
probability that one can observe two stars lensed at the same time at
the same position is practically zero, since $\int_{\Opsf} (
d^2\Gamma_{\rm tot} / dx\,dy ) \;dx \;dy \ll 1 $ holds.

\begin{figure}
  \epsscale{1.0}\plotone{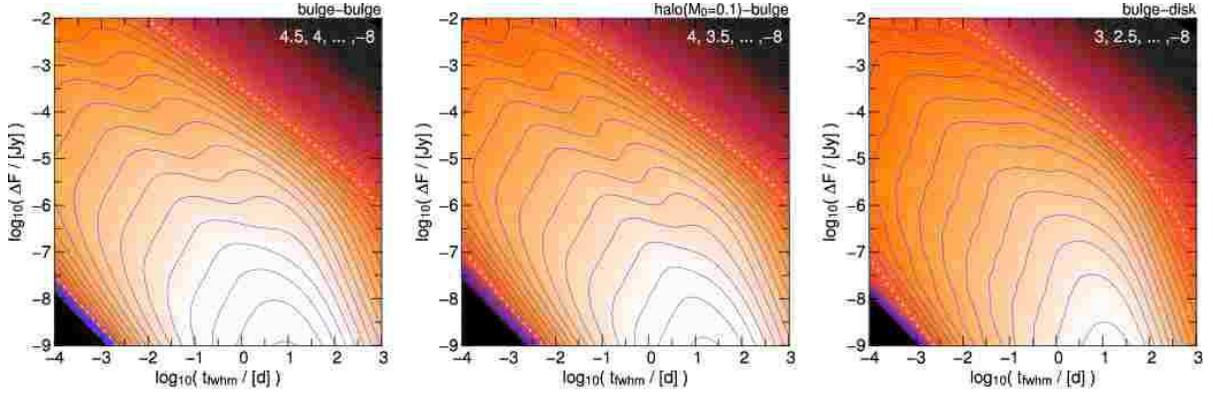}
  \caption{Event rate per area, per FWHM time and per flux excess,
    $d^4 \Gamma/(dx \;dy \;d\lgt\tfwhm\,\,d\lgt\DF)$, obtained
    from equation (\ref{eq.n_dgam_dt12_dMlum}) by mass and magnitude
    integration. The contours are shown in units of
    $\M{y}^{-1}\E{arcmin}^{-2}$, timescales and flux excess have been
    inserted in units of $\M{days}$ and $\M{Jy}$.  The equations are
    evaluated at $(x,y)=(1\arcmin,0\arcmin)$ in the intrinsic M31
    coordinate system (see Figure~\ref{fig.tau_wecapp}), i.e., at a
    distance of $1\arcmin$ along the disk major axis. We show
    bulge-bulge lensing ({\it left}), halo-bulge lensing with
    $\Mlens=0.1\,\Msun$ lenses, and bulge-disk lensing ({\it right}).
    The contour levels are given as inserts in each diagram.  The red
    dashed line marks the $10^{-3} \E{y}^{-1}\E{arcmin}^{-2}$ level in
    each diagram, brighter areas correspond to higher values. The
    double-wave shape of the contours with bulge stars as sources is
    caused by the shape of the PMS luminosity function of the bulge
    sources (see Figure~\ref{fig.dGamma_PMS_MS}). For the results
    shown in these panels all sources have been treated as
    point sources.  For finite source effects, see
    \S~\ref{ssec.finite_source2}.}
  \label{fig.dGamma}
\end{figure}

We carry out mass and magnitude integration of equation
(\ref{eq.n_dgam_dt12_dMlum}) for the position
$(x,y)=(1\arcmin,0\arcmin)$ in the intrinsic M31 coordinate system
(see Figure~\ref{fig.tau_wecapp}), i.e., at a distance of $0.22\E{kpc}$
along the disk major axis and show the results for bulge-bulge,
halo-bulge, and bulge-disk lensing in
Figure~\ref{fig.dGamma}. Compared to
Figure~\ref{fig.dGamma_dA0_dtfwhm} the contours are smeared out in the
$\Delta F$-direction, since they come from convolving those in
Figure~\ref{fig.dGamma_dA0_dtfwhm} with the source luminosity
function.

\begin{figure}
  \epsscale{1.0}\plotone{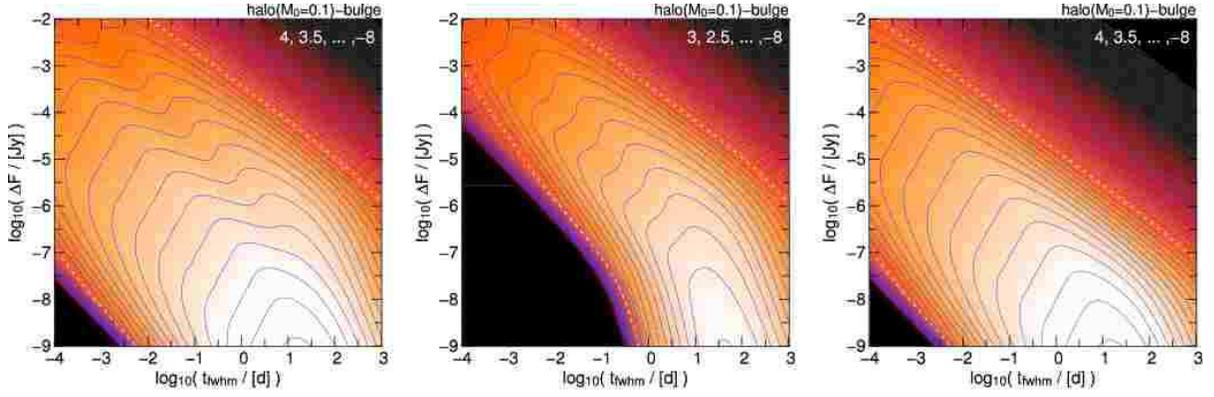} 
  \caption{Event rate per area, per FWHM time and per flux excess,
    $d^4 \Gamma/dx\;dy\;d\lgt\tfwhm\,\,d\lgt\DF)$, for
    halo-bulge lensing with $\Mlens=0.1\,\Msun$ calculated at the
    position $(x,y)=(1,0)\E{arcmin}$.  {\it Left panel:} Distribution
    for PMS and MS sources. {Middle panel:} Distribution for PMS
    sources alone. {Right panel:} Distribution for MS sources
    alone. The contour levels are shown as inserts in each
    diagram. The {red} line marks the $10^{-3}
    \E{y}^{-1}\E{arcmin}^{-2}$ level in each diagram. The double -wave
    shape of the PMS distribution is due to the two peaks in the PMS
    luminosity function of the bulge sources (see Figure~\ref{fig.LF}
    in \S~\ref{app.model}). For the results shown in these
    panels all sources have been treated as point sources.}
  \label{fig.dGamma_PMS_MS}
\end{figure}

In Figure~\ref{fig.dGamma_PMS_MS} we demonstrate for the halo-bulge
lensing case in Figure~\ref{fig.dGamma} that the ``double-wave shape''
in the contours in the two left panels of Figure~\ref{fig.dGamma}
indeed is caused by the luminosity function of the PMS stars. We split
the source stars into post--main-sequence (PMS) and main-sequence (MS)
stars and plot the corresponding contours into the middle and right
panels of that figure.  The double-wave shape appears only in the PMS
figure.  Besides that, it becomes obvious that PMS stars cannot be
lensed into events with short timescales and small flux excess. This
is because the faintest PMS stars in the M31 bulge have an unamplified
flux of $8\times\,10^{-9}\E{Jy}$ and thus need an amplification of only
a factor of $2$ to yield a flux excess of $\DF \approx
10^{-8}\E{Jy}$. Magnifications that small are incompatible with short
timescales according to Figure~\ref{fig.dGamma_dA0_dtfwhm}. In
contrast, MS stars need very high amplifications to reach a flux
excess comparable to that typical for PMS stars. According to the
right panel in Figure~\ref{fig.dGamma_PMS_MS}, ultra-short, large
excess flux events with MS source stars would be more common [compare,
e.g., the contour levels at $\lgt( \tfwhm /[\M{d}]) = -3$ and $\lgt
(\Delta F/ [\M{Jy}]) ={-4}$] than events with PMS source stars.

\subsection{Including Color Information in the Event Rate}
\label{ssec.Gamma_color}

The color of a point source remains unchanged during a lensing event,
since the lensing amplification does not depend on the frequency of
the source light.  In practice, {microlensing} events with blending by
nearby stars, and {any} event with finite source signatures may show
chromaticity in the light curve (see e.g.
\citet{1995lssu.conf..326V,1995ApJ...449...42W,2000MNRAS.316...97H}).
The difference imaging technique eliminates all blended light from the
lensing light curve. For lensing events without finite source effects
the color of the event therefore equals that of the source and can be
used to constrain the source-star luminosities.

Replacing $\LF(\Mlum)$ with $\pcmd(\Mlum,\Col)$, and $d\Mlum$ with
$\,d\Mlum \,d\Col$ (see \S~\ref{sec.color}), we obtain
\begin{equation}
  \frac{d^7 \Gamma_{s,l}}{\,dx \,dy \,d\tfwhm \,d\DF \,d\ml \,d\Mlum \,d\Col}  
  =  \frac{2}{\tfwhm^3}
  \,\pcmd_\iss(\Mlum,\Col) \,\xi_l(\ml)
  \Intninf n_\iss(x,y,\Dos) \,\frac{\Psi}{\F} 
  \Int_0^\Dos \rho_l(\Dol)  \,\RE^{3} 
  \,\pv{\left(\frac{\RE\Upsilon}{\tfwhm},v_0\right)} 
  \,d\Dol \,d\Dos 
  .
   \label{eq.n_dgam_dt12_dCol}
\end{equation}
We derive lens mass estimates starting from equation
(\ref{eq.n_dgam_dt12_dCol}) in \S~\ref{sec.massprob}. We also
demonstrate there that including the color information leads to
considerably smaller allowed lens mass intervals than for the case in
which color information is ignored (i.e., the case in which lens mass
probability functions are derived from
eq. [\ref{eq.n_dgam_dt12_dMlum}]).

Equation (\ref{eq.n_dgam_dt12_dCol}) allows to reconstruct the mass
function of the lenses and the MACHO fraction in the dark halo (see
\citet{1991MNRAS.250..348D,1994PhLB..323..347J,1994ApJ...432L..43J,
  1996ApJ...473...57M,1996ApJ...467..540H,1996PASP..108..465G}).  In
this way one can obtain the optimal parameterization for the mass
function $\xi_l(\ml)$ using a maximum-likelihood analysis for a set of
measured lensing events.  If the ingredients for the kernel (i.e., all
but the pre-factor $\xi_l(\ml)$ in eq. [\ref{eq.n_dgam_dt12_dCol}])
are accurately provided by theory and the number of lensing events is
large, then the mass distribution can be derived solving the Fredholm
integral equation of the first kind.  Inversely, a certain ensemble of
lenses allows conclusions on the based distribution functions.

\subsection{Event Rate Taking into Account Finite Source Effects}
\label{ssec.finite_source2}

As described in \S~\ref{sec.def_fs} the point-source approximation is
no longer valid, if the impact parameter $\uub$ is smaller than
$\ufs$, i.e., half the source radius projected onto the lens plane
(equation (\ref{eq.fs})).  In this case, the maximum amplification and
thus the flux excess stays below the value for the point-source
approximation, and timescales of events are enlarged (see
Eqs.~\ref{eq.t12_fin} and \ref{eq.A0_fin}).
\cite{2000ApJ...530..578B} already accounted for the upper limit in
magnification and obtained the correct value for the total number of
events (i.e., events with and without finite source signatures) as a
function of magnification threshold.  Their approximation, however, is
limited to high amplifications and ignores the change of magnification
and event timescale.\footnote{\citeauthor{2000ApJ...530..578B}'s (\citeyear{2000ApJ...530..578B}) eq.~(26) with
  eqs.~(20) and (22) can be written in our notation as (see
  footnote~\ref{fn.baltz})
  \begin{displaymath}
    \frac{d\Gamma}{d\tfwhm}
      =  \frac{\Dos}{\Mlens} \Int_{-\infty}^\infty \LF(\Mlum) \,\Int_0^1
      \,\theta{\left( \left(1+ \frac{\Rstar^2 c^2}{16 G\ml \Dos
              \uT^2}\right)^{-1} - \frac{\Dol}{\Dos}\right)} \,\rho(\Dol)
      \,\Int_0^\infty \,\theta{\left(2\frac{\RE}{\tfwhm} \sqrt{12} \uT -
          \vt\right)} \,\vt^2 \,\pv(\vt) \,\left\{w'{\left[w^{-1}{\left(\frac{\tfwhm
                  \vt}{2\RE}\right)}\right]}\right\}^{-1} \,d\vt
      \,d\left(\frac{\Dol}{\Dos}\right) \,d\Mlum .
  \end{displaymath}
}
Thus, the flux excess and timescale distributions of the events are
not predicted accurately.

We have shown in \S~\ref{sec.def_fs} that finite source effects are
likely already for small maximal magnifications and that the timescale
changes due to finite source effects can be large. Therefore, we
derive precise relations and account for the finite source sizes as
follows: 
\\

1. Events with $\uub> \ufs$, i.e., those for which the finite source
sizes are irrelevant, are treated as before; we redo all calculations
starting from equation (\ref{eq.dNdb}), and if the impact parameter
$b$ is involved in an integral we multiply the integrand with
$\theta{\left(b-\RE\ufs\right)}$; the step function allows only
contributions in the integrand, if $b \ge \RE \ufs$ holds.  To see how
this transports into the $d\Dol$-integration if the variables are
changed from $b$ and $\vt$ to $\tfwhm$ and $\Delta F$ in the
Eqs.~\ref{eq.dgam_dt12_dMlum}, \ref{eq.n_dgam_dt12_dMlum} and
\ref{eq.n_dgam_dt12_dCol}
\begin{equation}
    \theta{\left(b-\RE\ufs\right)} = \theta{\left(\ANfs - \A \right)}
    = \theta{\left( \Dos \left\{1 + \frac{c^2 \,\Rstar^2
    \frac{\DF}{\F} \left(2+\frac{\DF}{\F}\right)}{16 \,G\ml\,\Dos}\right\}^{-1} -
    \Dol\right)} = \theta{\left( \Dolfs - \Dol\right)}
  \label{eq.dgam_dt12_dFlum_nfs}
\end{equation}
where  we are using the following relations:
\begin{displaymath}
\Dolfs := \Dos \left(1 + \frac{\DF(2\F+\DF)}{\Cfs\,\Dos} \right)^{-1} 
\approx \Dos \left(1 + \frac{\DF^2}{\Cfs\,\Dos} \right)^{-1},
\end{displaymath}
and
\begin{displaymath}
  \Cfs:=\frac{16 \,\F^2 \,G\ml}{c^2 \,\Rstar^2}
  \,\left(\M{flux}^2\E{length}^{-1}\right).
\end{displaymath}
Multiplying the integrand of equations (\ref{eq.dgam_dt12_dMlum}),
(\ref{eq.n_dgam_dt12_dMlum}), and (\ref{eq.n_dgam_dt12_dCol}) with
equation (\ref{eq.dgam_dt12_dFlum_nfs}) extracts only those light
curves, where finite-source effects can be neglected.
\\

2. For events where the finite source sizes are relevant, i.e., events
with $\uub<\ufs$, we use the approximations for the maximum
amplification and the FWHM time given in equations (\ref{eq.Afin}) and
(\ref{eq.t12_fin}).  This means that we just replace the relations for
the impact parameter and the maximum magnification and the FWHM
timescale relations of events by equations (\ref{eq.A0_fin}) and
(\ref{eq.t12_fin}) when switching from the point source to the finite
source regime.  We then can derive the equations for the event rates
with finite source effects from equation (\ref{eq.GAMb}) analogously
to the point-source approximation, but this time with a step function
of $\theta{\left(\RE\ufs-b\right)}$ in the integrands allowing only
small impact parameters. With $\ANfs(\Dol,\Dos,\Rstar,\ml)$,
$\ufs(\Dol,\Dos,\Rstar,\ml)$ and $\vt = \left(\RE/\tfwhm\right)
\Upsfs(\ANfs,\uub)$ and its derivative
$\left|d\tfwhm(\vt)/d\vt\right| = \left(\RE/\vt^2\right)
\Upsfs$ we obtain
\begin{equation}
  \darray
  \begin{array}{rl}
   \frac{d^6 \Gamma_{s,l}}{\,dx \,dy \,d\tfwhm \,d\DF \,d\ml \,d\Mlum} 
    = & \Intninf n_\iss(x,y,\Dos)  \,\LF_\iss(\Mlum) \Int_0^\Dos \Intninf \Intninf  \GAMb \,\delta{\left(\DF -\F (\ANfs -1) \right)} \\
    & \times 
    \,\delta{\left(\tfwhm - \frac{\RE}{\vt} \Upsfs\right)} 
    \,\theta\left(\RE\ufs-b\right) 
    \,db \,d\vt\,d\Dol\,d\Dos \\ 
    =  & \Intninf n_\iss(x,y,\Dos) \,\LF_\iss(\Mlum) 
     \Int_0^\Dos\rho(\Dol) \Intninf  \Intninf   \,\xi(\ml) \,\pv(\vt) \,2 \vt    \\  
     &  \times \,\delta{\left(\DF -\F (\ANfs -1) \right)} 
    \,\frac{\delta{\left(\vt - (\RE/\tfwhm) \Upsfs\right)}}
		       {(\RE/\vt^2) \Upsfs}
    \,\theta\left(\RE\ufs-b\right) 
   \,d\vt \,db  \,d\Dol \,d\Dos \\ 
    =  & \Intninf n_\iss\, 2\,\LF_\iss(\Mlum)   
      \Int_0^\Dos\rho(\Dol) \,\xi(\ml)  \,\delta{\left(\DF -\F (\ANfs -1) \right)} 
		 \Int_0^{\RE\ufs} 
		 \,\pv{\left( \frac{\RE}{\tfwhm} \Upsfs\right)} 
		 \, \frac{\RE^2{\Upsfs}^2}{\tfwhm^3} \,db
    \,d\Dol \,d\Dos \\ 
     =  & \Intninf n_\iss(x,y,\Dos)    \frac{2}{\tfwhm^3}   \,\LF_\iss(\Mlum) \,\xi(\ml) 
		 \,\rho(\Dolfs)   
     \,\Omegafs
		 \,\RE(\Dolfs)^3 \Int_0^{\ufs}  
		 \,\pv{\left( \frac{\RE(\Dolfs)}{\tfwhm} \Upsfs\right)} 
		 \,{\Upsfs}^2 \,d\uub
     \,d\Dos ,
\end{array}
\label{eq.dgam_dt12_dFlum_wfs}
\end{equation}
with
\begin{displaymath}
    \Omegafs := \left|\frac{d\Dolfs(\DF,\Dos)}{d\DF}\right| 
     = 2 \Cfs \Dos^2 (\F + \DF) \left(\Cfs \Dos + \DF(2\F+\DF)\right)^{-2} 
     \approx  2 \Cfs \Dos^2 \DF \left(\Cfs \Dos + \DF^2\right)^{-2}.
\end{displaymath}
Alternatively a transformation inverting $\tfwhm(b)$ is
possible.\footnote{Using $b = \left\{\left[\RE\,\uFStfwhm\right]^2 -
    \vt^2\left(\frac{\tfwhm}{2}\right)^2\right\}^{1/2}$ and its derivative
  $\left|\frac{db}{d\tfwhm}\right| = \vt^2 \left\{\left[\frac{2\RE}{\tfwhm}
      \uFStfwhm\right]^2 - \vt^2\right\}^{-1/2}$ we obtain
  \begin{displaymath}
    \frac{d^6 \Gamma_{s,l}}{\,dx \,dy \,d\tfwhm \,d\DF \,d\ml \,d\Mlum}
    = \Intninf n_\iss(x,y,\Dos) 2 \,\LF_\iss(\Mlum) \,\xi(\ml)
    \left|\frac{d\Dol}{d\DF}\right| \,\rho(\Dol)
    \,\Int_{v_\M{t,min}}^{v_\M{t,max}} \pv \vt \left|\frac{db}{d\tfwhm}\right|
    \,d\vt \,d\Dos 
  \end{displaymath}
  with $v_\M{t,min} := \frac{2\RE}{\tfwhm} \sqrt{\uFStfwhm^2 - \ufs^2}
  \approx \frac{\sqrt{3} \,\Rstar\,\Dol(\DF,\Dos)}{\tfwhm\,\Dos}$ and
  $v_\M{t,max} := \frac{2 \RE}{\tfwhm}\uFStfwhm \approx \frac{2 \Rstar
    \Dol}{\tfwhm\,\Dos}$.  }
We use the values for the source radius, luminosity, and color
relations $\Rstar(\Mlum,\Col)$ summarized in Appendix~\ref{app.model} 
(\S\S~\ref{sec.LFmodel} and
\ref{sec.radius_flux}).
\begin{figure}
  \epsscale{1.0}\plotone{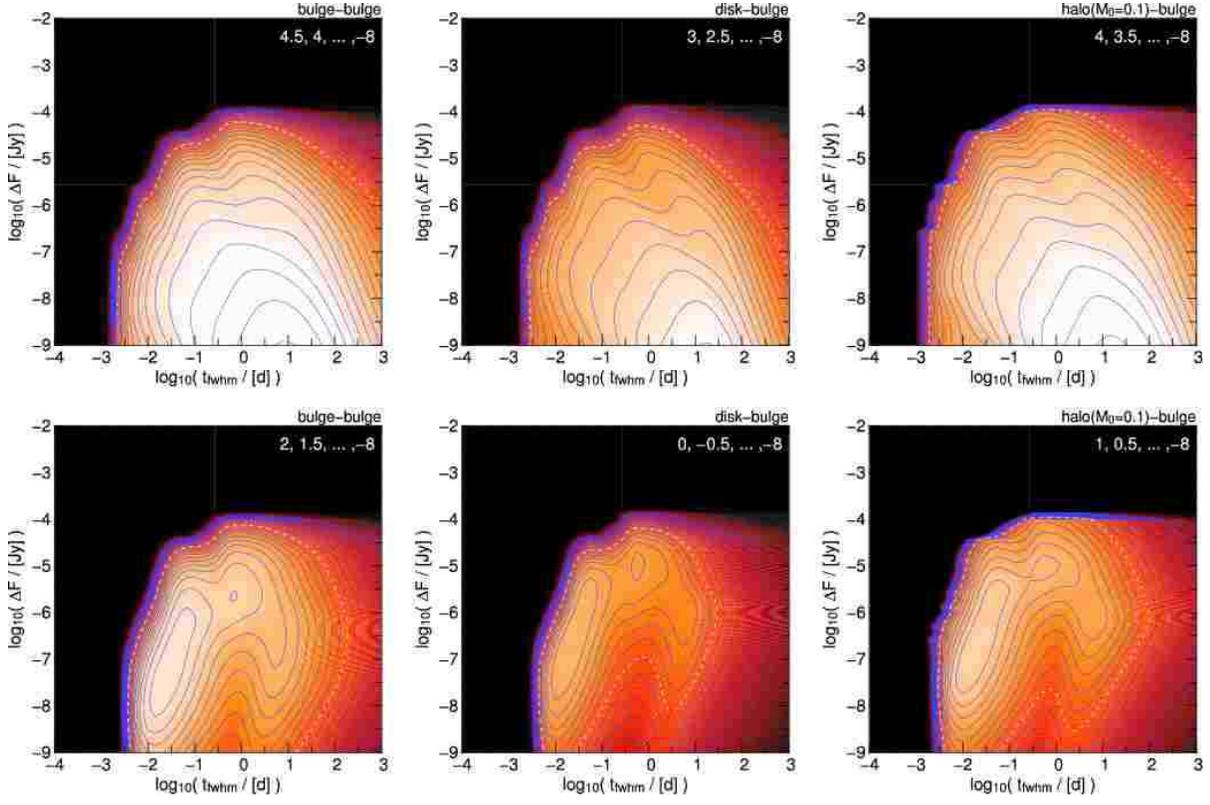} 
	\caption{$d^2 \Gamma/\left(d\lgt\tfwhm\,d\lgt\DF\right)$
    [$\E{arcmin^{-2}}\E{y^{-1}}$] at $(x,y)=(1\arcmin,0\arcmin)$ in
    the $\lgt(\DF)$-$\lgt(\tfwhm)$ plane, for bulge-bulge ({\it
      left}), disk-bulge ({\it middle}), and halo-bulge ({\it right})
    lensing with $0.1\,\Msun$ MACHOs (columns 1-3).  The upper panels
    show the distribution for light curves not affected by the finite
    source sizes.  The contours have been obtained from inserting
    eq.~(\ref{eq.n_dgam_dt12_dMlum}) into
    eq.~(\ref{eq.dgam_dt12_dFlum_nfs}) and carrying out the mass and
    source luminosity integral.  The lower panels show the
    distribution for light curves with finite source signatures (mass
    and source star luminosity integral of
    eq.~(\ref{eq.dgam_dt12_dFlum_wfs}).  The contour levels can be
    read off from the inserts in each diagram.  The {dashed} line
    marks the $10^{-3} \E{arcmin}^{-2}\E{y}^{-1}$ level, areas with
    brighter colors correspond to higher contour values.  Taking into
    account the finite source sizes implies an upper limit for $\DF$
    and a lower limit for $\tfwhm$ for {all} light curves, i.e., for
    light curves with and with out finite source signatures (see
    text).  For the source-lens configurations shown here there are no
    lensing eventslight curves with excess fluxes $\DF>5\times
    10^{-4}\E{Jy}$. The results shown here have been obtained by
    taking into account the source sizes of lensed stars.}
\label{fig.dGamma_fs_1}
\end{figure}
\begin{figure}
  \epsscale{1.0}\plotone{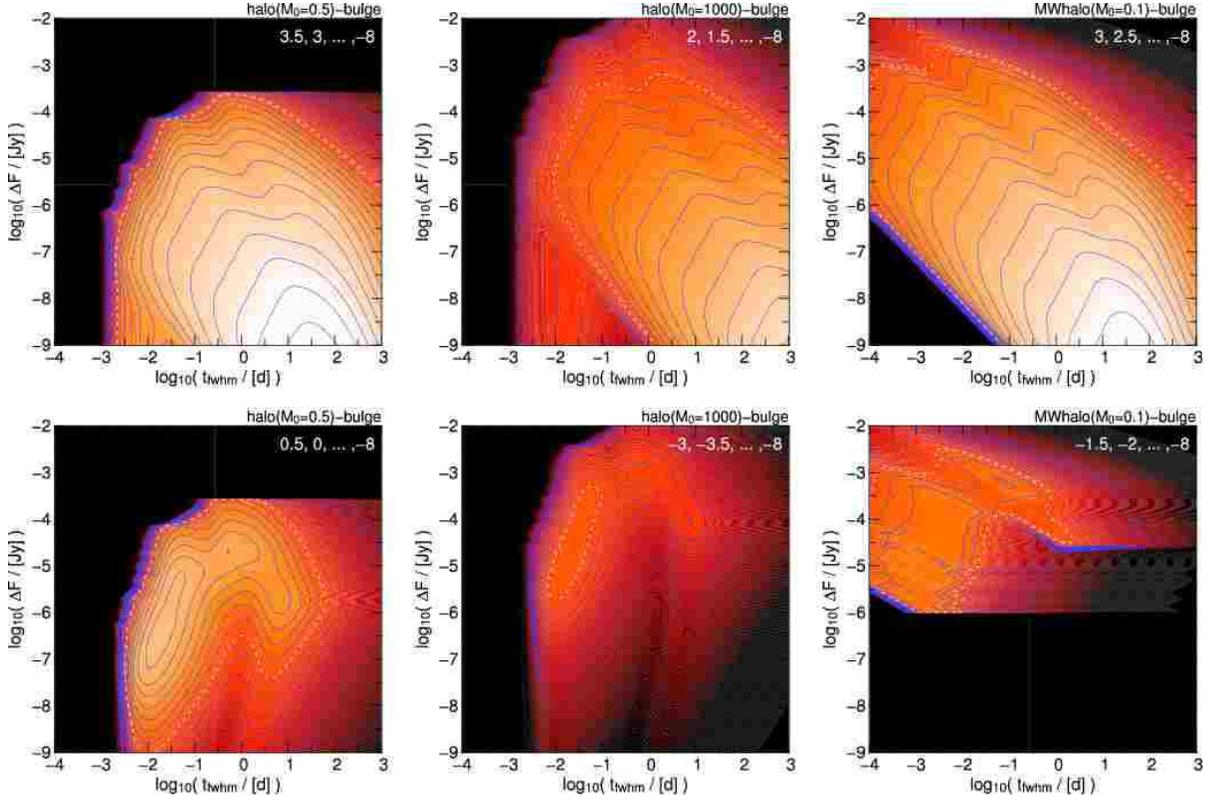}
  \caption{Same as Fig.~\ref{fig.dGamma_fs_1} for halo-bulge lensing
    with larger MACHO masses ($0.5\,\Msun$ and $1000\,\Msun$ MACHOs in the
    first and second column) and for Milky Way halo - M31 bulge
    lensing (for $0.1\,\Msun$ MACHOs in the third column) As before, the
    {dashed} line marks the $10^{-3} \E{arcmin}^{-2}\E{y}^{-1}$ level
    in each diagram, and areas with brighter colors correspond to
    contour higher values.  {The results shown here have been obtained
      by taking into account the source sizes of lensed stars.}  }
\label{fig.dGamma_fs_2}
\end{figure}

Figures~\ref{fig.dGamma_fs_1} and \ref{fig.dGamma_fs_2} show contours
of the event rate per $\tfwhm$ timescale and flux excess, per year and
square arcminute, with finite source effects taken into account. We
use the same position as before, at $(x,y)=(1\arcmin,0\arcmin)$ in the
intrinsic M31 coordinate system (see Figure~\ref{fig.tau_wecapp}) or
at a disk major axis distance of $1\arcmin$.  The upper panels show
the distribution for light curves showing no finite source effects
(eq.~[\ref{eq.n_dgam_dt12_dMlum}] with
eq.~[\ref{eq.dgam_dt12_dFlum_nfs}]), whereas the lower panels show the
distribution obtained from mass and source luminosity integration of
equation (\ref{eq.dgam_dt12_dFlum_wfs}), i.e., for light curves
affected by finite source effects.

The black areas indicate the event parameter space, which is not
available to source stars once their real sizes are taken into
account: as finite source effects mainly occur at large
amplifications, large $\DF$ and small $\tfwhm$ values are suppressed.
Events in the point-source approximation, which fall into the black
areas in the upper panels of Figures~\ref{fig.dGamma_fs_1} and
\ref{fig.dGamma_fs_2}, end up with longer timescales and lower excess
fluxes ({\it lower panels}) if the sources sizes are taken into
account.  The sharp cutoff at large flux excesses arises, since there
is an upper limit in $\DF$ depending on source luminosity and size
(see equation (\ref{eq.largest_deltaF})) and since the luminosity
function of the stars has a steep cutoff at giant luminosities of
$\Mlum_R = -0.83\E{mag}$ (bulge) and $\Mlum_R = -2.23\E{mag}$ (disk).
The maxima with vertical contours for finite source effects in the
lower panels come from shifting events for which the point-source
approximation ``just'' fails at longer times scales (see
eq.~[\ref{eq.t12_fin}]).  Light curves with finite source effects have
(depending on their flux excesses) most likely FWHM timescales of
about $0.01$days, or 15 minutes, and the sources lensed with that
timescales are MS stars.  The secondary maxima around $1$ day and flux
excesses of $5\times 10^{-6}$ to $2\times 10^{-5}\E{Jy}$ for
bulge-bulge, disk-bulge, and $0.1\,\Msun$ halo-bulge lensing, and of
about $10^{-3}\E{Jy}$ for $1000\,\Msun$ halo-bulge lensing, are due to
lensing of PMS stars.

\begin{figure} 	
  \epsscale{0.45}\plotone{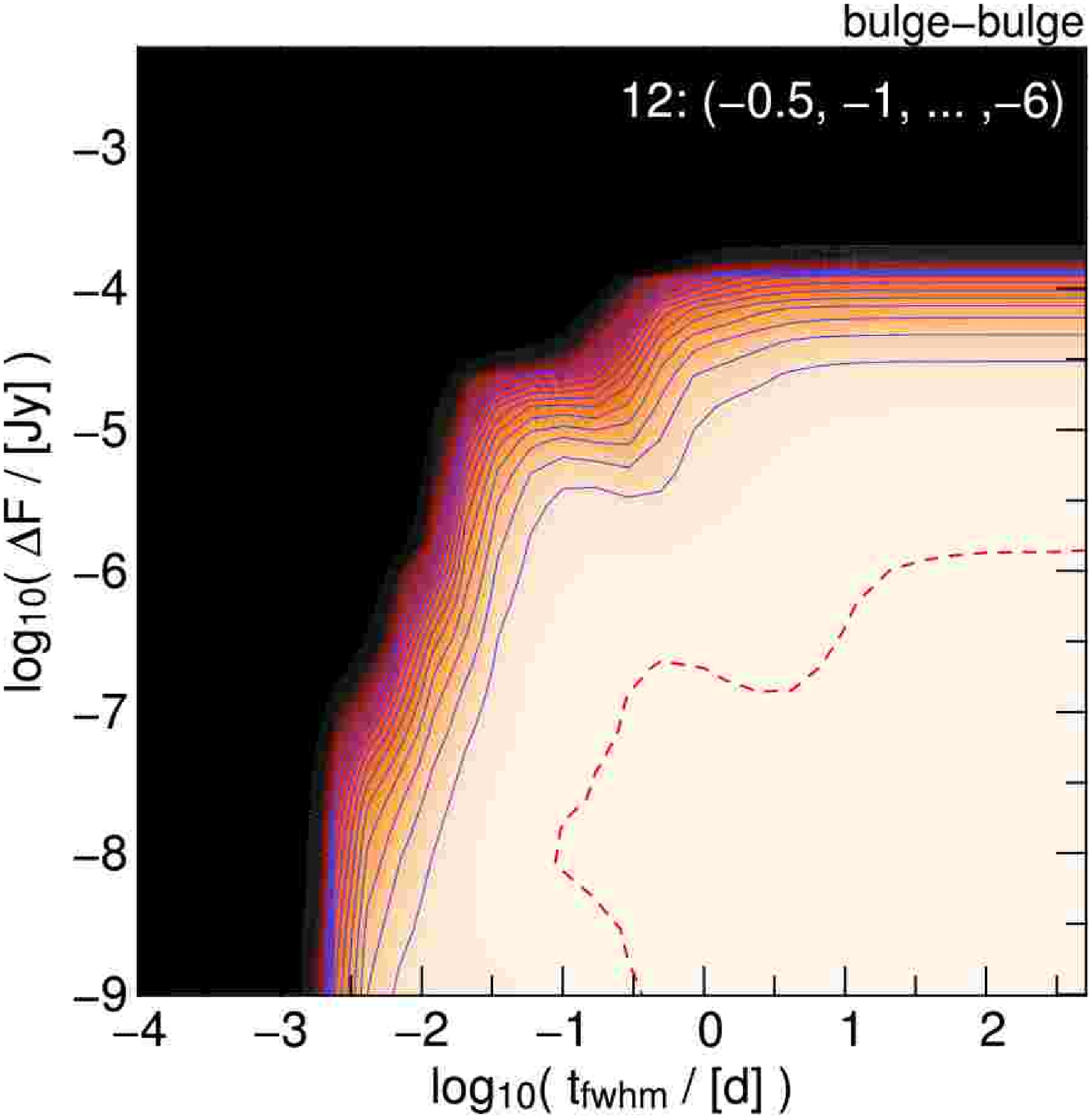} \hspace{5mm}
  \epsscale{0.45}\plotone{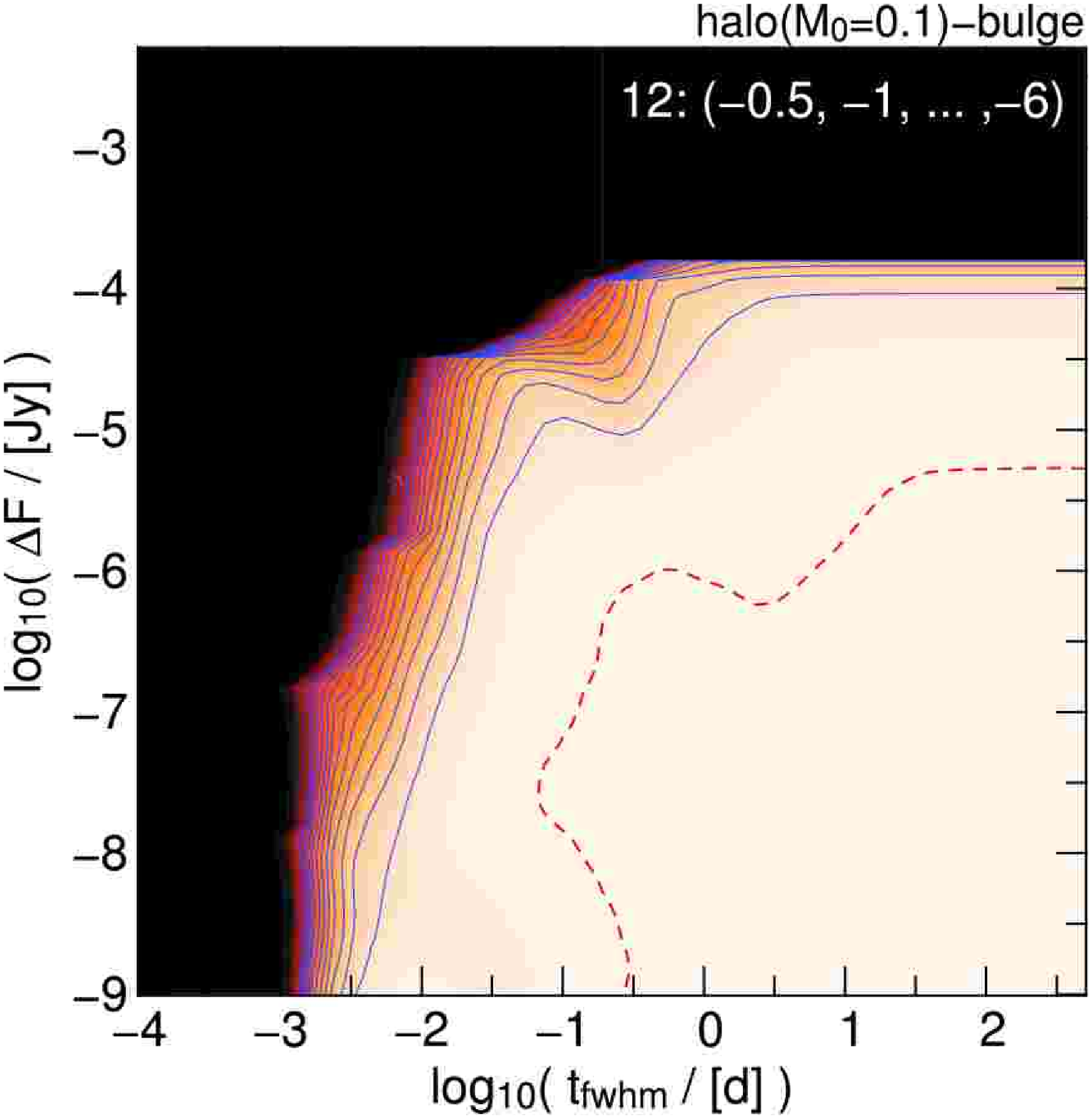}
	\caption{ Ratio of the event rate distribution for extended sources
    counting events not showing finite source signatures (equation
    (\ref{eq.dgam_dt12_dFlum_nfs}) and Figure~\ref{fig.dGamma_fs_1})
    to the event rate assuming pure point sources (equation
    (\ref{eq.n_dgam_dt12_dMlum}) and Figure~\ref{fig.dGamma}), as a
    function of flux excess and timescale of the events:
    $\left[d^2\Gamma/(d\lgt\tfwhm\,d\lgt\DF)\right]_\M{finite\
      sources} \left/\left[d^2\Gamma/(d\lgt\tfwhm\,d\lgt\DF)\right]_\M{point\
        sources}\right.$ for bulge-bulge ({\it left}) and $0.1\,\Msun$
    halo-bulge ({\it right}) lensing at $(x,y)=(1\arcmin,0\arcmin)$ in
    the $\lgt(\DF)$-$\lgt(\tfwhm)$ plane. The dashed red line marks a
    ratio of 0.999. The finite source sizes cause a strong suppression
    of the brightest lensing events relative to the point-source
    predictions.  }
	\label{fig.ratio_bb_ps_fs}
\end{figure}

In general, the ratio of lensing events with and without finite source
signatures is minute for $\tfwhm \ga 0.5$d and $\Delta F < 10^{-6}$Jy,
and raises to about an order of unity for bright lensing events with
$\Delta \F \approx 1.6 \times 10^{-5}$ Jy (corresponding to a magnitude
of the excess flux of $m_R=20.7\E{mag}$) for bulge-bulge-lensing and
$\Delta \F \approx 5 \times 10^{-5}$ ($m_R=19.5\E{mag}$) for halo-bulge
lensing with $0.1\,\Msun$ lenses.  We compare the first column,
bulge-bulge lensing, with results for the same lens-source
configuration in Figure~\ref{fig.dGamma}, which had been obtained
assuming the full validity of the point-source approximation. The
ratio of these contours is shown in Figure~\ref{fig.ratio_bb_ps_fs}.
The parameter space of interest for current surveys are flux excesses
$>10^{-5}\E{Jy}$ (excess magnitude of $m_R=21.2\E{mag}$) and timescales
between 1 and 200 days. One can see that the true event rate
can differ strongly from that for the point-source approximation
depending on the flux excess limit of the survey. The brightest events
are preferentially suppressed.  This means that taking into account
the source sizes is essential for predicting the correct number of
lensing events.

Furthermore, one has to be aware that a fair fraction of the brightest
lensing events show finite source signatures in their light curves and
might be missed when using event filters with a classical lensing
event shape in a stringent way. For the detection of finite-source
events or even of binary lensing events less stringent thresholds or
modified filters are needed, which, however, enhance the risk of a
mismatch with variable source detections.

Finally, Figure~\ref{fig.dGamma_fs_2} compares halos with different
MACHO masses in its first and second row. An increase in MACHO mass
dramatically reduces the event rate and increases the event
timescales. This explains the shift in the contours toward longer
timescales (compare the change of the $\A$-$\tfwhm$ contours in
Fig.~\ref{fig.dGamma_dA0_dtfwhm}) and the decrease in the contour
levels.  For larger MACHO masses, Einstein radii do increase, and one
expects finite source effects to become less important: the largest
possible flux excess $\DFmax$ for the lensing events indeed increases;
the size of the shift is as expected, since the maximum flux excess is
proportional to the square root of the MACHO mass according to
equation (\ref{fig.dGamma_dA0_dtfwhm}).  The contours in the last row
of Figure~\ref{fig.dGamma_fs_2} show MW-halo lensing with $0.1\,\Msun$
MACHOs. Finite source effects are
unimportant. Figures~\ref{fig.dGamma_fs_1} and \ref{fig.dGamma_fs_2}
make it obvious that lensing events above the maximum flux excess
predicted for self-lensing would be a clear hint for either massive
MACHOs in M31 or MACHOs with unconstrained masses in the Milky Way.

\begin{figure}
  \epsscale{0.99}\plotone{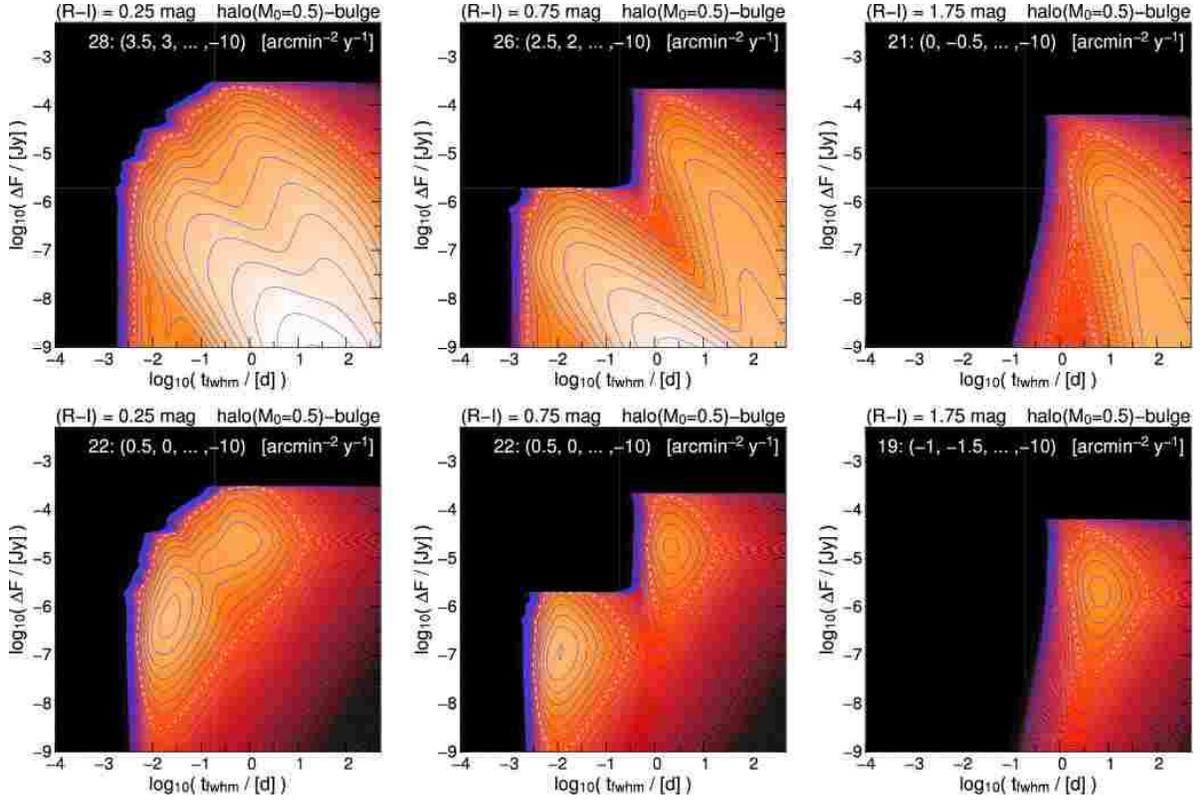}
  \caption{Distribution $d^2 \Gamma/(d\lgt\tfwhm\,d\lgt\DF)$
    $[\M{arcmin^{-2}}\E{y}^{-1}]$ for $0.5\,\Msun$ halo-bulge lensing at
    $(x,y)=(1,0)\E{arcmin}$ in the $\lgt(\DF)$-$\lgt(\tfwhm)$
    plane. We show the distribution for bulge sources within
    different color intervals, $(R-I) \in [0,0.5]\E{mag}$ (first
    column), $(R-I) \in [0.5,1.0]\E{mag}$ (second column), and $(R-I)
    \in [1.5,2.0]\E{mag}$ (third column). The {upper panels} show the
    distributions for light curves showing no finite source effects,
    whereas the {middle panels} show the distributions for light
    curves with finite source signatures.  The {lower panels} show the
    sum of both distributions.  The contour levels are shown as
    inserts in each diagram.  The {dashed} line marks the $10^{-3}
    \E{arcmin}^{-2}\E{y}^{-1}$ level in each diagram, brighter areas
    correspond to higher values. See text for more details.}
\label{fig.dGamma_fs_col}
\end{figure}

Figure~\ref{fig.dGamma_fs_col} shows the distribution for bulge-bulge
lensing splits in color space. The selected color intervals are
$0.0<R-I<0.5$, $0.5<R-I<1.0$, and $1.5<R-I<2.0$. In the bluest color
interval ({\it first column}) we find MS stars close to the MS turnoff
as well as SGB, red clump and some RGB stars.  The medium red sample
contains MS, RGB. and AGB stars, and the reddest sample ({\it last
  row}) contains stars in the RGB and AGB phase and no MS stars.  As
expected, the timescale of the most likely finite source lensing
events changes with color: for the bluest color interval MS stars are
responsible for the most likely finite source signature events and the
event timescales are very short. The secondary maximum is caused by
red clump and SGB stars, which are brighter, need less magnification
and therefore have longer event timescales.  The color interval of
$0.5<R-I<1.0$ contains the central part of the MS, and RGB and AGB
stars. The MS stars are fainter (in $R$) and have smaller radii than
those in the blue sample, and therefore, the maximally probable event
caused by the MS stars is at lower flux excess and timescale than that
for the bluer sample. The PMS stars are brighter (which enhances the
possibility of longer timescale events) and have larger radii (which
leads to stronger peak-flux depression by finite source sizes) than in
the bluer sample, and therefore, the events have similar brightness
but take longer on average.  The reddest color interval, $1.5<R-I<2.0$
contains the reddest PMS stars and no MS stars. These PMS stars are
fainter (in $R$) and have larger radii than those contained in the
$0.5<R-I<1.0$ sample, and therefore suffer most strongly from finite
source effects causing events with even longer timescales than for the
bluer PMS stars.

\newpage

\section{Application to Experiments: Total Event Rates and Luminosity Function of Lensed Stars}
\label{sec.exp}

We now apply our results from \S\S~\ref{sec.app} and
\ref{sec.app_source} to difference imaging surveys. The goal of this
section is to predict realistic event rates that take into account
observational constraints (like timescales of events and the
signal-to-noise ratios of the light curves, e.g., at maximum). These
event rates can be taken for survey preparations or for a first-order
comparison of survey results with theoretical models.  Exact survey
predictions and quantitative comparisons with models can be obtained
with numerical simulations of the survey efficiency.

\subsection{``Peak-Threshold'' for Event Detection}
\label{sec.peakandevent}

In order to identify a variable object at position $(x,y)$, its excess
flux $\DF$ has to exceed the rms flux $\sigma_F(x,y)$ by a certain
factor $\Q$:
\begin{equation}
   \darray
 \begin{array}{l}
   \DFmin(x,y) = \Q  \sigma_F(x,y)  .
 \end{array}
\label{eq.dfmin}
\end{equation}
The parameter $Q$ characterizes the significance of the amplitude of a
lensing event, but not of the event itself, since that also depends on
the timescale (and the sampling) of the event. We will call events
characterized by the signal-to-noise ratio at maximum light
``peak-threshold-events'' in the following
\citep{2000ApJ...530..578B}. Considering only the maximum flux excess
of an event (and not its timescale) of course can lead to an
over-prediction of lensing events, since events might be too fast to
be detected. In addition, long timescale events with low excess flux
can have many data points with low significance for the excess flux,
which all together make a significant lensing candidate.  The
detectability of events therefore depends on both its amplitude (flux
excess at maximum) and its timescale. This is the reason, why we
derived the contribution to the event rate as a function of flux
excess and FWHM timescale in \S~\ref{sec.app_source}.

The flux excess threshold that a source with intrinsic flux
$\F(\Mlum,\Dos,\ext)$ must achieve in order to be identified as an
event can be translated to thresholds in maximum magnification and
relative impact parameter using equations (\ref{eq.lc_diff}) and
(\ref{eq.u0_A0}):
\begin{equation}
  \AT(x,y,\F(\Mlum,\Dos,\ext)) := \DFmin(x,y) \,\F^{-1} + 1 \approx 
\DFmin(x,y) \,\F^{-1} ,
\label{eq.AT}
\end{equation}
\begin{equation}
  \uT(x,y,\F(\Mlum,\Dos,\ext)) :=
  \sqrt{\frac{2\AT}{\sqrt{\AT^2-1}}-2} \approx \frac{1}{\AT}  ;
\label{eq.uT}
\end{equation}
in both cases we have also given the high-magnification approximations
in the last step.

\begin{figure}[b]
  \epsscale{0.50}\plotone{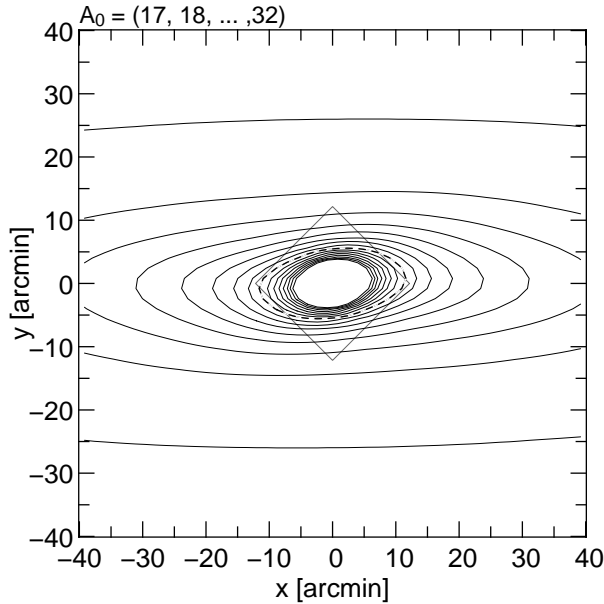} 
  \caption{Contours show the minimum magnifications $\AT$ that stars
    with $\Mlum_R=0\E{mag}$ at a distance of $\Dos=770\E{kpc}$ need to
    exceed the M31 rms flux by a factor of $Q=\SN=10$ in the $R$ band,
    for an experiment (with respect to, e.g., pixel size and seeing)
    like WeCAPP.  The contour levels are $\AT=17,18,...,32$. The
    dashed line marks the $\AT=25$ level; the square shows the field
    observed by WeCAPP given in the intrinsic M31 coordinate system
    (see Fig.~\ref{fig.tau_wecapp}).  }
  \label{fig.A_0min}
\end{figure}

In contrast to the microlensing regime (where $\uT$ is assumed to be
constant), $\uT$ depends on the local noise value via $\DFmin(x,y)$
and the luminosity $\Mlum$ of the source star being lensed.  In
Figure~\ref{fig.A_0min} we show contours of the minimum magnification
required to observe an event at a distance of $\Dos=770\E{kpc}$,
source luminosity of $\Mlum_R=0\E{mag}$ and a signal-to-noise
threshold of $Q=10$ for a survey like WeCAPP in the $R$ band.  Since
the M31 surface brightness and thus also the rms photon noise
increases toward the center, magnifications of $50$ or larger are
needed in the central part. The M31 rms photon noise and rms flux
within a PSF in the $R$ band had been estimated using
equations~(\ref{eq.noisephoton}) and (\ref{eq.noiseflux}) below.

To obtain an upper limit for the event rate, we assume that all events
with flux excesses above the peak-threshold can be identified,
irrespective of their timescales.  In previous event rate estimates
the $\tfwhm$ timescales have only been considered correctly in
Monte-Carlo simulations.  Ignoring the event timescales in analytical
estimates the event rate predictions are much more alike the upper
limit we present here (eq. [\ref{eq.Gamma_uT}]).  In this case one can
simply use the transformation from minimum flux excess at maximum
magnification to the threshold relative impact parameter $\uT$ in
equations~(\ref{eq.AT}) and (\ref{eq.uT}) and integrate equation
(\ref{eq.dNdu0}) over mass, lens distance, and relative velocities,
multiplying it with the relative impact parameter threshold
$\uT(x,y,\F(\Mlum,\Dos,\ext))$ and the number density of sources
with brightness $\Mlum$, $n_\iss(x,y,\Dos)\,\LF_\iss(\Mlum)$, and
finally integrate along the line-of-sight and source luminosity,
(\S~\ref{sec.LF}):

\begin{equation}
  \begin{array}{rl}
    \frac{d^2\Gamma_{s,l}}{dx\,dy} 
    = &  \Intninf \Intninf n_\iss \,\LF_\iss(\Mlum) 
    \Intninf  \Int_0^\Dos \Intninf
    \,\Int_0^{\uT(x,y,\F(\Mlum,\Dos,\ext))\RE}  \GAMb 
     db \,d\vt \,d\ml \,d\Dol  \,d\Mlum \,d\Dos\\
    =  & \Intninf \Intninf n_\iss(x,y,\Dos) \,\LF_\iss(\Mlum) 
    \,\uT(x,y,\F(\Mlum,\Dos,\ext))\,\Gamma_{1,l}(\Dos) \,d\Mlum \,d\Dos.
  \end{array}
  \label{eq.Gamma_uT}
\end{equation}
In this equation, the subscript ``s'' indicates the different stellar
populations (bulge, disc) and their sum yields the upper limit for the
total event rate.  This upper limit for the event rate can therefore
be also obtained as a product of the single-star event rate $\G_{1,l}$
(equation (\ref{eq.gamma_ml})), [using
$\uT(x,y,\F(\Mlum,\Dos,\ext))$] and the number density of sources
with luminosity $\Mlum$ on the line of sight. Equation
(\ref{eq.Gamma_uT}) is similar to the equations of \citet[eq. (2.5)]{1996ApJ...472..108H} and \citet[eq. (2.2.1)]{1996ApJ...473..230H}.\footnote{With $\beta_\Max(F_{0,i})
  \equiv \uT$, $\Gamma_0 \equiv \Gamma_{1,l}(\Dos)$,
  $\frac{\Sigma}{\sum_i \phi(F_{0,i}) \,F_{0,i}} \equiv \Intninf n_\iss
  \,d\Dos$ .}

Up to now we have not discussed the value of $\sigma_F(x,y)$, i.e., the
value of the rms flux that appears in the equations for the detection
thresholds.  This value can in principle be taken from the error
propagation in the reduction process.  Due to varying observing
conditions (seeing, exposure time of the co-added images per night),
the errors can differ from day to day by a factor of up to 10; to
obtain predictions for the most typical situation, one therefore
should use the median error at each image position of a survey to
predict the rms flux $\sigma_F(x,y)$.

\citet{2001A&A...379..362R} showed that using our reduction pipeline
(that propagates true errors through all reduction steps) errors in
the light curves are dominated by the photon-noise contribution of the
background light.  Therefore the typical error can be estimated from
the surface brightness\footnote[100]{
Note that we neglect the correct indizes refering to the 
band $\XX$ and define
$\F \equiv F_{0,\XX}$,
$\FVega\equiv{F_{\M{Vega},\XX}}$,
$\Flum\equiv{\mathcal{F}_\XX}$,
$\FsunX\equiv\Flum_{\odot,\XX}$,
$\DF\equiv{\Delta_{F_\XX}}$,
$\Mlum\equiv{\mathcal{M}_\XX}$,
$\Col\equiv{\mathcal{C}_\CX}$,
$\SB\equiv\mu_\XX$,
$(M/L) \equiv (M/L)_\XX$,
$\ext\equiv A_\XX$.} profile $\SB(x,y)$ of M31 and the typical,
i.e., median, observing conditions of the survey. Using analytically
predicted rms-values, one can study the impact of the observing
conditions on event rates and optimize survey strategies.

To measure the variability of objects one has to perform
(psf-)photometry defining the angular area of the psf $\Opsf:=\pi
\,\theta_\M{psf}^2 / \ln{4}$ and the FWHM of the psf
$\theta_\M{psf}$. For a given experimental setup the rms photon noise
$\sigma_\M{photon}(x,y)$ within an area $\Opsf\, [\M{arcsec}^2]$ at a
position $(x,y)$ is
\begin{equation}
  \sigma_\M{photon}(x,y):= 
          \left[ \left(10^{-0.4(\SB(x,y)+ \kappa \,\M{AM})} + 10^{-0.4\,\SB_\M{sky}}\right) 
            10^{-0.4 (- \M{ZP})} \,\texp\,\Opsf\right]^{1/2} ,  
\label{eq.noisephoton}
\end{equation}
where $\SB_\M{sky}$ [mag arcsec$^{-2}$] is the sky surface brightness,
$t_\M{exp}$ is the exposure time in seconds, $\M{AM}$ is the air mass of the
observation, $\M{ZP}$ is the photometric zero point of the telescope camera
configuration in photons per second and $\kappa$ is the atmospheric extinction
for the observing site.\footnote{We have neglected readout noise of the
detector because it is negligible compared to the photon noise.}

The rms photon noise can be translated to the rms flux (in Jy) using
the flux of Vega, $\FVega$, and its magnitude $m_\M{Vega}=0$:
\begin{equation}
  \sigma_{F}(x,y):=  \frac{\FVega}{\texp\,10^{-0.4
          (m_\M{Vega} + \kappa \,\M{AM} -\M{ZP})}} \;
          \sigma_{\rm photon} 
            =  \FVega
          \left[ \left(10^{-0.4(\SB(x,y)+ \kappa \,\M{AM})} + 10^{-0.4\,\SB_\M{sky}}\right) 
            10^{0.8 \,\kappa \,\M{AM}} \,10^{- 0.4 \,\M{ZP}}
          \,\frac{\Opsf}{\texp}\right]^{1/2}. 
\label{eq.noiseflux}
\end{equation}
The last equation shows that the rms flux within an aperture is proportional
to $t_\M{exp}^{-1/2}$, making the signal-to-noise $Q$ proportional to
$t_\M{exp}^{1/2}$, as expected for background noise-limited photometry of
point-like objects.
 
The extincted surface brightness profile $\SB(x,y)$ in
Eqs.~\ref{eq.noisephoton} and \ref{eq.noiseflux} can be taken either
from very high signal-to-noise measurements of M31 or from analytical
models that are constructed to match the observed SFB-profile and
dynamics of M31. In the latter case, the extincted surface brightness
$\SB(x,y)$-model combines the luminous matter density
$\rho_\iss(x,y,\Dos)$ with the mass-to-light ratio for each source
components ($s$=bulge,disk) and accounts for Galactic and intrinsic
extinction $\ext(x,y,\Dos)$ along the line-of-sight:
\begin{equation}
  \begin{array}{l}
 \SB(x,y) = -2.5 \lgt \left( \FVega^{-1} \Sum_\iss 
   \,\Intninf  10^{-0.4 \,\ext(x,y,\Dos)} \,\frac{\rho_\iss(x,y,\Dos)}
               {\ML_\iss \,\frac{\Msun}{\FsunX}}
   \dfak \left(\frac{2\pi}{360 \times 3600}\right)^2 \Dos^2 \,d\Dos  \right)
  \end{array}   , 
\end{equation}
where the units are $\M{mag}\E{arcsec}^{-2}$.

\subsection{``Event-Threshold'' for Event Detection}
\label{sec.event}

Gould \& Han
\citep{1996ApJ...470..201G,1996ApJ...472..108H,1996ApJ...473..230H}
introduced an ``event threshold'', where the detectability of events
depends on the total excess light of the light curves.  They obtained
an implicit equation for the threshold $\uT $ of the relative impact
parameter,
\begin{equation}
\frac{\zeta(\uT)}{\uT} =
\frac{t_\M{cyc}\,\sigma_F(x,y)^2}{\overline{\tE}(x,y,\Dos)\,\pi\F^2}
\Qgou^2  ,
\end{equation}
where $\sigma_F(x,y)$ is the rms flux at that position and
$\overline{\tE}(x,y,\Dos)$ is the mean Einstein time of the events (equation
(\ref{eq.tE_bar})); $\zeta$ is defined by 
\begin{displaymath}
  \zeta(\uub) := \frac{\int
    \left[A(t)-1\right]^2 \,dt}{\int \left[(t/\tE)^2+\uub^2\right]^{-1}\,dt},
\end{displaymath}
$\F(\Mlum,\Dos,\ext)$ is the unlensed source flux and $t_\M{cyc}$ is the
(equidistant) difference between observations.  

This equation assumes equidistant sampling of the light curves and is
therefore most readily applied to space-based experiments. In
addition, it takes into account the mean Einstein timescale of events
only, although the relative impact parameter threshold depends on the
individual timescale of the event. For realistic event rate estimates,
however, one has to to take into account the timescale distributions,
as well.

One can in fact obtain an analog relation for flux excess $\DF$ and
$\tfwhm$ timescale of the events (i.e., the actual observables),
\begin{equation}
\DFmin^2 \,\tilde{\zeta}{\left(\frac{\F}{\DFmin},\tfwhm\right)} = 
  \frac{\sqrt{12} \,t_\M{cyc} \,\sigma_F(x,y)^2}{\pi \,\tfwhm} \Qgou^2 , 
  \label{eq.DF_event}
\end{equation}
with 
\begin{displaymath}
\tilde{\zeta}(\A-1, \tfwhm) := \frac{\int \left[A(t)-1\right]^2
\,dt}{\int (\A-1) \left[12 (t/\tfwhm)^2+1\right]^{-1}\,dt}.
\end{displaymath} 
Equation (\ref{eq.DF_event}) can be numerically inverted to obtain the
peak flux threshold $\DFmin(\tfwhm,x,y,\Mlum)$ as a function of the
event timescale.  Therefore, it is obvious that the peak threshold and
event threshold criteria are related assuming equidistant sampling and
that the event threshold criterion is a special case of the peak
threshold plus a $\tfwhm$ threshold criterion, which is evaluated in
equation (\ref{eq.totevrate}) (see
\S~\ref{eq.gamma_DF_tfwhm}).\footnote{To be able to roughly compare
  the event rate predictions of \citet{1996ApJ...473..230H}, who used
  the event threshold criterion, we can assume $\tfwhmmin \approx
  t_\M{cyc}$ and $\tilde{\zeta} \approx 1$ and obtain $\Q \approx
  12^{0.25}\pi^{-0.5} \,\Qgou \approx 0.6 \,\Qgou$.}

\subsection{Total Event Rate with Excess Flux Threshold $\DFmin$ and Timescale $\tfwhmmin$ Threshold}
\label{eq.gamma_DF_tfwhm}

The upper limit derived in \S~\ref{sec.peakandevent} still includes
numerous events that cannot be detected in finite time resolution
experiments. At this point, where not only must the flux excess
(maximum magnification or relative impact parameter) of the event be
considered, but also the timescale of the event, the transformation of
the event rate from the ``theoretical quantities'' to the
``observational quantities'' in \S~\ref{sec.app_source} becomes most
relevant.  Using equation (\ref{eq.n_dgam_dt12_dMlum}) we can simply
integrate from the lower limits $\DFmin$ and $\tfwhmmin$ to infinity
(or any other value specified by the experiment):
\begin{equation}
  \frac{d^2 \Gamma_{l,s}}{dx \,dy} := 
  \Int_{\tfwhmmin(x,y)}^\infty \, \Int_{\DFmin(\tfwhm,x,y)}^\infty
  \frac{d^4 \Gamma_{s,l}}{dx \,dy \,d\tfwhm \,d\DF}  \,d\DF  \,d\tfwhm 
  \label{eq.totevrate}
\end{equation}
with
\begin{displaymath}
  \frac{d^4 \Gamma_{s,l}}{dx \,dy \,d\tfwhm \,d\DF} := \int \int \frac{d^6
    \Gamma_{s,l}}{dx \,dy \,d\tfwhm \,d\DF \,d\ml \,d\Mlum} \,d\ml \,d\Mlum.
\end{displaymath}
The thresholds $\tfwhmmin(x,y)$ and $\DFmin(\tfwhm,x,y)$ (see
eqs.~[\ref{eq.dfmin}] and [\ref{eq.DF_event}]) are set by the experiment
and the detection process.\footnote{For completeness we can also
  introduce the color thresholds $\Colmin$ and $\Colmax$, which may
  also depend on the experiment and use the distribution derived in
  eq.~(\ref{eq.n_dgam_dt12_dCol}).}

\begin{table}
  \centering
  \caption{Observational setups for the WeCAPP survey and a potential 
    experiment using the ACS on board of {\it HST}}
  \begin{tabular}{lcc}
    \hline\hline
       Parameter                              & WeCAPP & {\it HST} with ACS WFC \\
    \hline                                                                                         
    $\texp$ (s)                               & 500    & 1000 \\
    $\pixelsize$  (arcsec)                    & 0.5    & 0.049 \\
    Field of view  (pixels)                   & $2048 \times 2048$ & $4096 \times 4096$  \\
    Field of view  (arcmin)                   & $17.2 \times 17.2$ & $3.37 \times 3.37$ \\
    Filter                                    & Johnson R & F625W \\
    $\ZP$ (mag)                               & 23.68     & 25.73 (\cite{2004STScI_08_ISR}, Table~3)  \\
    Average $\skymag$                         & 20.0      & 22.5   \\
    FWHM of the psf $\theta_\M{psf}$ (arcsec) & 1.5       & 0.12 (\cite{2003STScI_06_ISR}, p.~13) \\
    $\Opsf$ (arcsec$^2$)                      & 5.1      & 0.033   \\
    AM                                        & 1.0    & 0    \\   
    Atmospheric extinction $\kappa_R$         & 0.1    & 0    \\ 
    $\Dt$             (days)                  & 200    & 30    \\ 
    $\tfwhmmin$       (days)                  & 1      & 1     \\   
    $\tfwhmmax$       (days)                  & 200    & 20     \\   
    inner saturation radius (arcsec)          & 20   & 0  \\   
    CCD orientation angle (deg)               & 45   & 0  \\
    \hline
  \end{tabular}
  \label{tab.observations}
\end{table}

For the WeCAPP experiment (see Table~\ref{tab.observations}) toward
M31 it turned out that the efficiency can easily be evaluated using
Monte-Carlo simulations.  As in the WeCAPP experiment errors are
propagated through all reduction steps \citep{2001A&A...379..362R};
the final errors in the light curve $\sigma(x,y,t_i)$ include the full
reduction procedure. For a simple set of detection limits,
i.e., $\DFmin \sim \sigma$ and $\tfwhmmin = \const$, the efficiency
$\eff(x,y,\tfwhm,\DF)$ for a survey can easily be evaluated as a
function of the directly observable parameters $x$, $y$, $\tfwhm$, and
$\DF$ (in contrast to the variables $\tE$ and $\A$).  This and more
sophisticated thresholds (as used in \cite{2001ApJS..136..439A}) and
efficiency simulations for WeCAPP we will present in a forthcoming
paper.

Using this efficiency we can generalize equation (\ref{eq.totevrate})
to
\begin{equation}
  \frac{d^2 \Gamma_{l,s}}{dx \,dy} := 
  \int \int 
  \frac{d^4 \Gamma_{s,l}}{dx \,dy \,d\tfwhm \,d\DF} \,\eff(x,y,\DF,\tfwhm) \,d\DF  \,d\tfwhm 
  .
  \label{eq.totevrate_eff}
\end{equation}
As the total event rate depends on the model parameters of the
luminous and dark component, precise measurements of the event numbers
and event rate's spatial variation can in principle constrain the
source and lens densities [$\rho_l(x,y,\Dol)$, $n_\iss(x,y,\Dos)$], the
lens mass functions [$\xi_l(\ml)$], the distribution of the
transversal velocities [$\pv(\vt,v_0(x,y,\Dol,\Dos))$], the luminosity
function of the sources [$\LF_\iss(\Mlum)$ or $\pcmd_\iss(\Mlum,\Col)$],
and finally the MACHO fraction in the halo. There are, of course other
valuable parameters, like event duration, flux excess distribution,
color of the lensed stars, and finite source effects, which make the
lensing analysis much more powerful than the pure counting of events.

Table~\ref{tab.Gammatot_WeCAPP} summarizes the event rate predictions
for the WeCAPP experiment toward the bulge of M31, using different
realistic thresholds\footnote{This is equivalent to an efficiency
  $\eff(x,y,\DF,\tfwhm) = \theta(\tfwhm-\tfwhmmin)
  \,\theta(\DF-\DFmin(x,y))$ of the experiment.} for the
signal-to-noise threshold necessary to derive ``secure'' events, and
for $\tfwhmmin$. These numbers do not take into account that events
cannot be observed when M31 is not visible (one-third of the year),
that in the remaining time some -- in particular short-term events --
escape detections because of observing gaps, and that some of the area
is not accessible for identification of lensing events due to
intrinsically variable objects. We calculated the predictions for
signal-to-noise thresholds of $Q=10$ and $Q=6$; these thresholds
correspond to flux excess thresholds of $6.2\times 10^{-6}\E{Jy}$
(Q=10) and $3.7\times 10^{-6}\E{Jy}$ (Q=6) in the edges and $2.4\times
10^{-5}\E{Jy}$ (Q=10) and $1.5\times 10^{-5}\E{Jy}$ (Q=6) 20\arcsec\
off center (outside saturation) of the WeCAPP field.\footnote{A value
  of $10^{-5}\E{Jy}$ correspond to an ''excess magnitude'' of
  $21.2\E{mag}$ in the $R$ band} The $Q>10$ events are events like
those published in the past (e.g., WeCAPP-GL1 and WeCAPP-GL2 have
values of $Q\approx 85$ and $Q\approx 16$), whereas $Q=6$ should be
more similar to the medium bright event candidates of MEGA.  For the
$Q=10$ cases we have separated events that do not show finite source
effects in the light curves (``without fs'') from those which show
finite source effects (``with fs''). Finite source events are
relatively more important for high signal-to-noise, short timescale
self-lensing events.  In most current pixel-lensing surveys, light
curves with finite source effects are not specially searched for and
may preferentially get lost in the detection process, unless one
allows for a less good fit for bright events.

For the $Q=6$, $\tfwhm=2\E{days}$ case we split the predictions into
the near and far side of M31.  Within our field, the predicted
halo-bulge asymmetry is small, but the bulge-disk and halo-disk
asymmetry are on a noticeable level. (Note that the disk-bulge lensing
does show the reversed asymmetry).  It has been pointed out in the
past (\cite{2004MNRAS.351.1071A}) that dust lanes in the M31 disk are
an additional source of asymmetry; this is obvious if one considers
the spatial distributions of variables found in pixel-lensing
experiments (see \cite{2004MNRAS.351.1071A},
\cite{2004A&A...421..509A} and \cite{2006A&A...445..423F}).  These
can, however, be taken to quantitatively account for extinction, in
addition to extinction maps. The values given in our table do not
account for the small spatial dependence of extinction and thus place
lower limits to the observed far-near asymmetry of the individual
lens-source configuration.

The comparison for different timescale thresholds (cases III, IV, and
V) shows that (except high mass halo lensing) the majority of events
has timescales smaller than 10 days.  A clustering of event candidates
with short and long timescales as \cite{2004A&A...417..461D} observed
for the MEGA analysis of the POINT AGAPE survey (they obtained 6
candidates with timescales smaller than 10 days and 8 candidates with
timescales larger than $20\E{d}$) can be hardly explained for the
WeCAPP field.  This is because, even for supermassive MACHOs, one
would expect roughly as many events between 2 and 20 days than above
20 days (compare case III and case V in
Table~\ref{tab.Gammatot_WeCAPP}).  \cite{2004A&A...417..461D} argue
that their long-term events arise in the outskirts of M31, where the
photon noise is smaller, and could be understood from selection
effects. This would still lack to explain the bimodality of
timescales.  At the moment it is not excluded that these long-term
event candidates are still misidentified variable
objects.\footnote{See \cite{2006A&A...446..855D} for recent results.}
In the last line we add the analogous numbers for halo lensing
resulting from Milky Way halo lenses of 0.1 $\Msun$.  The number of
MACHO events caused by the MW MACHOs should be roughly one-third of
that caused by M31 MACHOs.

\begin{table}[t]
\begin{center}    
\caption{Total event rate $\Gamma_{s,l}$ $[\E{y}^{-1}]$ for the WeCAPP
      experiment for self-lensing and halo-lensing}
\begin{tabular}{r|rcl|rcl|rcl|rcl|rcl|rcl}
\hline\hline
 {\scshape Model}                   & \mc{3}{c|}{I}    & \mc{3}{c|}{II}   &   \mc{6}{c|}{III}                      &  \mc{3}{c|}{IV}             & \mc{3}{c}{V}                \\
                       &      & &         &      & &         &   \mc{3}{c}{Near Side} & \mc{3}{c|}{Far Side}&      & &             &       & &                    \\
\hline
b-b             &      1.2 &+&       1.9 &      0.57 &+&      0.68 &       1.4 &+&      0.98 &       1.4 &+&      0.99 &      0.16 &+&     0.046 &     0.026 &+&    0.0062 \\  
h0.1-b          &      8.2 &+&       5.4 &         4 &+&       1.7 &       6.3 &+&       1.6 &       7.1 &+&       1.9 &      0.92 &+&     0.071 &      0.16 &+&    0.0094 \\  
h0.5-b          &      7.4 &+&       2.7 &       4.4 &+&         1 &       5.5 &+&      0.72 &       6.3 &+&      0.82 &       1.8 &+&     0.051 &      0.47 &+&    0.0074 \\  
h1000-b         &      0.7 &+&    0.0013 &      0.61 &+&    0.0011 &      0.51 &+&   0.00051 &      0.59 &+&   0.00056 &      0.75 &+&   0.00025 &       0.6 &+&    $5\tV$ \\  
d-b             &     0.57 &+&      0.34 &      0.26 &+&     0.098 &      0.89 &+&      0.16 &     0.087 &+&     0.026 &     0.057 &+&    0.0029 &    0.0072 &+&   0.00031 \\  
h$_\M{MW}$0.1-b &      3.9 &+&    0.0046 &       1.9 &+&    0.0019 &       2.7 &+&     0.001 &       2.7 &+&     0.001 &      0.47 &+&    $4\tV$ &     0.088 &+&   $5\tVI$ \\  
h$_\M{MW}$0.5-b &      2.4 &+&    0.0009 &       1.5 &+&    0.0002 &       1.8 &+&    0.0001 &       1.8 &+&    0.0001 &      0.76 &+&   $2\tVI$ &      0.23 &+&  $3\tVII$ \\  
\hline
b-d             &      2.3 &+&       1.6 &       1.4 &+&      0.77 &      0.43 &+&      0.22 &       3.8 &+&       1.3 &      0.43 &+&     0.049 &     0.082 &+&    0.0055 \\ 
h0.1-d          &       11 &+&       4.3 &       6.6 &+&         2 &       4.9 &+&      0.72 &        12 &+&         2 &       2.1 &+&       0.1 &       0.5 &+&     0.014 \\ 
h0.5-d          &      8.6 &+&       1.2 &       5.8 &+&      0.65 &       3.6 &+&      0.21 &       9.5 &+&      0.55 &       3.1 &+&     0.069 &         1 &+&     0.011 \\ 
h1000-d         &     0.61 &+&   0.00015 &      0.56 &+&   0.00014 &      0.26 &+&    $5\tV$ &       0.7 &+&    $9\tV$ &       0.7 &+&    $8\tV$ &      0.55 &+&    $2\tV$ \\ 
d-d             &      0.2 &+&      0.13 &      0.14 &+&     0.075 &       0.2 &+&     0.056 &      0.19 &+&     0.056 &     0.095 &+&     0.018 &     0.036 &+&    0.0055 \\ 
h$_\M{MW}$0.1-d &      3.1 &+&    0.0019 &         2 &+&   0.00049 &       2.2 &+&   0.00025 &       2.2 &+&   0.00024 &      0.95 &+&   $6\tVI$ &       0.3 &+&  $7\tVII$ \\ 
h$_\M{MW}$0.5-d &      1.9 &+&   0.00022 &       1.4 &+&   0.00013 &       1.4 &+&    $7\tV$ &       1.4 &+&    $7\tV$ &      0.97 &+&   $4\tVI$ &      0.43 &+&  $5\tVII$ \\ 
\hline
\end{tabular}
\end{center}
    \tablecomments{(d) disk; (b) bulge; (h0.1) halo
      consisting of 0.1 $\Msun$ lenses; (h0.5) halo consisting of 0.5
      $\Msun$ lenses (h$_\M{MW}$ for MW-halo); (h1000) halo consisting
      of 1000 $\Msun$ lenses. The numbers give the event rate in
      events per year integrated over the observed field and applying
      the following peak-flux signal-to-noise thresholds: (I) $Q=10$
      and $\tfwhmmin=1\E{day}$; (II) $Q=10$ and $\tfwhmmin=2\E{days}$;
      (III) $Q=6$ and $\tfwhmmin=2\E{days}$ for the near and far side;
      (IV) $Q=6$ and $\tfwhmmin=10\E{days}$; (V) $Q=6$ and
      $\tfwhmmin=20\E{days}$.  These (S/N)-limits (at the light curve
      peak) are more realistic than a flux threshold $\DFmin$, which
      is constant over the observed field, since the central region
      shows a strong gradient in the surface brightness and photon
      noise values.  We have also separated events that do not show
      finite source effects in the light curves from those with finite
      source effects by a plus sign. Note that light curves with
      finite source effect signatures might be missed when using event
      filters with a classical lensing event shape in a stringent
      way. For the ($Q=6$, $\tfwhm=2$) case we split the predictions
      in those for the near and far side of the M31 (disk and
      bulge). Within our field, the predicted halo-bulge asymmetry is
      small, but the bulge-disk and halo-disk asymmetries are on a
      noticeable level. The comparison for different timescale
      thresholds (columns labeled ''III'', ''IV'', and ''V'') shows
      that (except for high-mass halo lensing) the majority of events
      has timescales smaller than $10\E{days}$. A peak of events with
      timescales of $20\E{days}$ or larger can only be understood with
      supermassive MACHOs or miss-identifications of variable objects.
      In the last line we add the analogous numbers for halo lensing
      resulting from Milky Way halo lenses of 0.1 $\Msun$.  The MACHO
      events caused by the MW MACHOs should be roughly one-third of
      that caused by M31 MACHOs. }
\label{tab.Gammatot_WeCAPP}
\end{table}
\begin{figure}
   \epsscale{1.0} \plotone{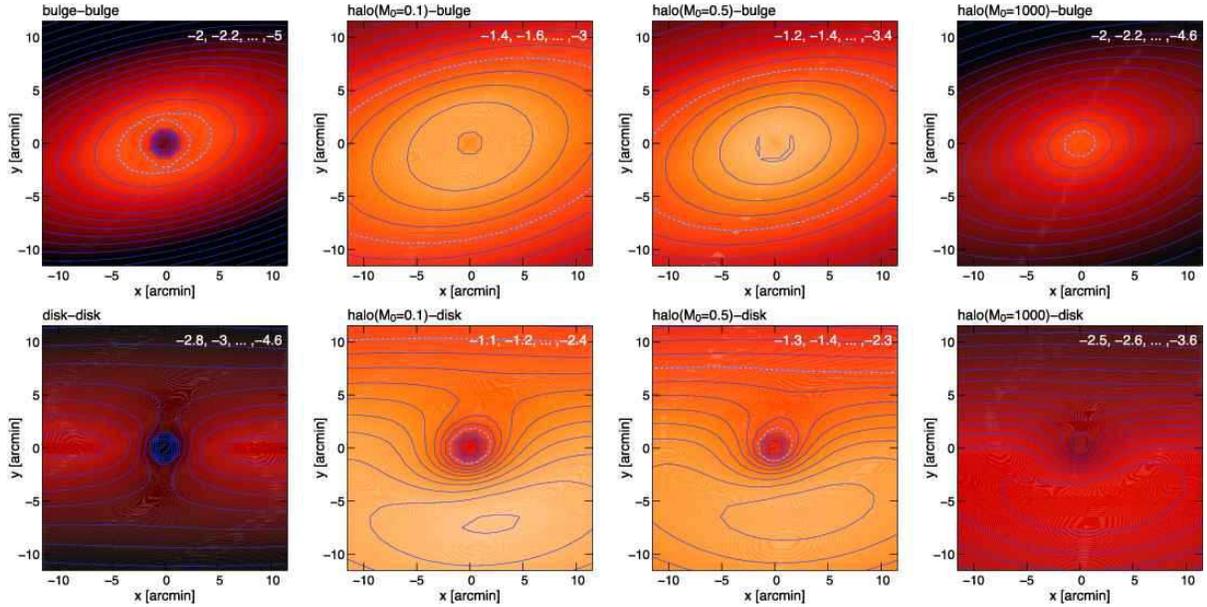}
   \caption{Event rate maps in logarithmic units (in units of events
     $\M{yr}{^-1}\E{arcmin}^{-2}$) for the WeCAPP survey for $\Q=10$
     and $\tfwhmmin=1\E{d}$ (column labeled ''I'' in
     Table~\ref{tab.Gammatot_WeCAPP}). Coordinates are given in the
     intrinsic M31 coordinate system (see
     Figure~\ref{fig.tau_wecapp}). The contour levels are shown in
     inserts in the upper right corners of each diagram. The dashed
     line marks the $0.01\E{ev}\E{y}^{-1}\E{arcmin}^{-2}$ level in
     each diagram. The event rate decreases near the center of M31 due
     to higher noise combined with finite source effects.  Note that
     the maximum lensing (maximum optical depth) region defined by
     \cite{1992ApJ...399L..43C} and \cite{1996AJ....112.2872T} is
     predicted at about $1.5\E{kpc}$ ($7.5\arcmin$) from the nucleus
     for a simple halo model. For the calculations we have taken into
     account the finite stellar source sizes; the numbers shown,
     however, include only those among all events that do not show
     finite source signatures in their light curves, i.e., those which
     are usually searched for in lensing experiments.}
    \label{dGamma_dxy}
\end{figure}

Figure~\ref{dGamma_dxy} shows the predictions for the spatial
distribution of the lensing events for the WeCAPP survey, evaluated
for the $Q=10$ and $\tfwhm=1\E{day}$ thresholds (column labeled ``I''
in Table~\ref{tab.Gammatot_WeCAPP}). One can see that the event rate
density becomes maximal close to the M31 center for bulge-bulge and
halo-bulge lensing configurations.  That seems counterintuitive to the
results about the lensing optical depth and single-star event rate in
Figures \ref{fig.tau_wecapp} and \ref{fig.dis_u0}, where the maximum
is attained on the M31 far side, significantly offset from the
center. This difference is due to the density of source stars, which
rises toward the center much more than the single-star event rate and
the detectability of the events drops. As can also be seen in
Table~\ref{tab.Gammatot_WeCAPP}, a far to near side asymmetry ({\it
  lower and upper part in the figure}) is not present for bulge-bulge
lensing, is modest for halo-bulge lensing, and is stronger for the
disk-bulge lensing. This is because the disk effectively cuts the
bulge in one part in front and the other behind the disk, and only the
stars in the second part can contribute to disk-bulge lensing. The
bulge-disk self-lensing shows the opposite far to near side asymmetry
and attains its maximum event rate per area in the far side of the
disk. The same is true for the halo-disk lensing (main maximum on far
side of disk), which shows a secondary maximum close to the M31 center
caused by the increase of the disk-star density. The disk-disk lensing
event rate per area is symmetric with respect to the near and far side
of the disk.  The fact that the maximum for bulge-bulge and disk-disk
lensing is not located exactly at the M31 center is caused by the
increased photon noise combined with finite-source effects.

The total self-lensing (disk-bulge + bulge-disk + bulge-bulge +
disk-disk) shows an asymmetry arising from the different luminosity
functions and mass functions of the bulge and disk population which
leads to different event characteristics for disk-bulge and bulge-disk
lensing. Therefore, lens and source populations cannot easily be
exchanged. The fact that the near side is closer to us -- lensing
strength and apparent magnitude of sources change by a few percent --
than the far side of M31 plays a minor role for the asymmetry of
self-lensing event rates.

\begin{figure}
  \epsscale{1.15}
  \plotone{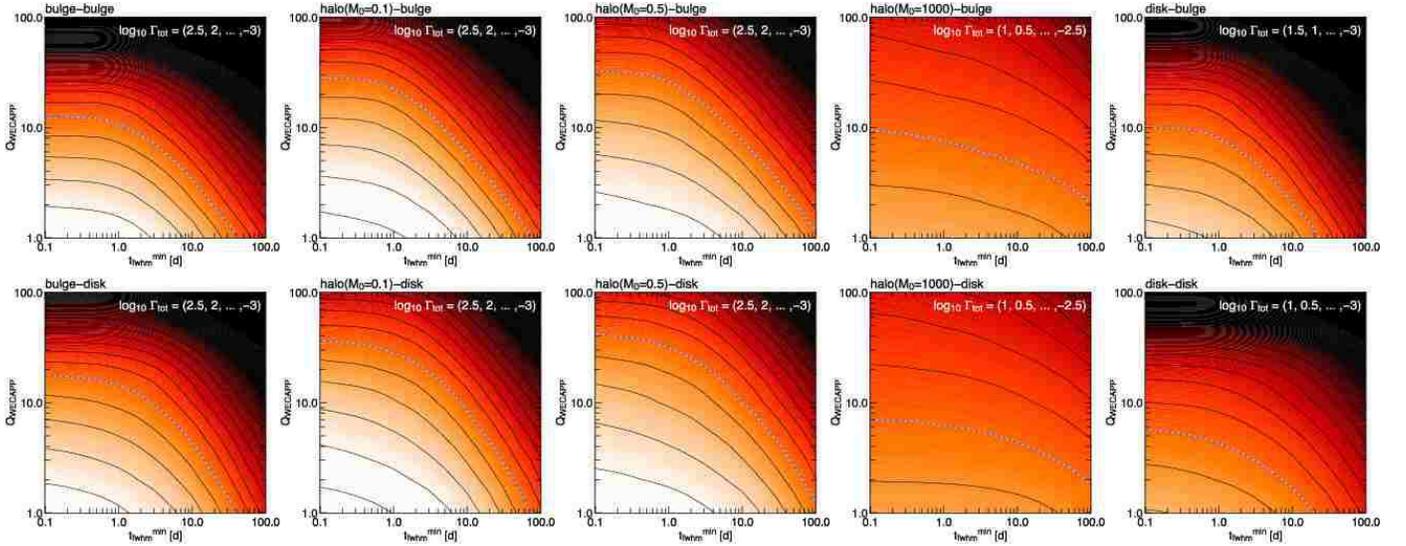}
  \caption{Contours of the logarithm of the lensing event rate per
    year within the $17.2\arcmin\times 17.2\arcmin$ WeCAPP field as a
    function of the signal-to-noise threshold $Q$ for the peak-flux
    excess and as a function of the timescale threshold of the events.
    For these numbers, eq.~(\ref{eq.totevrate}) was integrated
    over the WeCAPP field, and the noise was estimated for the WeCAPP
    survey (characterized by Table~\ref{tab.observations}).  We show
    results for different lens-source configurations, from left to
    right: bulge-bulge self-lensing, halo-bulge lensing with $0.1
    \,\Msun$ lenses, halo-bulge lensing with $0.5 \,\Msun$ lenses, and
    disk-bulge self-lensing. The {blue dashed} line marks the 1~event
    per year level.  For the calculations we have taken into account
    the finite stellar source sizes; the numbers shown, however,
    include only those among all events that do not show finite source
    signatures in their light curves, i.e., those which are usually
    searched for in lensing experiments. For signal-to-noise ratios of
    $Q>10$, the rates for events with finite source effects can be of
    the same order as the rates for events with point-source light
    curves. For lower signal-to-noise ratios, events with finite
    source effects become much less important.}
  \label{fig.Gammatot}
\end{figure}

The last figure (Fig.~\ref{fig.Gammatot}) in this section shows the
total event rates $\Gamma_{l,s}(\tfwhmmin,Q) = \int \int \frac{d^2
  \Gamma_{l,s}}{dx \,dy} \,dx \,dy $ in the WeCAPP field depending on
the peak flux threshold and the timescale threshold of the survey.
We have taken into account the finite source sizes but show only the rate for
those events that do not show any finite source signature in their light
curves.

For high signal-to-noise events (e.g., $Q=20$), all configurations do
show more or less flat contours in the $\tfwhm$ direction for $\tfwhm$
values between $0.1$ and $1$ days.  This indicates that there are
relatively few very high signal-to-noise events with timescales around
$0.1$ days compared to events with timescales of about $1$ day. That
this is true is confirmed by Figures~\ref{fig.dGamma_fs_1} and
\ref{fig.dGamma_fs_2}, which show that the highest signal-to-noise and
thus highest flux excess events occur with timescales between $0.6$
and a few days and that events with timescales of about $0.1$ days are
significantly fainter. Only for smaller flux excesses, the events with
timescales of $0.1$ days can be as common as events with timescales of
a couple of days.  This implies that if one can measure only lensing
events that have a $(S/N)$ ratio of $Q\ge10$ in the WeCAPP setup, one
cannot greatly increase the number of observed events by increasing
the sampling (the largest increase for sampling below 1 day would
occur for halo-disk lensing). The detectability of events with
timescales of hours can therefore be increased effectively only (in
the central M31 field), if the noise level of the observations is
lowered. It can also be seen in all panels of
Figure~\ref{fig.Gammatot} that one expects the number of events to
decrease strongly for $\tfwhm$-thresholds larger than several days.
The quantitative differences in the different lensing configurations
in the subpanels of Figure~\ref{fig.Gammatot} can be easily understood
(by combining the dependence of the event rate on event timescale and
magnification with the luminosity function of source stars, and
accounting for the difference in the importance of finite source
effects) and are discussed for the $0.1\,\Msun$ MACHO bulge and the
$0.5\,\Msun$ MACHO bulge lensing case. Table~\ref{tab.Gammatot_WeCAPP}
already suggests that these two lensing configurations are very
similar for the event numbers that do not show finite source
signatures (the $0.1\,\Msun$ MACHOs do cause more high signal-to-noise
finite source events). One expects that the increase of the MACHO mass
decreases the total event rate per line-of-sight (fewer lenses) but
also increases the events' timescales. On the basis of the event rate
per timescale and event magnification
(Figure~\ref{fig.dGamma_dA0_dtfwhm}) one would therefore expect longer
timescales but fewer events for the stellar mass MACHO case. This
simple picture is altered by the finite source effects, which limit
the maximal flux excess and thus signal-to-noise of an event.  The
question, whether one expects more events for brown dwarf or stellar
mass MACHOs depends therefore on the combination of signal-to-noise
threshold and timescale threshold.

The predicted event rates rise strongly for lower $Q$ thresholds. This
means that if one lowers the (S/N) threshold -- or equivalently
increases the signal or decreases the noise level by changing the
experiment -- one could dramatically increase the event rates.  This
can be achieved by an increase of telescope area and integration time,
but much better with a decrease of the PSF of the experiment. This
makes the space experiments most promising. In addition, the
comparison of the subpanels in Figure~\ref{fig.Gammatot} shows that
the bulge-bulge lensing rate will profit much stronger from a decrease
of the noise than any other lensing configuration.

Assume that the center of M31 is monitored with the ACS on 30
consecutive days, with three 6 minute exposures in the F625W band and
two 6 minute exposures in the F555W band, which would need one orbit
per day altogether (see more details in
Table~\ref{tab.observations}). Assume furthermore that the background
light has the level of the smoothed M31 SFB isophotes (in reality a
fraction of the brightest stars gets resolved lowering the background
light in between the resolved objects). Using that background level,
we predict the event rates with timescales between 1 and 20 days for
bulge-bulge self-lensing and halo-bulge lensing in
Table~\ref{tab.Gammatot_ACS}. If one assumes a halo fraction of about
25\% then the halo-lensing events do not contribute more than 10\%
relative to the bulge-bulge lensing rate.

\begin{table}[t]
  \begin{center}
\caption{Total event rates $\Gamma_{s,l}$ $[\E{y^{-1}}]$ for a 30
      day, 1-orbit-per day {\it HST} ACS experiment (see
      Table~\ref{tab.observations})}
    \begin{tabular}{p{5cm}|cc}
      \hline\hline
      {\scshape Model}  &   Without fs    &    With fs       \\ 
      \hline
      b - b         &  1350   &  100    \\
      h0.1 - b      &   620   &  10  \\
      \hline
    \end{tabular}
  \end{center}
 \tablecomments{The rates are shown for different
      lens-source configurations: (b) bulge; (h0.1) - halo consisting
      of 0.1 $\Msun$ lenses.  The numbers give the rate in events per
      year integrated over the whole ACS field and applying $Q=6$,
      $\tfwhmmin = 1\E{day}$ and $\tfwhmmax=20\E{days}$ thresholds.
      ''Without fs'' describes events, that show no finite source
      signatures in the light curve; ''with fs'' gives the number of
      light curves with finite source signatures.  }
    \label{tab.Gammatot_ACS}
\end{table}

Current measurements of extragalactic mass functions reach masses down
to 0.6 solar masses (for the LMC \cite{2005ApJ...623..846G}).
Microlensing allows a measurement of the low-mass end of the stellar
mass function, while not relying on the luminosity of those low-mass
stars. The mass of these stars becomes visible by their lensing effect
on (in general) brighter stars.  Therefore, M31 bulge-lensing combined
with space observations makes it possible to test an extragalactic
mass function well below $0.5\,\Msun$.

\subsection{The Luminosity Function Sensitivity}

\begin{figure}    
  \begin{center}
    \epsscale{1.0}\plotone{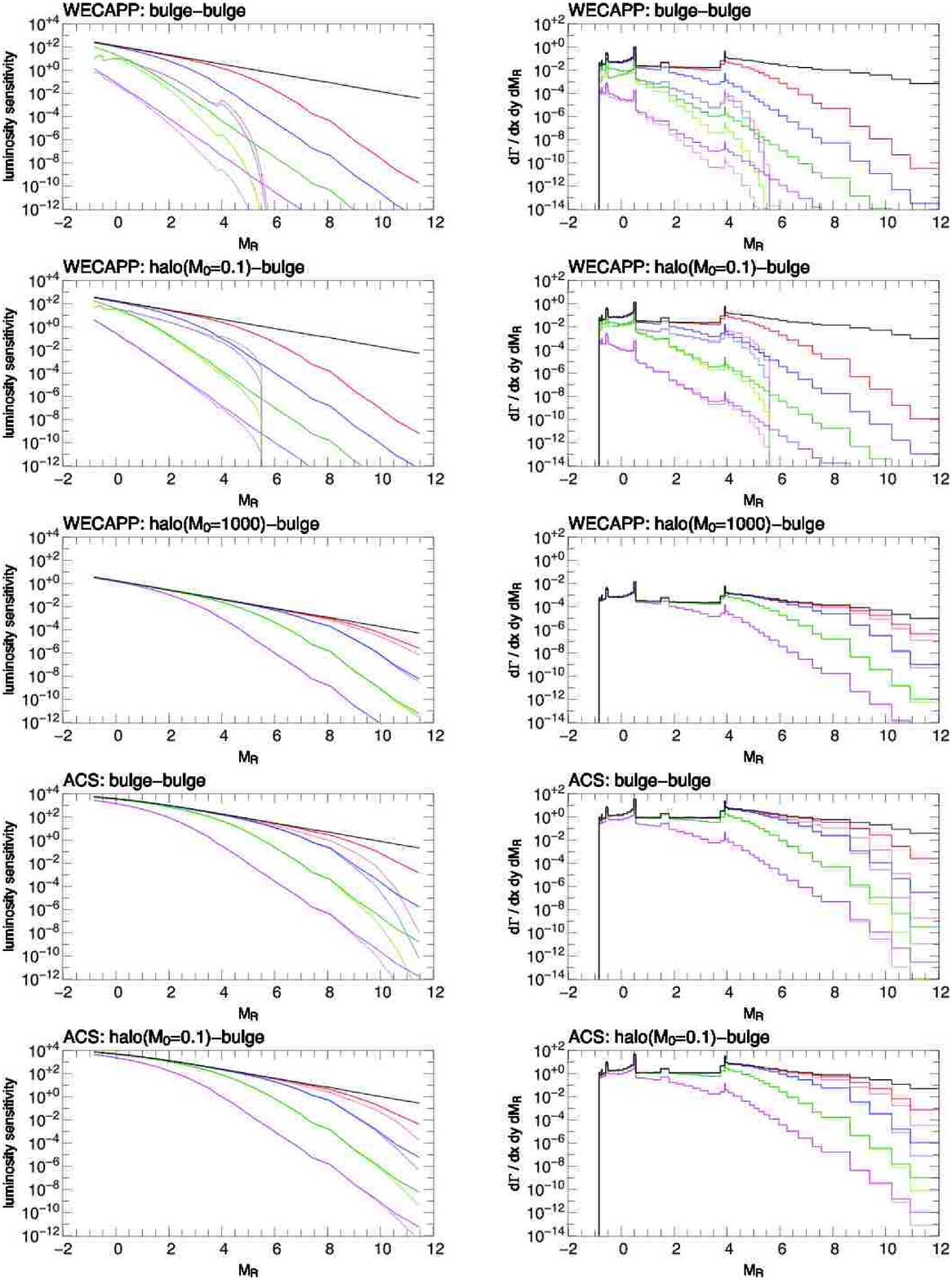}
    \caption{{\it Left panel:} Luminosity function sensitivity
      (eq.~[\ref{eq.lum_sens_mag}]) $[\M{arcmin}^{-2}\E{y}^{-1}]$ for
      bulge-bulge, $0.1\,\Msun$ halo-bulge, $1000\,\Msun$ halo-bulge
      lensing in the WeCAPP and ACS experiment (see
      \S~\ref{app.model}) with $\Q=12$ at $(x,y)=(1,0)\E{kpc}$
      (corresponding $\DFmin=1.7\tV\E{Jy}$ for WeCAPP and
      $\DFmin=3\tVII\E{Jy}$ for ACS) {\it Black line:} no $\tfwhm$
      threshold; {\it magenta line}: $\tfwhmmin=10\E{days}$; {\it
      green line}: $\tfwhmmin=1\E{day}$; {\it blue line}:
      $\tfwhmmin=0.1\E{day}$; {\it red line}: $\tfwhmmin=0.01\E{day}$.
      The curves for the dark colors have been obtained with the
      point-source approximation, the curves with the light colors
      account for the extended source sizes, which further suppresses
      the luminosity sensitivity (using $\Dolfs$ instead of $\Dos$ as
      integration limit in eq.~[\ref{eq.lum_sens_mag}]) . Stars
      fainter than $\Mlum_R = 5.5\E{mag}$ cannot be lensed at all to
      an event with a signal-to-noise ratio $Q$ larger than 10. {\it
      Right panel:} Luminosity function of the sources for lensing
      events $d^3\Gamma_{s,l}/(dx\,dy \,d\Mlum)$
      $[\M{arcmin}^{-2}\E{y}^{-1}\E{mag}^{-1}]$.  For $\tfwhmmin$
      thresholds of $1\E{day}$ that are typical for current
      experiments the probability to have an MS star among the lensed
      stars is very low ( $<3\tVI$ for bulge-bulge lensing, $<3\tV$
      for $0.1\,\Msun$ halo-bulge lensing, $<0.0002$ for $0.5\,\Msun$
      halo-bulge lensing, $<0.2$ for $1000\,\Msun$ halo-bulge lensing).
      }
    \label{fig.lum_sens}
  \end{center}
\end{figure}

Whereas the probability for a star to be lensed does not depend on its
luminosity, the probability that the event can be detected strongly
depends on the luminosity of the source star. This implies that the
luminosity function of the source stars of lensing events is biased
toward high-luminosity stars.  The selection probability of a star
with luminosity $\Mlum$, which we call ``luminosity function
sensitivity'', is obtained with equation (\ref{eq.n_dgam_dt12_dMlum})
as

\begin{equation}
    \frac{1}{\LF_\iss(\Mlum)} \frac{d^3 \Gamma_{s,l}}{dx\,dy \,d\Mlum} = 
    \Int_\tfwhmmin^\infty \Int_\DFmin^\infty 
    \frac{2}{\tfwhm^3}
    \Int_0^\infty 
    \,\xi_l
    \Intninf n_\iss \,\frac{\Psi}{\F}
    \Int_0^\Dos \rho_l  \,\RE^{3} 
    \,\pv{\left(\frac{\RE\Upsilon}{\tfwhm},v_0\right)} 
    \,d\Dol \,d\Dos 
      \,d\ml \,d\DF \,d\tfwhm ,
  \label{eq.lum_sens_mag}
\end{equation}
with the parameters and relations used for
$\LF_\iss{\left(\Mlum\right)}$, $\xi_l(\ml)$, $n_\iss(x,y,\Dos)$,
$\rho_l(x,y,\Dol)$, $\RE(\Dol,\ml,\Dos)$, $\F(\Mlum,\Dos)$,
$\Psi(\A(\F,\DF))$, $\Upsilon(\A(\F,\DF))$.  The luminosity function
sensitivity gives the event rate per area per source star luminosity
bin, normalized by the luminosity function $\LF_\iss(\Mlum)$.  In
Figure~\ref{fig.lum_sens} we show results for the luminosity function
sensitivity using equation (\ref{eq.lum_sens_mag}) for several minimal
detectable timescales $\tfwhmmin$ (0, 0.01, 0.1, 1, and 10 days) using
the configuration of the WeCAPP and ACS experiments and the model of
M31 presented in \S~\ref{app.model}.

Figure~\ref{fig.lum_sens} shows that the sensitivity strongly
increases with decreasing timescale thresholds. Applying no $\tfwhm$
threshold for the total event rate (equivalent to evaluating equation
(\ref{eq.Gamma_uT})) overestimates the luminosity sensitivity for
faint MS stars.  In consequence, the total event rate is overestimated
as well. Accounting for the timescale thresholds for microlensing
surveys (i.e., using equation (\ref{eq.totevrate}) for the total event
rate) suppresses the contribution of faint stars and yields a much
more realistic estimate of the total event rate.  In
Figure~\ref{fig.lum_sens} ({\it right}) we show the luminosity
function of the sources for lensing events, which is obtained as the
product of the luminosity function sensitivity and the luminosity
function $\LF_\iss(\Mlum)$. The results differ for an experiment like
WeCAPP and an experiment with small PSF noise like the suggested ACS
imaging campaign.

An experiment like WeCAPP induces a cutoff of $\Mlum_R\approx
6\E{mag}$ in the luminosity of the lensed stars, because one would
need magnifications larger than the finite source size magnification
limit to obtain an observable flux excess for source luminosities
below that value. This cutoff is valid for all lens-source
configurations within M31, with the exception of supermassive M31
MACHOs, and it does not hold for lensing by MW MACHOs.  With an ACS
experiment, the minimum measurable flux excess is much smaller than
for WeCAPP, and therefore even the faintest MS stars can act as
sources for detectable lensing events (no cutoff in the luminosity
function of lensed stars). For events with timescales above 1 day, the
luminosity function of lensed stars becomes almost flat for magnitudes
brighter than $\Mlum_R\approx 4\E{mag}$ ({\it green curve}).

\clearpage

\section{The Lens Mass Probability Distribution for Individual Lensing Events}
\label{sec.massprob}

The lens mass probability distribution is one central goal of the
analysis of lensing events. We show how this function is extracted
from the individual events, depending on whether $\tE$ and $\A$
(microlensing), or $\tfwhm$ and $\DF$ (pixel-lensing) can be measured.

\subsection{The Lens Mass Probability Distribution, Obtained from the
  Observable $\tEmeas$}

For completeness we recall in this section the classical microlensing
formalism, where the intrinsic source flux $\F$ is supposed to be
known, to derive the mass probability distribution from direct
measurements of $\tEmeas$ and $\uubmeas$, or equivalently $\tEmeas$
and $\Ameas$.

Starting from the integrand in equation (\ref{eq.dG_du0_dtE}), without
carrying out the mass integral and averaging over all source distances
yields

\begin{equation}
  \left<\frac{d^3 \Gamma(\ml,\tE)}{d\ml \,d\tE \,d\uub}\right>_\iss
  = \int p_\iss \,\frac{2}{\tE^3}  \,\xi(\ml) \,\Int_0^\Dos \rho(\Dol)  \, \pv{\left(\frac{\RE}{\tE}\right)} 
    \,\RE^3 \,d\Dol  \,d\Dos    , 
\label{eq.tEmeas}
\end{equation}
which can also be converted to
\begin{equation}
 \left<\frac{d^3 \Gamma(\ml,\tE,\A)}{d\ml \,d\tE \,d\A} \right>_\iss =
   \int p_\iss \,\frac{d\uub(\A)}{d\A}
    \frac{2}{\tE^3} \,\xi(\ml) \Int_0^\Dos 
\rho(\Dol)   \,\pv{\left(\frac{\RE}{\tE},\Dol\right)} \,\RE^3  \,d\Dol  \,d\Dos 
  .
\label{eq.tEmeas2}
\end{equation}
In the right-hand side of equation (\ref{eq.tEmeas2}) the maximum
magnification $\A$ appears only as a pre-factor (see
eq.~[\ref{eq.du0dA0}]), and the right-hand side of equation
(\ref{eq.tEmeas}) is independent of the relative impact parameter
$\uub$. This implies that the magnification at maximum or equivalently
the relative impact parameter does not enter the mass probability
function of the lenses. This is expected, since these quantities
depend on the lens-source trajectory but do not contain any
information about the lens (unless the impact parameter $b$ could
measured by other means). The lens mass probability function thus
solely depends on the measured Einstein timescale, and becomes
 \begin{equation}
   p(\ml; \tEmeas) \sim \left. \left<\frac{d^3 \Gamma(\ml,\tE)}{d\ml \,d\tE \,d\uub}\right>_\iss \right|_{\tEmeas}  
   = \int \left<\frac{d^3 \Gamma(\ml,\tE)}{d\ml \,d\tE \,d\uub}\right>_\iss
   \,\delta(\tE-\tEmeas) \,d\tE  
   =   \left<\frac{d^3 \Gamma(\ml,\tEmeas)}{d\ml \,d\tE \,d\uub} \right>_\iss   ,
    \label{eq.tEmeas3}
\end{equation}
if $\tE$ can be measured without any error.

The measurement error of $\tE$ can be accounted for by replacing the
$\delta$-function with the probability for the Einstein time $\tE$ for
a measured value of $\tEmeas$. For a Gaussian probability with width
$\sigma_{\tEmeas}$ one obtains
\begin{equation}
  \hat p(\ml;\tEmeas,\sigma_{\tEmeas} ) 
  \sim \int \left<\frac{d^3 \Gamma(\tE,\ml)}{d\tE \,d\uub \,d\ml}  \right>_\iss 
  \, g(\tE;\tEmeas,\sigma_{\tEmeas}) \,d\tE
   ,
  \label{eq.tEmeasgauss}
\end{equation}
with 
\begin{displaymath}
  g(t;t^\M{meas},\sigma_{t^\M{meas}}) := \frac{1}{\sqrt{2}\sigma_{t^\M{meas}}}
   \,\exp{\left(-\frac{(t-t^\M{meas})^2}{2\sigma_{t^\M{meas}}^2}\right)}.
\end{displaymath}

Our result in equation (\ref{eq.tEmeas3}) is proportional to the
result of \citet{1994PhLB..323..347J}\footnote{
Converting \citet{1994PhLB..323..347J} to our notation
with $P(\mu,T) \equiv \frac{1}{\xi(\ml)}  
\frac{d^2\Gamma(\ml,\tE)}{d\ml\,d\tE}$, $T \equiv \tE$,  
$\mu \propto \ml$, $x \equiv \Dol/\Dos$, $\rho_0 H(x) 
\equiv \rho(\Dol)$, 
$\frac{dn_0(\mu)}{d\mu} \propto \xi(\ml)$, $r_\M{E}^2 \mu x(1-x) \propto \RE^2$,
$v_H \equiv \sqrt{2} \sigma$ gives 
\begin{displaymath}
  \frac{1}{\xi(\ml)} \frac{d^2\Gamma(\ml,\tE)}{d\ml\,d\tE} \propto 
  \frac{\ml^2}{\tE^4} \Int_0^\Dos d\Dol \left(\frac{\Dol(\Dos-\Dol)}{\Dos}\right)^2 \rho(\Dol)
  \,\exp\left(-\frac{\RE^2}{2 \sigma^2 \tE^2}\right) .
\end{displaymath}
} 
\citep[see also][]{1994ApJ...432L..43J} but differs from the result of
\cite{1998A&A...330..963D}.
\footnote{Converting Dominik's (1998) eq.~(21) to our notation with
$\mu \equiv \ml/\Msun$, $\omega_0 \equiv \uT$, $x \equiv \Dol/\Dos$,
$\rho_0 H(x) \equiv \rho(\Dol)$, $v_c \equiv \sqrt{2} \sigl$, $\zeta \equiv
\vt/(\sqrt{2} \sigl)$, $\tilde{K}(\zeta) \equiv \pv(\vt)\sqrt{2} \sigl$,
$\frac{dn_0(\mu)}{d\mu}=\alpha \mu^p \equiv \xi(\ml)$, $r_0\sqrt{x(1-x)}
\equiv \RE/\sqrt{\ml/\Msun}$ gives 
\begin{displaymath}
  \frac{d\Gamma}{d\ml} = \uT \int
  \xi(\ml) \,\RE \,\rho(\Dol) \,\vt \,\pv(\vt) \,\delta{\left(\ml- \tE^2 \vt^2
      \left(\frac{4G}{c^2}\frac{\Dol(\Dos-\Dol)}{\Dos}\right)^{-1} \right)} \,d\Dol
  \,d\vt. 
\end{displaymath}
}.
The lens mass probability function has also been calculated by 
\indent \citet{1991MNRAS.250..348D} for events with measured
maximum magnification $\Ameas$ and Einstein timescale
$\tEmeas$. We could not match the result published by them with ours
once we converted their notation to ours.
\footnote{Converting \citet[eq. (25)]{1991MNRAS.250..348D} to our notation with
$\mu \propto \ml$,
$\tau' \propto \tE$,
$\tau \propto \tEmeas$,
$u_\Min' \equiv \uub$,
$u_\M{TH} \equiv \uT$,
$A' \equiv \A(u_0)$,
$A_\Max \equiv \Ameas$,
$A_\M{TH} \equiv \AT(\uT)$,
$x \equiv \frac{\Dol}{\Dos}$, 
$\rho_0 H(x) \equiv \rho(\Dol)$ gives 
\begin{displaymath}
  \frac{1}{\xi(\ml)} \left.\frac{d\Gamma(\ml)}{d\ml}\right|_{\tEmeas,\Ameas} \propto 
  \int_0^\uT d\uub \int \int \int \,\delta(\tE-\tEmeas) \,d\tE
  \,\delta{\left(\Ameas-\A(\uub)\right)} \,\RE(\Dol)\, \rho(\Dol) \,d\Dol \,\pv(\vt) \,d\vt.
\end{displaymath}
This equation disagrees in some powers in $\RE$ and $\tE$ to our result.}

\subsection{The Lens Mass Probability Distribution, Obtained from the
  Observable $\tfmeas$ and $\uubmeas$}

Now we discuss the case in which the FWHM $\tfwhm$ and the relative
impact parameter $\uub$ are the available observables.  Starting from
the integrand in equation (\ref{eq.dN_dtfwhm_du}), without carrying
out the mass integral and averaging over all source distances yields
\begin{equation}
  \left<\frac{d^3 \Gamma}{d\ml d\tfwhm \,d\uub} \right>_\iss  =
   \int p_\iss \,\frac{2 \,w^2(\uub)}{\tfwhm^3} \xi(\ml)  \Int_0^\Dos 
  \rho(\Dol) \,\pv{\left(\frac{\RE}{\tfwhm} \,w(\uub)\right)} 
  \,\RE^{3} \,d\Dol  \,d\Dos   .
\label{eq.tfwhmmeas}
\end{equation}
The mass probability functions $p(\ml;\tfmeas,\uubmeas)$ and
$\hat{p}(\ml;\tfmeas,\sigma_{\tfmeas},\uubmeas,\sigma_{\uubmeas})$ can
then be obtained analogously to equations~\ref{eq.tEmeas3} and
\ref{eq.tEmeasgauss}.  Of course, equations (\ref{eq.tEmeas}),
(\ref{eq.tEmeas2}) and (\ref{eq.tfwhmmeas}) are equivalent and can be
converted into each other as long as $\tfwhm$ and $\uub$, and thus
$\tE$ are known.

Equation~(\ref{eq.tfwhmmeas}) nicely illustrates the transition to the
pixel-lensing regime: As soon as $\tE$ is not an observable anymore
(but only $\tfwhm$), the relative impact parameter enters in the
integral in equation (\ref{eq.tfwhmmeas}), and the mass probability
function becomes dependent on the maximum magnification of the source.
In pixel-lensing one is often in the situation where the $\tfwhm$ is
known quite accurately and $\uub(\A)$ is known to certain limits (if
finite source effects and/or space observations can rule out certain
magnifications and constrain the magnification interval, equation
(\ref{eq.tfwhmmeas}) leads to more realistic results than equation
(\ref{eq.tEmeas})).  In this case, the mass probability function can
roughly be obtained with
$\hat{p}(\ml;\tfmeas,\sigma_{\tfmeas},\uubmeas,\sigma_{\uubmeas})$
including the errors of $\tfmeas$ and $\uubmeas$.  But it is not
appropriate in this case to convert $\tfmeas$ to $\tEmeas$ (using
$\uubmeas$) and then to obtain $\hat{p}(\ml;\tEmeas,\sigma_{\tEmeas})$
from equation (\ref{eq.tEmeas3}), since then the error for $\tEmeas$
derived from a measured $\tfmeas$ also depends on $\uubmeas$.

\subsection{The Lens Mass Probability Distribution, Obtained from the
  Observables $\tfmeas$, $\DFmeas$, and $\Colmeas$}
\label{sec.masstfwhmDF}

Finally, we discuss the situation most relevant for pixel-lensing,
i.e.,  the case where only the flux excess $\DF$, the FWHM timescale
$\tfwhm$, and the color of the event $\Colmeas$ are determined
accurately from the light curve with coordinates $\xmeas$ and
$\ymeas$.

With the use of equation (\ref{eq.n_dgam_dt12_dCol}) one obtains
\begin{equation}
  \begin{array}{rl}
    p_{l,s}(\ml; \tfmeas, \DFmeas, \Colmeas) \sim &
    \Gamma_{l,s}(\ml,\tfmeas,\DFmeas,\Colmeas) \\
    = & \xi_l \,\frac{2}{\left(\tfmeas\right)^3} 
    \Intninf \,\pcmd_\iss 
    \Intninf n_\iss \,\frac{\Psi}{\F}
    \Int_0^\Dos \rho_l  \,\RE^{3} 
    \,\pv 
    \,d\Dol \,d\Dos  \,d\Mlum 
    ,
  \end{array}
  \label{eq.massprob_col}
\end{equation}
with 
\begin{displaymath}
  \Gamma_{l,s}(\ml,\tfwhm,\DF,\Col) := 
  \int \frac{d^7 \Gamma_{l,s}(\xmeas,\ymeas,\ml,\tfwhm,\DF,\Col,\Mlum)}
  {d\ml \,dx \,dy \,d\tfwhm \,d\DF \,d\Col \,d\Mlum } 
  \,\,d\Mlum ,
\end{displaymath} 
ignoring the (tiny) errors for the location of an event.
The functional dependence of the arguments in the integrand are 
$\xi_l(\ml)$,
$\pcmd_\iss(\Mlum,\Colmeas)$, 
$n_\iss(\xmeas,\ymeas,\Dos)$,
$\Psi(\A)$, 
$\A(\F,\DFmeas)$,
$\F(\Mlum,\Dos,\ext)$,
$\rho_l(\xmeas,\ymeas,\Dol)$,
$\RE(\Dol,\ml,\Dos)$, 
$\pv{\left(\frac{\RE\Upsilon}{\tfmeas},v_0\right)}$,
$\Upsilon(\A)$, 
and
$v_0(\xmeas,\ymeas,\Dol,\Dos)$.

Again, as outlined in equation (\ref{eq.tEmeasgauss}), one can include
the errors of the observables with a Gaussian measurement probability:
\begin{equation}
  \begin{array}{rl}
  \hat{p}_{l,s}(\ml; \tfmeas, \sigma_{\tfmeas},  \DFmeas,
\sigma_{\DFmeas}, \Colmeas,\sigma_{\Colmeas})
\sim & \int \int \int   \Gamma_{l,s}(\ml, \tfwhm, \DF, \Col) 
  \,g(\tfwhm;\tfmeas,\sigma_{\tfmeas}) \\
& \times g(\DF;\DFmeas,\sigma_\DFmeas)
\,g(\Col;\Colmeas,\sigma_{\Colmeas}) \,d\tfwhm \,d\DF \,d\Col  . 
  \end{array}
\end{equation}
If the light curve colors $\Colmeas$ can be measured very precisely,
the calculations can be simplified using a luminosity distribution
taken from the color-magnitude-diagram for a certain population.
Mathematically this can be written as
\begin{equation}
 \LF^\Col (\Mlum) \propto
 \int \pcmd(\Mlum,\Col) \,\delta{\left(\Col-\Colmeas\right)} \,d\Col .
\end{equation}
The modified luminosity distribution $\LF^\Col (\Mlum)$ replaces $\pcmd_\iss$ in 
equation (\ref{eq.massprob_col}).

\section{Conclusions and Outlook}
\label{sec.outlook}

Gravitational microlensing is a powerful method to detect compact
luminous and dark matter objects in the foreground of stars in nearby
galaxies.  It can thus be applied to measure the mass function of
stellar populations and dark halo objects (MACHOs).

One could infer the mass of an individual lensing object from the
lensing light curve directly, if the luminosity of the source, the
observer-lens-source distances, and velocities would be
known. However, at least the lens distance and velocity are
unfortunately almost never known. Hence, distribution functions for
the lens and source quantities (see \S\S~\ref{sec.prob_lens} and
\ref{sec.source_dis}) have to be used to finally obtain the
mass-probability function for individual lensing events.

We used these distribution functions to rederive well-known relations
like that for the optical depth, single-star event rate or mean
Einstein time of the events. These quantities were taken in the past
as ``back-of-the-envelope'' estimates of lensing frequencies to design
microlensing surveys and were evaluated for line-of-sight distances to
the plane of M31 only, i.e., simplifying the three dimensional
structure of M31.  We also accounted for the distance distribution of
the sources and obtained the line-of-sight distance-averaged
quantities for the optical depth, single-star event rate, and Einstein
time instead. We show their values as a function of line-of-sight
positions with contour plots in \S~\ref{sec.app}. The shape of the
total optical depth contours (Fig.~\ref{fig.tau_wecapp}, {\it third
  row, left}) obtained in this way deviates from earlier results (in a
way that is understood by the simplifications made;
\cite{2000ApJ...535..621G}).

Furthermore, we derived the distribution of the microlensing events
rate as a function of FWHM timescale and the magnification of the
event. We evaluated this function for a position [i.e.,
$(x,y)=(1\arcmin,0\arcmin)$] within the WeCAPP field and find the
following: the values of timescale and magnification are largely
confined to a linear region within the time magnification-FWHM
timescale plane (\S~\ref{sec.dis_tfwhm},
Fig.~\ref{fig.dGamma_dA0_dtfwhm}); an observing frequency of once per
day is sufficient to identify the majority of events with
magnification of the order 30--100; and higher magnification events
will have smaller timescales on average. Progress in the number of
detected lensing events can made by lowering the magnification
threshold for the event detection or, less efficiently, by further
improving the time sampling. The lowering of the noise per PSF can be
best achieved by small PSF and pixel sizes, i.e., by space
observations.

We then discussed the pixel-lensing or difference imaging regime,
which is the situation where the majority of stars is hardly or not at
all resolvable anymore. One then has to include the source luminosity
function to account for the additional unknown variable, the intrinsic
source flux. With that, we derived the distribution of the lensing
events (at a fixed position in the central M31 field) as a function of
the two main observables in the pixel-lensing regime, the excess flux
and the FWHM timescale. The values of these two quantities are not as
confined as those in the magnification-FWHM timescale plane
anymore. Due to the broad luminosity function there exists a variety
of combinations of magnification and intrinsic source flux which
yields the value for the flux excess. Events with high flux excess are
dominated by PMS source stars.

It had been noticed before
\citep{1994ApJ...421L..71G,2001ApJ...553L.137A} that measuring or
excluding finite source effects is useful to tighten constraints on
the masses of lensing objects. But finite source effects also change
the number and characteristics of events: In the presence of finite
source effects, the event timescales are increased and the maximum
magnification saturates below the maximum for the point-source
approximation (Fig.~\ref{fig.fs_Az}).  This shifts events to longer
timescales, but also suppresses the number of high-magnification
events, and therefore the number of observable events.  Since events
that are ultra-short (of order 0.001 days) in the point-source
approximation are mostly high-magnification events
(Figure~\ref{fig.dGamma_dA0_dtfwhm}), they all do show finite source
effects (if the lenses are residing in M31) and thus have larger
timescales than 0.001 days if the source sizes are taken into
account. This explains the absence of ultra-short events for
configurations with lenses in M31 (see Figs.~\ref{fig.dGamma_fs_1} and
\ref{fig.dGamma_fs_2}).

Using equation (\ref{eq.n_dgam_dt12_dCol}) and a flux excess
threshold, one can predict the time scale distribution of the events
in Figures~\ref{fig.dGamma_fs_1} and \ref{fig.dGamma_fs_2}.  At
different locations within M31 the amplitudes of the contours change,
and some details of the contours can be changed and moved in the flux
excess-FWHM timescale plane. However, in any case shown here one
expects many more short-term events with timescales of 1 to several
days, than long-term lensing events with timescales of 20 days or
longer. Even supermassive MACHOs with $1000\,\Msun$ have about roughly
the same number of events within 1 and 20 days as above 20 days. A
bimodal distribution of event timescales, with most events between 1
to 5 days, none between 10 and 20 days, and a second group of events
with timescales $20\E{d}$ and above is difficult to understand
(compare event candidates of \cite{2004A&A...417..461D}) on that
basis. De~Jong~et~al. argue that their result (many long-term events,
and the correlation of the event duration with the distance to the M31
center) can be understood, since the noise level is lower in the
outskirts, which would allow the detection of the long timescale
events. This does not explain the bimodality in the event timescales
(see Figs.~\ref{fig.dGamma_fs_1} and \ref{fig.dGamma_fs_2}).

Most searches for microlensing were started based on fairly simple
calculations of the expected event rates (see
\cite{1996ApJ...472..108H}). Their event-threshold criterion can be
translated to a peak-threshold criterion (see \S~\ref{sec.event}).
This yields about 200 events per year with a minimum signal-to-noise
of $Q \approx 6$ at maximum flux, and 15 events with $Q \approx 50$
(for their model survey, assuming 100\% efficiency).

The event rates measured up to now in M31 pixel-lensing surveys are
below the expectation values for pure self-lensing (using simple
estimates of survey efficiencies), while for microlensing surveys
toward the LMC and the Galactic bulge the numbers of detected
self-lensing events satisfy the predictions.  The apparent lack of M31
events can be due to an overestimated detection efficiency or
previously overestimated lensing rates.  We used the event
distributions as function of flux excess and fwhm timescale, and the
light distribution of M31 to finally derive the number of halo-lensing
and self-lensing events within the WeCAPP field that exceed a given
signal-to-noise ratio at the light curve maximum and have timescales
of 1 day or larger (see Table~\ref{tab.Gammatot_WeCAPP}).

For minimum signal-to-noise ratios of $Q=10$ and a minimum timescale
of 1 day one expects about 4.3 (bulge-bulge, disk-bulge, and
bulge-disk) self-lensing events per year that have light curves as for
point like sources and about 4.0 with finite source signatures in
their light curves. For timescales above 2 days these numbers decrease
to about 2.4 for point-source and 1.6 finite source signature events
per year. Since M31 cannot be observed more the two-thirds of a year,
the total efficiency will be not larger than 50\% (WeCAPP), even for a
survey with good time coverage. This means that there are not much
more than a couple of self-lensing events with $Q=10$ and timescales
larger than 2 days in a WeCAPP field per year. A decrease of the
``acceptable'' (S/N) ratio at maximum light to $Q=6$ does increase
the number of point-source events (roughly) less than a factor of 5
and has little impact on the events with finite source signatures. In
addition, at this variation level, a considerable fraction of the area
is occupied by intrinsically variable objects, which makes the
detection of lensing events even less effective.

These values are much below the already mentioned previous estimates.
The identification of the WeCAPP-GL1 and WeCAPP-GL2 event with a
signal-to-noise ratio of $Q\approx 85$ and $Q \approx 16$ at peak flux
and a FWHM timescale larger than 1 day in the WeCAPP 2000/2001 data is
in good agreement with our theoretical expectations.  We will present
WeCAPP results on lower signal-to-noise events in a forthcoming paper
and compare these numbers with expectations in more detail.

The most efficient way to increase the number of lensing events is to
lower the noise level. We investigate the number of self-lensing
events that can be obtained with a 30 day survey of the M31 center
using the ACS on board {\it HST} (1 orbit of total integration time
per day). Since bulge-bulge self-lensing profits more from lowering
the noise than the halo lensing (see Figure~\ref{fig.dGamma_fs_1}), a
decrease of the noise level increases the self-lensing relative to the
halo lensing.  During this campaign we expect of order 120 bulge-bulge
self-lensing events with a peak signal-to-noise ratio of 6 and
timescales between 1 and 20 days. Halo lensing with $0.1\,\Msun$ lenses
would cause an additional 50 events if the halo is composed of MACHOs
by 100\%. If the halo fraction is not more than 25\%, then the
halo-lensing events would drop to a 10\% of the total lensing
events. The analysis of the lensing events (frequency and timescale)
would provide a measurement of the low-mass end of the mass function
in the bulge of M31, i.e., the first measurement of the mass function
of stars at low masses outside our Galaxy.

Finally, we investigated the luminosity function of the stars that are
lensed.  The result is very sensitive to the timescale threshold of
the survey.  MS stars can only be seen if they are highly magnified,
which implies (Fig.~\ref{fig.dGamma_dA0_dtfwhm}) extremely short event
timescales. Present day surveys with minimum timescales of one day
therefore do not see any main sequence stars (for self-lensing in the
central bulge field). One can turn that result around: if one could
identify a modestly bright event with $\DF \ge 10^{-5} -
10^{-4}\E{Jy}$ with an MS source star and timescale of 1 day or larger
within the WeCAPP field (e.g., with spectroscopy by an ``instantaneous
alert''), it would point to a MACHO. This MACHO would have to be very
massive if it was within M31 and could be less massive within the
Galaxy.

Another interesting observable is the flux excess of the brightest
events. One can infer from equation (\ref{eq.largest_deltaF}) and
Figures~\ref{fig.dGamma_fs_1} and \ref{fig.dGamma_fs_2} that the
inclusion of the source sizes yields to an upper limit of the excess
brightness of the events. The value depends on the flux-to-radius
ratio of the brightest PMS stars in the lensed population and the mass
of the lens, plus some source and lens distance factors. If the
radius-luminosity relation of the source population and the luminosity
of the brightest PMS stars are known, one can obtain for every event a
lower lens mass limit for each source-lens configuration considered.

\begin{acknowledgements}
We warmly thank Claudia Maraston for discussions concerning stellar
populations in M31. We also thank Joachim Puls for useful
conversations about mathematical problems. We thank Christof Wiedemair
and Aglae Keller for their help. We also thank the anonymous referee
for the very useful discussion. Part of this work was supported by the
Son\-der\-for\-schungs\-be\-reich, SFB 375 of the Deut\-sche
For\-schungs\-ge\-mein\-schaft (DFG).
\end{acknowledgements}

\begin{appendix}

\section{Standard Event Definition and Maximum Light Curve Event Definition}
\label{standard_vs_newdef}

\begin{figure}                                          
  \epsscale{0.5}\plotone{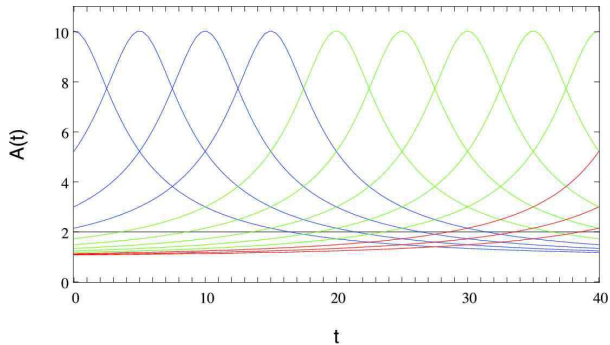} 
  \caption{Examples for the distributions of events within the time
    interval of the survey, for the standard definition and the
    maximum light curve definition. The magnification threshold is
    chosen as $\AT=2$. The green light curves are events for both
    definitions, the blue ones only for the maximum light curve
    definition, and the red ones only for the standard definition.  Of
    course, in both cases the times, when the events occur are
    homogeneously distributed within the survey time. The event time,
    however, is not equal to the time of light curve maximum in the
    standard definition case. It may happen, the light curve maxima is
    attained only after the survey has ended ({\it red curves}).  }
  \label{fig.bthresh}
\end{figure}

We motivate our alternative event definition in that section and
illustrate the differences from the standard definition.  The standard
definition, in which a lens becomes an event if it exceeds a threshold
in magnification $\AT$ (equivalent to entering the microlensing tube
with the corresponding radius $\bT$ for a given lens mass) has two
consequences:
\begin{enumerate}
\item Since only lenses that enter are counted, only the formal event
  times (when the magnification threshold is exceeded) but not the
  event maxima are homogeneously distributed within the survey time
  interval, if their event timescales are not much shorter than
  that. (see Figure \ref{fig.bthresh}, {\it red and green curves}; for
  the events with red light curves, the maximum will arise after the
  survey has ended) \hfill\break
\item The microlensing tube changes with the magnification threshold,
  and so does the spatial distribution of lenses that cause events
  within $\Dt$ (see Figure \ref{fig.bdist}, {\it red curves}, for a
  special example with $\Dvt = 1$).  Lenses that cause events with a
  higher magnification threshold within the survey time are not all a
  subset of those with a lower magnification threshold. Taking this
  event definition literally would make Monte-Carlo simulations
  time-ineffective, since high-magnification threshold subsamples
  could not be picked out from a more general sample.
\end{enumerate}
For event searches in data one usually requires to measure the light
curve around maximum (to check the light curve form, in particular its
symmetry), and of course, in practice, one would not exclude a light
curve from an event list, if it was above the magnification threshold
at the beginning of the survey.

This motivates the use of the maximum light curve definition, which
only accepts events that obtain their maximum within the survey time
interval $\Dt$. A threshold of the magnification $\AT$ at light curve
maximum, then, is equivalent to a maximum impact parameter $\bT$ of
the lens (for a given mass).
\begin{figure}                                                   
   \epsscale{0.5}\plotone{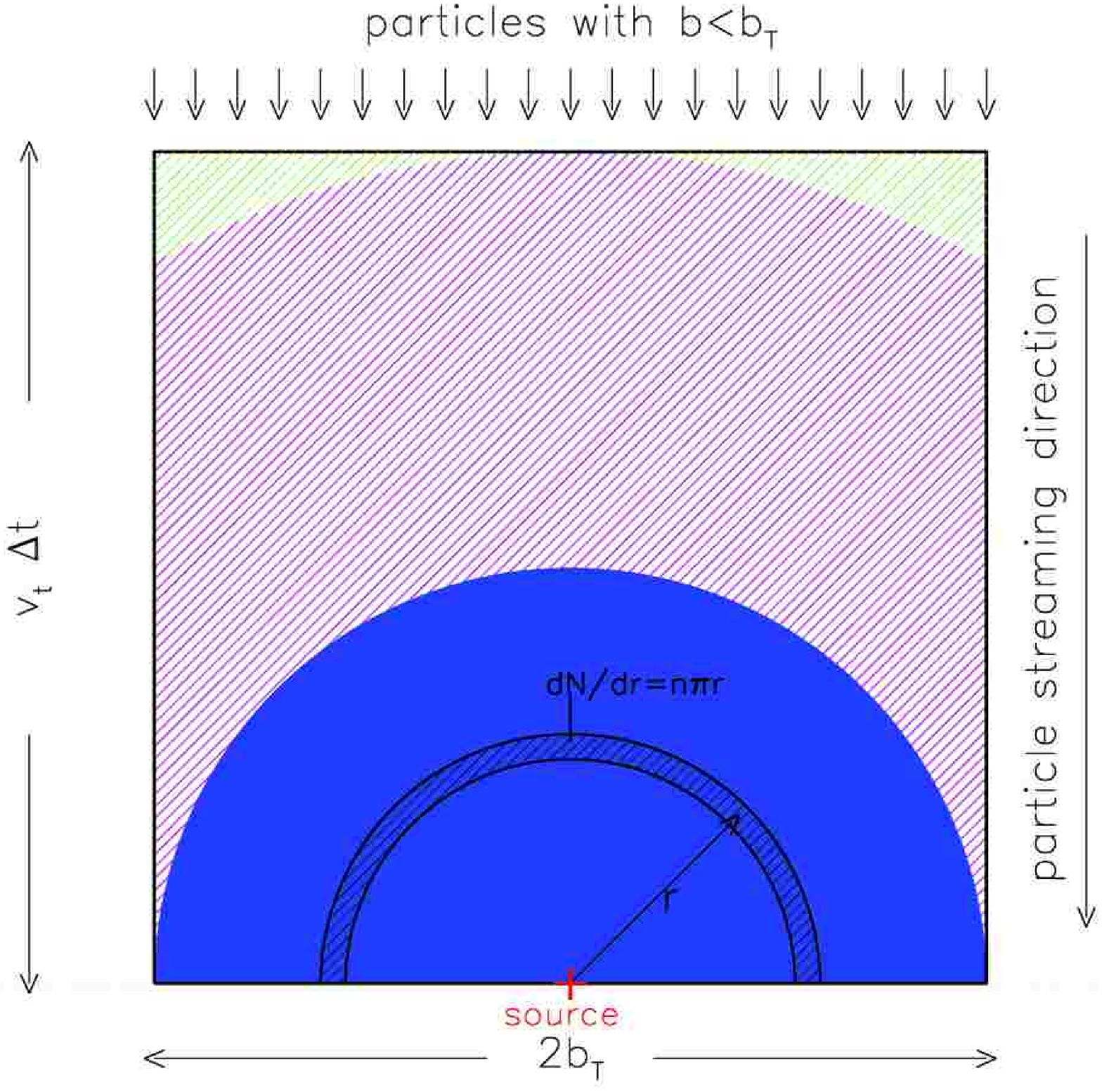} \,  
   \epsscale{0.5}\plotone{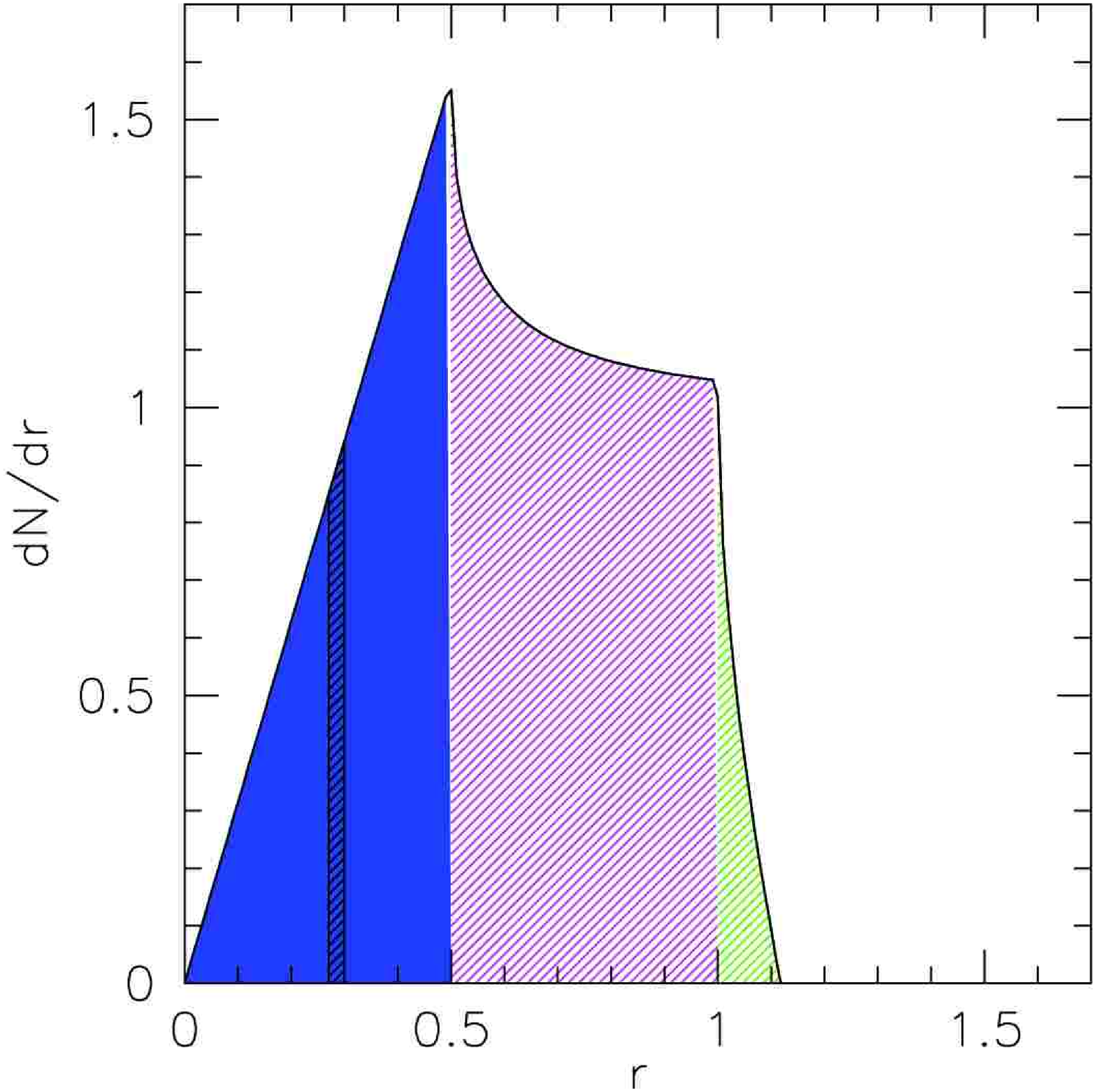} 
   \caption{These two figures illustrate how the impact parameter
     distribution and the radial distribution of lenses that become
     events in the maximum light curve definition are related. In the
     left panel we choose as an example a coherent particle flow with
     velocities $\vt$ in one direction.  Those particles that have an
     impact parameter $b \le \bT$ and will attain their minimum
     distance to the source within the survey time interval $\Delta t$
     are contained in the black box. Three areas have been shaded with
     blue, magenta, and green. The same colors have been used in the
     right panel to show in which part of the radial distribution
     function the events enter.}
  \label{fig.dNdr_events}                                                      
\end{figure}

\begin{figure}
  \epsscale{0.4}\plotone{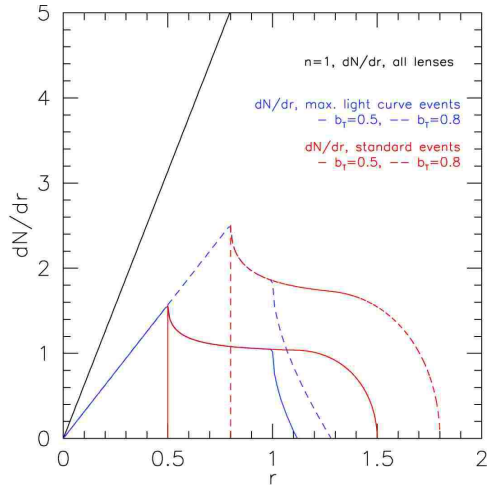} 
  \caption{Number of lenses per radius interval ({\it black}) and
    number of lenses causing events per radius interval. Curves for
    the standard definition are in red, and for the maximum light
    curve definition are in blue.  Length scales are in arbitrary
    units, the density is chosen such that it equals $n=1$ in these
    units, and the velocities are chosen such that $\Dvt=1$ holds.
    The thresholds for the microlensing tube radius and the maximum
    impact parameter have been chosen as $\bT=0.5$ ({\it solid lines})
    and $\bT=0.8$ ({\it dashed lines}).  The integral of the
    corresponding red and blue curves coincide and give the number of
    lenses that cause events within $\Delta t$.  For the standard
    definition (the locations of) the lenses that cause events with a
    higher magnification threshold are not all a subset of those with
    a lower magnification threshold. In addition, lenses that are
    already within the microlensing tube never will cause any event
    for the standard definition (implying the lower cutoff).  The red
    and blue curves shown here are straightforwardly obtained
    analytically (derivation not shown in this paper).  }
  \label{fig.bdist}                                                              
\end{figure}          
                                                         
We now consider events with a threshold $\bT$ for both event
definitions.  We calculate the location of those lenses at survey
begin that become events within the survey time interval.  We assume
the lenses to be distributed in a plane and to have velocities $\Dvt =
1 $. Length scales are given in arbitrary units and the density of
lenses is assumed to equal $n=1$ in these units.  The number of lenses
per radius interval is $\frac{dN}{dr}= 2 \pi r $. This curve is shown
in black in Figure~\ref{fig.bdist}.

The blue and red curves show the number of those lenses per radius
interval that become events within $\Dt$ in the new and the standard
definition, respectively.  For the standard definition, only lenses
with $\bT\le r \le \bT+\Dvt $ will become events within $\Dt$,
explaining the minimum and maximum radius in Figure~\ref{fig.bdist}
({\it red solid and red dashed curve}, for a threshold of $\bT=0.5$
and $\bT=0.8$, respectively).

In the maximum light curve definition lenses within $0< r <
[\bT^2+(\Dvt )^2]^{1/2}$ can cause events with $b<\bT$, explaining the
maximal radius in Figure~\ref{fig.bdist} ({\it blue solid and blue
  dashed curves}, for the threshold of $\bT=0.5$ and $\bT=0.8$,
respectively).  The relation between the features in the radial
distribution of lenses becoming events and the particles motion is
shown for the case of a coherent particle stream in
Figure~\ref{fig.dNdr_events}.

One can also see in Figure~\ref{fig.bdist} that lenses with the higher
magnification threshold (corresponding to $\bT=0.5$) are a subset of
those with the lower magnification threshold ($\bT=0.8$). In addition,
the lenses causing an event within $\Delta t$ are spatially more
confined than for the standard definition: the maximum radius from
which a lens can cause an event with an impact parameter $\bT$ within
$\Dt$ is $[\bT^2+(\Dvt)^2]^{1/2}$ for a relative velocity of $\vt$.

Hence, it is obvious that new event definition is more easy to use in
simulations, but also more directly linked to observations. On the
other hand, one can guess from Figure \ref{fig.bdist} that the
integral of the corresponding blue and red curves in Figure
\ref{fig.bdist} agrees. Therefore, the number of events for both
definitions is the same.

This implies that for both event definitions, a magnification
threshold $\AT$ or impact parameter threshold $\bT$ (for a given lens
mass) yields the same events (same number of events and same light
curve parameters for the events, with exception of the time of
maximum); only the lenses that cause the events are different in both
definitions.  Since it is not relevant where the lenses that cause
events have come from, one can conveniently switch definitions.

\section{Modeling of M31}
\label{app.model}

\subsection{Density Distribution}

This section contains our models for the bulge, disk and halo density
of M31 and comparison with observations.  We show that taking a bulge
with the same total mass as \cite{1989AJ.....97.1614K} and a disk with
the same total mass as \cite{2001MNRAS.323...13K} implies
mass-to-light ratios for the stellar populations of bulge and disk in
good agreement with expectations from population synthesis models.
Our bulge model matches the observed surface brightness values of M31
better than previously published analytical models, which is important
for the correct self-lensing prediction in the central part of
M31. The contributions of the bulge and disk to the rotation curve are
almost identical to that shown in \cite{2001MNRAS.323...13K}, which
allows us to assume the same density distribution for the dark halo as
they did.

In this section we use the disk major axis coordinate system $(x_0,
y_0, z_0$; see Fig.~\ref{fig.tau_wecapp}), which can easily be
transformed to the line-of-sight coordinate system using an
inclination $i$ of $77^\circ$ of M31 \citep{1998ApJ...503L.131S}.

\subsubsection{Bulge of M31}

\begin{figure}[b]
  \epsscale{1.0}\plottwo{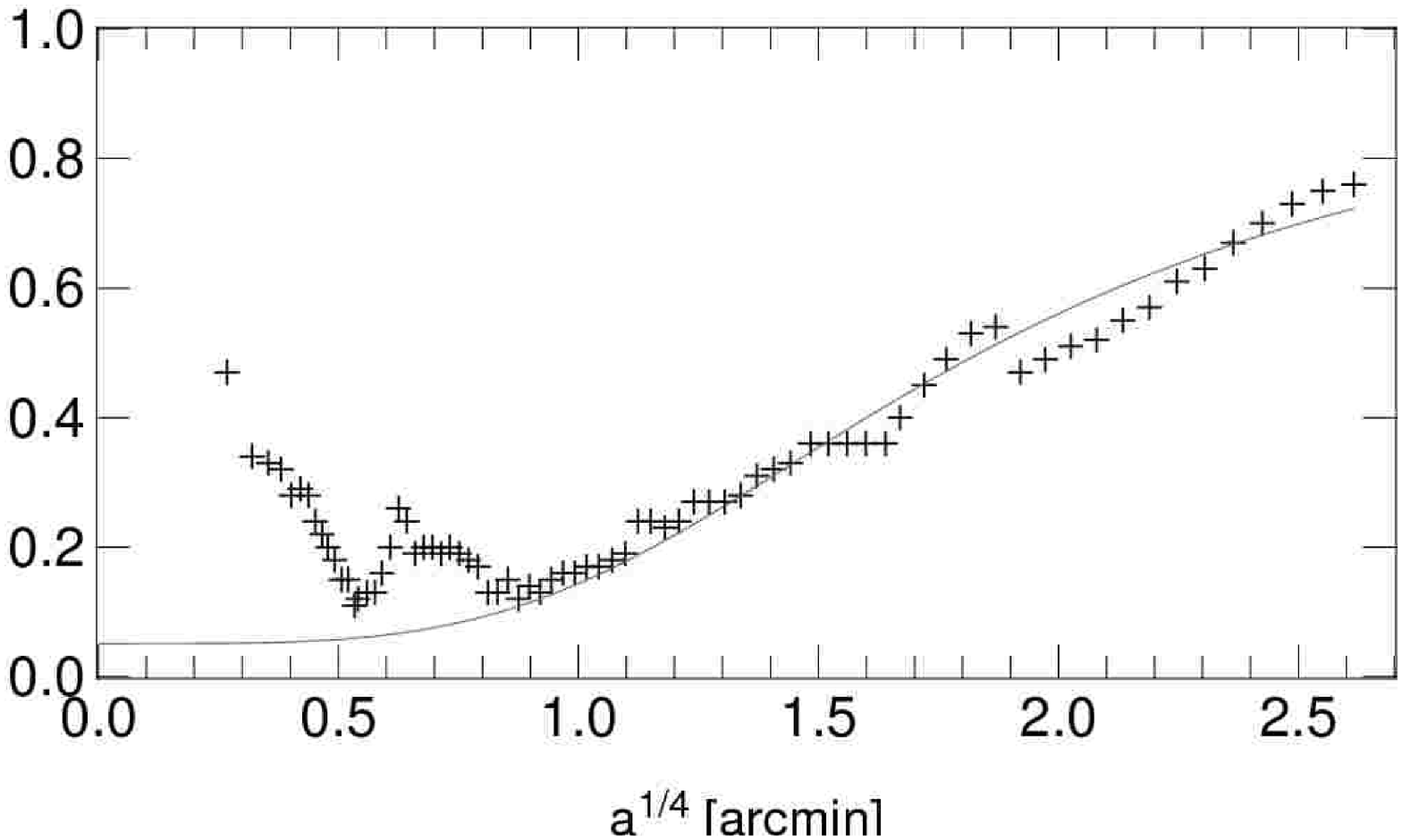}{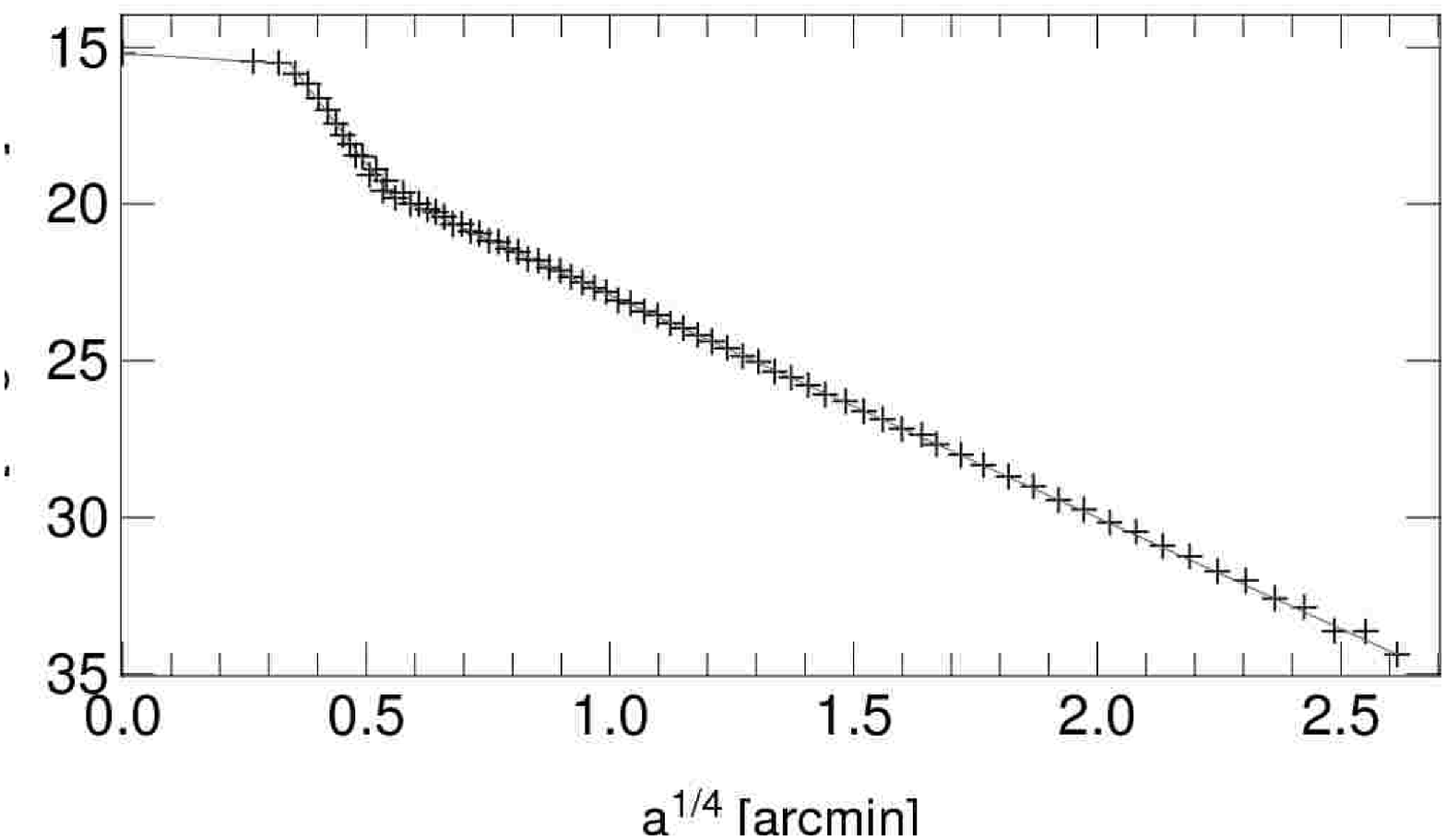} 
  \caption{({\it a}) The gray curve shows our approximation for
    $\epsilon (a)$ as defined in equation (\ref{eq.bulge_eps}), and
    the red crosses are the tabulated values of
    \cite{1989AJ.....97.1614K}; within $0.5\arcmin$ and $6\arcmin$ the
    agreement is excellent.  ({\it b}) Bulge surface brightness as
    tabulated by \cite{1989AJ.....97.1614K} is shown with crosses, and
    our approximation from equation (\ref{eq.bulge_rho}) as a gray
    curve.}
  \label{fig.bulge_kent}
\end{figure}

Our M31 bulge model starts from Table~I of \cite{1989AJ.....97.1614K},
containing the Gunn-r surface brightness and ellipticity values
$\rho^\M{Kent}_r(a)$ and $\epsilon^\M{Kent}(a)$ as a function of
major-axis distance $a$ to the center of M31.  We assumed $50^\circ$
for the position angle of the bulge.  Figure~\ref{fig.bulge_kent}
demonstrates that with

\begin{equation}
  \left[\frac{1}{1-\epsilon(a)}\right]^2 := 0.254 \frac{a}{\M{arcmin}}  + 1.11
  \label{eq.bulge_eps}
\end{equation}
the ellipticity $\epsilon (a)$ ({\it red curve}) becomes an excellent
approximation of $\epsilon_r ^{\rm Kent}(a)$ ({\it blue crosses}) between
$0.5\arcmin$ and $6\arcmin$.

With this relation we convert $(x_0,y_0,z_0)$ to $a$ by solving the
quadratic equation $a^2 = x_0^2 + y_0^2 + \left( 0.254 a + 1.11\right)
z_0^2$,
\begin{equation}
  a(x_0,y_0,z_0) = \frac{0.254 z_0^2+\sqrt{0.254^2 z_0^4 + 4 \left(x_0^2+y_0^2+ 1.11 
  z_0^2\right)}}{2} \hs [\M{arcmin}] ,
  \label{eq.bulge_axis}
\end{equation}
with $x_0$, $y_0$, $z_0$ and $a$ in arcminutes.

The three-dimensionally decomposed spatial brightness density profile
of the M31 bulge derived by Kent is well approximated by an $a^{1/4}$
law (see Figure \ref{fig.bulge_kent}b). With equation
(\ref{eq.bulge_axis}) the bulge mass density becomes
\begin{equation}
  \rho_\M{bulge}(x_0,y_0,z_0) := \left\{
\begin{array}{ll}
   \rho_{0} \, 10^{-0.4 ( 0.97  \,a^{1/4})},       & \quad a \le 0.014\arcmin, \\
   \rho_{0} \, 10^{-0.4 ( 20.4  \,a^{1/4}-6.68)},  & \quad 0.014\arcmin < a \le 0.09\arcmin, \\
   \rho_{0} \, 10^{-0.4 ( 7.1   \,a^{1/4}+ 0.61)}, & \quad a > 0.09\arcmin,                       
\end{array}
\right.
 \label{eq.bulge_rho}
\end{equation}
where
\begin{equation}
  \rho_{0} := 
  \ML_{\XX}  10^{-0.4\left[\rho^\M{Kent}_{0,r}-(\Mlum_r-\Mlum_\XX)-\ext_{\XX} - d_\M{mod}-\Mlum_{\odot,\XX}\right]} \,\frac{\Msun}{\M{arcsec}^3}
\label{eq.rhobulgenorm}
\end{equation}
is the central mass density derived from the central brightness
density in the $r$ band,
$\rho^\M{Kent}_{0,r}=15.19\E{mag}\E{arcsec}^{-3}$
(\cite{1989AJ.....97.1614K}, Table~I), and $(M/L)_\XX$ is the bulge
mass-to-light ratio in a fiducial filter $\XX$, and $(\Mlum_r-\Mlum_\XX)$ is the color of the bulge population; $\Mlum_{\odot,\XX}:=-2.5
\lgt({\FsunX}_\XX/\FVega_\XX)$ is the absolute brightness of the
Sun in that filter and $d_\M{mod}$ is the distance modulus to M31.

\begin{table}
\setlength{\tabcolsep}{1.9mm}
\begin{center}
\caption{Mass-to-Light Conversion Ratio for the M31 Bulge Model}
\begin{tabular}{ccccrccc}
\hline\hline
      & Distance  &  &   &  &               &  &     \\ 
 $\ml_\M{tot}/\Msun$       & (kpc) & Band $\XX$  & $(\Mlum_r-\Mlum_\XX)$   & $\ext_{\XX}$ &  $L_{\M{tot},\XX}/L_{\odot,\XX}$              & $(M/L)_{\XX}$  & Comment    \\  
\hline
 $4  \times 10^{10}$ & 690        & $r$ & 0    & 0    & $6.61 \times 10^9$ & 6.05  & Kent's model using eq. (\ref{eq.rhobulgenorm}) \\ 
 $4  \times 10^{10}$ & 770        & $R$ & 0.59 & 0.36 & $13.5 \times 10^9$ & 2.96  &  \\ 
\hline
\end{tabular}
\end{center}
\tablecomments{This table shows that a bulge mass of $M=4\times 10^{10}\,\Msun$ 
  as proposed by Kent is a good estimate, even for
  the more realistic value ($770\E{kpc}$) for the M31 distance.  
}
\label{tab.parabulge}
\end{table}

\cite{1989AJ.....97.1614K} fixes the bulge mass to $4 \times
10^{10}\,\Msun$, which for $d_\M{mod}=24.19\E{mag}$ ($690\E{kpc}$) and
without correcting for dust extinction implies a $(M/L)_{r}$-ratio of
$6.05$ (using our analytic approximation for $\rho_\M{bulge}$) and
$5.5-6.6$ (integrating the tabulated values of Kent and estimating the
maximal uncertainties due to the coarseness of the table\footnote{We
  derived the upper and lower limit for the total brightness of the
  bulge in Kents Table~I.  By summing over ellipses (with an area $A_i
  := \pi a_i^2 (1-\epsilon_i)$ at semi-major distance $a_i$, $A_1=0$,
  and with a surface brightness $l_{r,i} := L_{\odot,r} 10^{-0.4
    (\mu(a_i) - d_\M{mod} -\Mlum_{\odot,r})}$, we got
  $L_{r,\M{tot}}^\Min / L_{\odot,r} = \sum_{i=2}^{77} l_{r,i }
  \,(A_i-A_{i-1})$, and $L_{r,\M{tot}}^\Max / L_{\odot,r} =
  \sum_{i=2}^{77} l_{r,i-1} \,(A_i-A_{i-1})$.  These limits lead to a
  slightly higher $(M/L)_r$ between $5.5$--$6.6$ than the value given by
  \cite{1989AJ.....97.1614K}: $(M/L)_r=5\pm 0.5$.}). Using the favored
distance to M31 ($d_\M{mod}=770\E{kpc}$) and applying reasonable
extinction values, the dust corrected mass-to-light ratios reduce to
lower values (for a constant bulge mass of $4 \times10^{10}\,\Msun$, see
Table~\ref{tab.parabulge}).

The $R$ band values were obtained with $R_\odot=4.42\E{mag}$, $(\Mlum_r-\Mlum_R) =
0.43+0.15 (B-V) = 0.59$ \citep{2000A&AS..147..361M}, and a bulge color
of $(B-V) \approx 1.05$ \citep{1987A&AS...69..311W}.

According to \cite{1996ApJ...472..108H} the effect of an asymmetric
bulge light extinction caused by the highly inclined M31 disk is
negligible.  We therefore adopt his values for the mean internal
extinctions toward the bulge in the $V$ and $I$ bands of
$\ext_V=0.24\E{mag}$ and $\ext_I=0.14\E{mag}$ and interpolate to the
$R$ band which yields $\ext_R =0.19$.  With the foreground extinction
of $\ext_R=0.17$ \citep{1998ApJ...500..525S} the total extinction
becomes $\ext_R = 0.36$.  In this case, the mass-to-light ratio
corresponding to Kent's bulge mass becomes $(M/L)_R = 2.96$ (line 2 in
Table~\ref{tab.parabulge}).  This value is close to that
[$(M/L)_{R,\M{stellar}}=2.67$] one would obtain for a $12\E{Gyr}$ old,
$Z=2 Z_\odot$ metallicity single stellar population (SSP; see
\cite{2002A&A...391..195G}) for a \citet{2000ApJ...530..418Z} mass
function (MF) (see \S\S~\ref{sec.massfunc} and \ref{sec.LFmodel}).

We conclude that a normalization (equation (\ref{eq.rhobulgenorm})) of
\begin{equation}
  \rho_0=2.07\times 10^{6} \,\Msun\E{arcsec}^{-3} = 3.97\times 10^{4}
  \,\Msun\E{pc}^{-3}   ,
\end{equation}
which reproduces Kent's bulge mass of $M=4\times 10^{10}\,\Msun$, is a
reasonable assumption and represents an upper limit for the luminous
matter in the bulge.

\begin{figure}[b]  
  \epsscale{1.0}
  \plottwo{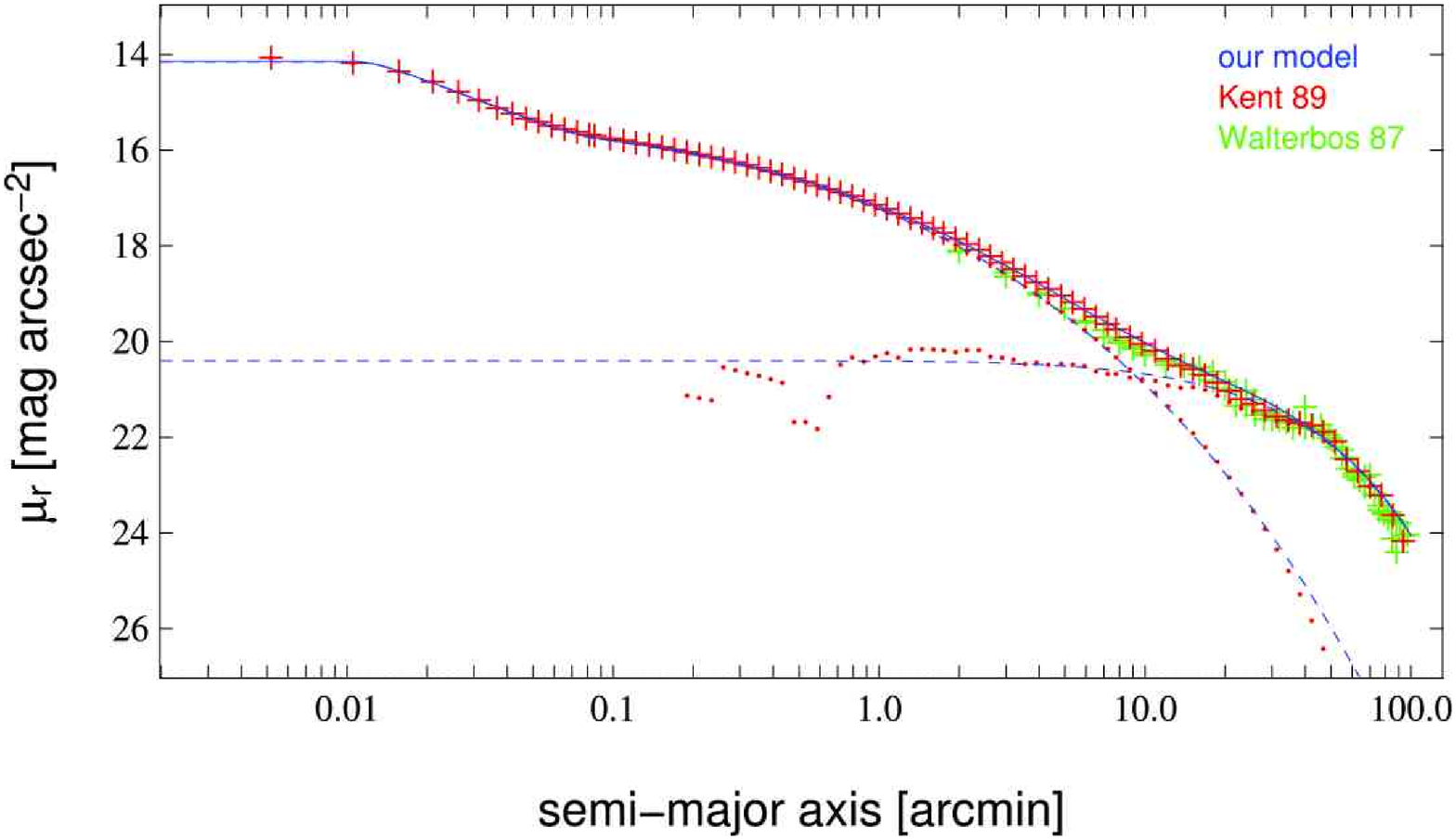}{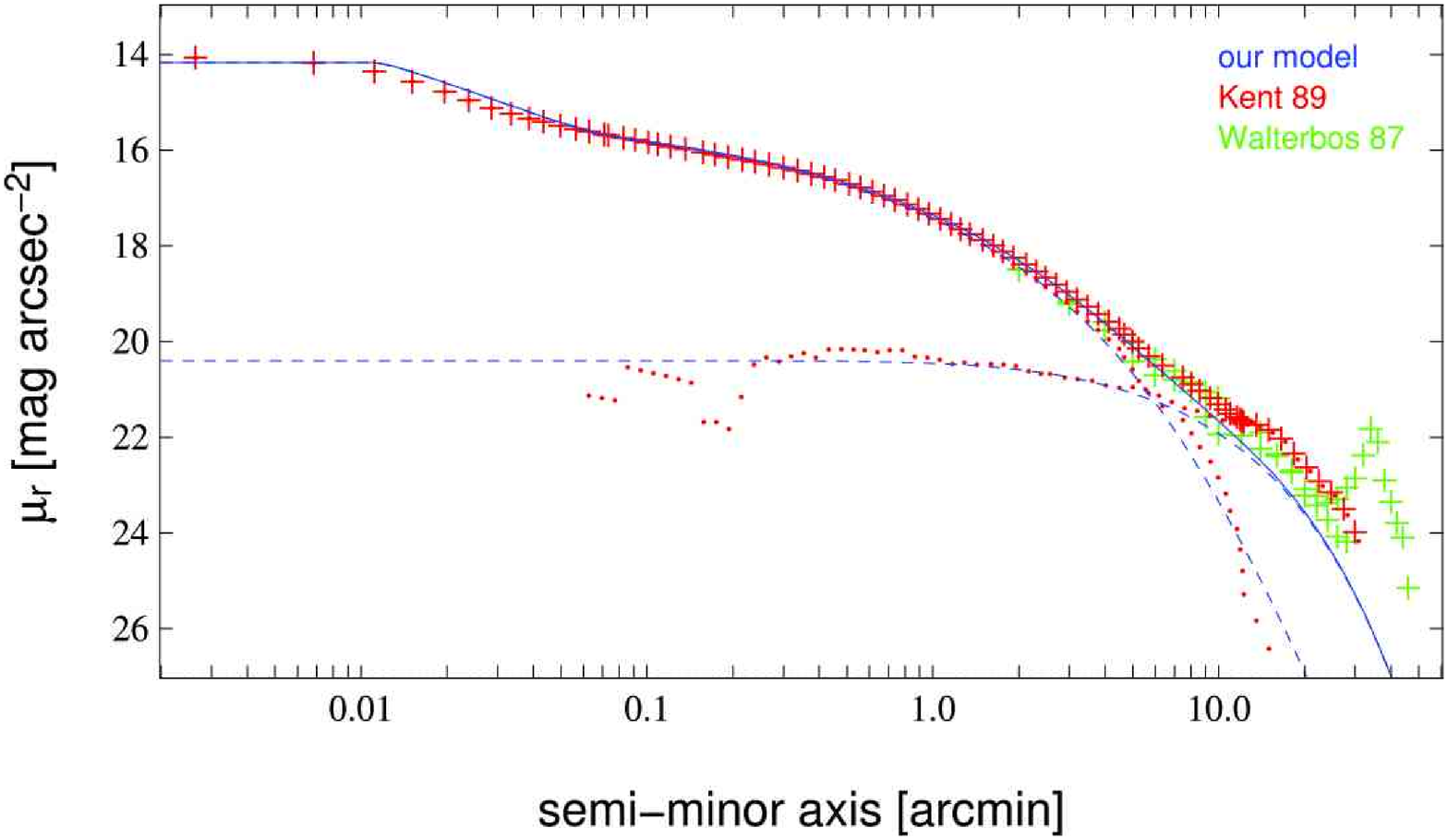} 
  \caption{Surface-brightness profile of M31 in the $r$ band: red
    crosses are Kent's $r$ band data for the central region of M31;
    green crosses are the \cite{1987A&AS...69..311W} data (their
    Table~V) transformed to the $r$ band.  The left and right panels
    show the profiles along semi-major and semi-minor axis,
    respectively.  Kent has decomposed the surface brightness profile
    into the bulge and disk component ({\it red dots}). For comparison
    we have superposed our bulge and disk surface brightness models
    from Eqs.~\ref{eq.bulge_rho} and \ref{eq.disk_rho}. With the
    exception of spiral arm imprints, they match the observations
    extremely well.  }
  \label{fig.SBwecapp}
\end{figure}

\subsubsection{Disk of M31}

Like \cite{2001MNRAS.323...13K}, we model the disk by a $\M{sech}^2$
law,
\begin{equation}
  \rho_\M{disk}(x_0,y_0,z_0)  = \rho_0 \exp{\left(-\frac{\sigma(x_0,y_0)}{h_\sigma}\right)} \,\M{sech}^2{\left(\frac{z_0}{h_z}\right)} ,
 \label{eq.disk_rho}
\end{equation}
with $\sigma(x_0,y_0)={\left(x_0^2+y_0^2\right)}^{1/2}$ being the
radial distance in the disk plane inclined by $77^\circ$; the radial
scale length $h_\sigma=28.57\arcmin$ and the vertical scale lengths
$h_z=1.34\arcmin$ are equivalent to \cite{2001MNRAS.323...13K} values
$h_\sigma=6.4\E{kpc}$ and $h_z=0.3\E{kpc}$ for a M31 distance of 770
kpc.  Adopting a central brightness density of the disk in the $r$
band $\rho^\M{Kent}_{0,r}=27.39\E{mag}\E{arcsec}^{-3}$ yields a
surface brightness profile that matches the data of
\cite{1989AJ.....97.1614K} on the major axis and that agrees well with
his central surface brightness of $\mu_0=20.4\E{mag}$ in the $r$ band.
Spiral arms and dust explain the discrepancies at the minor axis (see
Figure~\ref{fig.SBwecapp}).  We assumed $38^\circ$ for the position
angle of the disk.

As for the bulge, we transform the luminosity density to matter
density, using the disk color $(\Mlum_r-\Mlum_\XX)$, disk extinction $\ext_\XX$,
and disk mass-to-light ratio $(M/L)_\XX$
\begin{equation}
  \rho_{0} = \ML_\XX 10^{-0.4(\rho^\M{Kent}_{0,r}-(\Mlum_r-\Mlum_\XX)-\ext_\XX - 
  d_\M{mod}-\Mlum_{\odot,\XX})} \frac{\Msun}{\M{arcsec}^3},
 \label{eq.rhodisknorm}
\end{equation}
with the absolute brightness of the Sun $\Mlum_{\odot,\XX}$, and the
distance modulus $d_\M{mod}$ to M31.
\begin{table}
\setlength{\tabcolsep}{1.4mm}
\begin{center}
\caption{Mass-to-Light Conversion Ratio for the M31 Disk Model}
\begin{tabular}{cccccrcl}
  \hline\hline
  & Distance  &  &   &  &               &  &     \\ 
  $\ml_\M{tot}/\Msun$         & (kpc) & Band $\XX$ & $(\Mlum_r-\Mlum_\XX)$ & $\ext_{\XX}$ &  $L_{\M{tot},\XX}/L_{\odot,\XX}$   & $(M/L)_{\XX}$   & Comment    \\  
  \hline                  
  $16 \times 10^{10}$          & 690       & gunn $r$ & 0    & 0    & $1.4 \dots 1.7 \times 10^{10}$  & $11.3\dots9.6$ & for Kent's max. disk  mass \\
  $3.09 \times 10^{10}$        & 690       & gunn $r$ & 0    & 0    & $1.34\times 10^{10}$          & 2.31      &   for Kerins's disk mass  \\
  $3.09 \times 10^{10}$        & 770       &    $R$   & 0.54 & 0.68 & $3.5\times 10^{10}$          &  0.88    &       \\
  \hline
\end{tabular}
\end{center}
\tablecomments{This table shows in its last two lines the mass-to-light
  ratios resulting from the disk mass of \cite{2001MNRAS.323...13K},
  $3.09 \times 10^{10}\,\Msun$. The mass-to-light ratio for a
  realistic amount of extinction (last line), is close to a theoretical 
  $(M/L)_{R,\M{stellar}}=0.62$ for a $2\E{Gyr}$ old, solar metalicity SSP 
  disk population (based on \cite{1997ApJ...482..913G} and 
  \cite{2002A&A...391..195G}).
  Comparing the first and second lines shows that the maximum disk ---
  assumed by \cite{1989AJ.....97.1614K} (first line) --- would imply a
  much too large mass-to-light ratio, which is usually obtained for
  maximum disk models).}
  \label{tab.paradisk}
\end{table}

We normalize equation (\ref{eq.rhodisknorm}) with
\begin{equation}
\rho_0=10.4 \,\Msun\E{arcsec}^{-3} = 0.2 \,\Msun\E{pc}^{-3}
\label{eq.normdisk}
\end{equation}
to obtain the same disk mass as \citet{2001MNRAS.323...13K}
$M_{\mathrm{disk}}=\int \int \rho \,dz \,d\sigma = 4 \pi \rho_0 h_z
h_\sigma^2=3.09 \times 10^{10}\,\Msun$. Table~\ref{tab.paradisk}
demonstrates that this normalization results in a mass-to-light ratio
that is expected for the disk population.

With $E(B-V)=0.22$ \citep{2003AJ....125.2473S} we obtain $\ext_V = 3.1
E(B-V) = 0.682$ for the extinction in the M31 disk.  This translates
to $\ext_R = 0.748 \,\ext_V = 0.51$ \citep{1998gaas.book.....B}.
Adding the foreground extinction, $\ext_R=0.17$
\citep{1998ApJ...500..525S}, we obtain $\ext_R = 0.68$ for the total
extinction for sources residing in the disk of M31.  Using that
extinction, the M31 distance of 770 kpc and the central luminosity
density of $\rho^\M{Kent}_{0,R}=26.86\E{mag}\E{arcsec}^{-3}$ [obtained
from $\rho^\M{Kent}_{0,r}$ and $(\Mlum_r-\Mlum_R)=0.53$ for a disk color $(B-V)
\approx 0.7$; \cite{1987A&AS...69..311W}], we get a disk luminosity of
$L_{\M{tot},R}/L_{\odot,R}= 3.5\times 10^{10}$.  For the disk mass of
\citet{2001MNRAS.323...13K}, our $(M/L)_R$ ratio becomes 0.88.  This
mass-to-light ratio is well consistent with a theoretical
$(M/L)_{R,\M{stellar}}=0.61$ for a $2\E{Gyr}$ old, solar metalicity SSP
disk population (based on \cite{1997ApJ...482..913G} and
\cite{2002A&A...391..195G}).

We also summarize the maximum disk model of \cite{1989AJ.....97.1614K}
in Table~\ref{tab.paradisk} ({\it first row}). This model implies a 4
times higher $(M/L)_r$-ratio, which is hard to reconcile with
population synthesis models.

Note that the results from \citet{1996ApJ...473..230H} are not easy to
compare with ours: they used a double exponential disk with
$\rho_0=0.35 \,\Msun\E{pc}^{-3}$, $h_z=0.4$~kpc, and
$h_\sigma=6.4$~kpc corresponding to a disk mass of $7.2 \times 10^{10}
\,\Msun$. At the same time their bulge is also more massive than ours
($4.9 \times 10^{10} \,\Msun$).

\subsubsection{Halo of M31}

\begin{figure}[b]
   \epsscale{0.5}\plotone{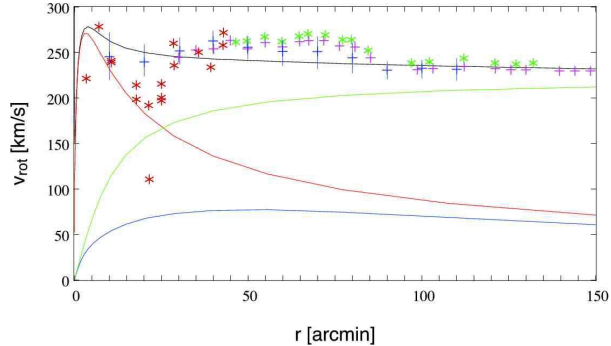}
   \caption{Overall rotation curve of our model ({\it black curve})
     and its contributions of the bulge ({\it red}), disk ({\it
       blue}), and halo ({\it green}).  These rotation curves match
     with Fig. 3{\it b} in \cite{2001MNRAS.323...13K}.  In {red}
     crosses we show the data points derived from CO measurements of
     \cite{1995A&A...301...68L}; in green, HI measurements from
     \cite{1984A&A...141..195B}; in blue, averaged data points from
     \cite{2003ApJ...588..311W} (based on \cite{1989PASP..101..489K},
     and \cite{1991ApJ...372...54B}); in magenta, the data points of
     \cite{2001MNRAS.323...13K} (based on \cite{1989AJ.....97.1614K}).\\
   }
  \label{fig.vrot}
\end{figure}

Our density models for the bulge and disk differ only slightly
(e.g., in the central region) from that of \cite{2001MNRAS.323...13K}.
The contributions to the rotation velocity resulting from the
different populations are therefore very much the same as in the
\cite{2001MNRAS.323...13K} model. This implies that we can use the
halo density distribution from \cite{2001MNRAS.323...13K} to obtain a
halo model consistent with the observed M31 rotation curve.  This halo
density distribution is that of an isothermal sphere with a core
radius of $r_\M{c}=2\E{kpc}$:
\begin{equation}
  \rho_\M{halo}(x_0,y_0,z_0) =
  \frac{\rho_0}{1+\left(r/r_\M{c}\right)^2},\quad  r\le 200\E{kpc},
\end{equation}
with $r = (x_0^2+y_0^2+z_0^2)^{1/2}$, $r_\M{c}=2\E{kpc}$, and $\rho_0 =
0.23 \,\Msun\E{pc}^{-3}$. Figure \ref{fig.vrot} shows the overall
rotation curve of our model.

In the model of \citet{1996ApJ...473..230H} the core radius of the
halo is much larger ($r_\M{c}=6.5\E{kpc}$) to compensate for their
higher disk and bulge mass in order to match the rotation curve of
M31.

\subsubsection{Halo of the Milky Way}

The halo of the Milky Way (MW) is also modeled as a cored isothermal
sphere,
\begin{equation}
  \rho_\M{MW}(\Dol) = \frac{\rho_0}{1+\left(r/r_\M{c}\right)^2},   \quad  r<200\E{kpc} ,
\end{equation} 
where we choose a core radius of $r_\M{c}=2\E{kpc}$ as used in
\citet{1996ApJ...473..230H} and \citet{2000ApJ...535..621G}. 

The central density is taken from \citep{1996ApJ...473..230H}:
\begin{equation}
  \rho_0 = 0.0079 \,\Msun\E{pc}^{-3}
  \,\left[1+\left(\frac{r_\odot}{r_\M{c}}\right)^2\right] = 0.1343
  \,\Msun\E{pc}^{-3} .
\end{equation}
We convert the Galactocentric distance $r$ to our line-of-sight
coordinate system according to
\begin{equation}
r(\Dol)=\sqrt{r_\odot^2 - 2 r_\odot \Dol \cos(l) \cos(b) +
\Dol^2} ,
\end{equation}
using the M31 Galactic coordinates $l=121.14988^\circ$,
$b=-21.61707^\circ$ and the solar Galactocentric distance
$r_\odot=8\E{kpc}$ \citep{1983ApJ...265..730B}.

\subsection{The Mass Function}

\subsubsection{The Mass Function for the bulge and disk sources}
\label{sec.massfunc}

For the M31 bulge we take the mass function (MF) $\xi \sim M^{-1.33}$
of \cite{2000ApJ...530..418Z}, which was derived for the Galactic
bulge.  The MF is cut off at $0.01\,\Msun$ at the lower end and at the
MS turnoff $1.01\,\Msun$ at the upper end for a 12 Gyr old SSP with
$Z=2 Z_\odot$.

We describe the disk with a Gould MF, $\xi \sim M^{-2.21}$, which has
a flattening $\xi \sim M^{-0.56}$ below $0.59\,\Msun$
\citep{1997ApJ...482..913G}.  We cut the disk MF at $0.01$ and
$1.71\,\Msun$ (2 Gyr old SSP with $Z=Z_\odot$), respectively. Of
course, the number of stars with a given mass changes for different
cut off values or for alternative mass functions
(e.g., \citet{2003PASP..115..763C}).  The investigation of halo-lensing
and self-lensing rates for different MFs is not a subject of that
paper.

\subsubsection{The Mass Function for the Halo}

The mass function $\xi(\ml)$ for the potential MACHO population
residing in the halo of M31 is of course unknown. In this paper we
simply assume that the halo consists of one mass objects $\Mlens$
only,
\begin{equation}
\xi(\ml) = \frac{\delta(\ml-\Mlens)}{\Mlens} ,
\label{eq.mfHalo}
\end{equation}
satisfying the normalization constraint
\begin{equation}
\int \ml\,\xi(\ml)\,d\ml=1 .
\label{eq.mfHaloNorm}
\end{equation}

\subsection{The Luminosity Function and CM Diagram}
\label{sec.LFmodel}

We use a stellar LF obtained from isochrones of the Padova database of
stellar evolutionary tracks and isochrones given by
\cite{2002A&A...391..195G} (based on \cite{2001A&A...377..132M}).

The luminosity function can be extracted from the mass function
$\xi(\ms)$ discussed in \S~\ref{sec.massfunc}.  Using the
mass-magnitude relation provided by theoretical stellar isochrones
each mass bin $[\ms_i,\ms_{i+1}]$ of stars is connected to a absolute
brightness\footnote[100]{
Note that we neglect the correct indizes refering to the 
band $\XX$ and define
$\F \equiv F_{0,\XX}$,
$\FVega\equiv{F_{\M{Vega},\XX}}$,
$\Flum\equiv{\mathcal{F}_\XX}$,
$\FsunX\equiv\Flum_{\odot,\XX}$,
$\DF\equiv{\Delta_{F_\XX}}$,
$\Mlum\equiv{\mathcal{M}_\XX}$,
$\Col\equiv{\mathcal{C}_\CX}$,
$\SB\equiv\mu_\XX$,
$(M/L) \equiv (M/L)_\XX$,
$\ext\equiv A_\XX$.} bin $[\Mlum_i,\Mlum_{i+1}]$:
\begin{equation}
  \begin{array}{l}
    \Int_{\Mlum_i}^{\Mlum_{i+1}} \Phi(\Mlum) \,d\Mlum
\stackrel{!}{=} \Int_{\ms_i}^{\ms_{i+1}} \xi(\ms) \,d\ms
  \end{array}
\end{equation}
and therefore,
\begin{equation}
  \Phi(\Mlum) \approx
  \frac{\Int_{\ms_i(\Mlum_i)}^{\ms_{i+1}(\Mlum_{i+1})} \xi(\ms) \,d\ms} 
       {\Mlum_{i+1}-\Mlum_i} \hs \Mlum_i \le \Mlum \le \Mlum_{i+1} .
  \label{eq.lf_ms}
\end{equation}
For the bulge we assumed a 12 Gyr old SSP with $Z=2 Z_\odot$
({isoc\_z040s.dat}\footnote{See \cite{2002A&A...391..195G} and
  http://pleiadi.pd.astro.it.\label{fn.girardi}}), which leads to good
results for the stellar content of the bulge (C. Maraston 2004,
private communication).

For the disk we used for simplicity a 2 Gyr old SSP with $Z=Z_\odot$
({isoc\_z019m.dat}; see footnote \ref{fn.girardi}) leading to
acceptable results for the disk data shown in
(\cite{2002MNRAS.331..293W}; fields INNER, NGC224-DISK, NGC224-POS2,
G287, G11, G272, G87, K108, and G33).

With the mass function $\xi(\ms)$ and the luminosity function
$\Phi(\Mlum)$ we obtain the mass-to-light ratio
\begin{equation}
  \ML = \frac{\Int_{\ms_\Min}^{\ms_\Max} \ms\,\xi(\ms) \,d\ms
      \,/\ms_{\odot}} {\Intinf \FVega 10^{-0.4 \Mlum} \Phi(\Mlum) \,d\Mlum
      \,/\FsunX} 
      = \frac{\Int_{\ms_\Min}^{\ms_\Max} \ms\,\xi(\ms)
      \,d\ms \,/\ms_{\odot}}{<\Flum> \,\Intinf \Phi(\Mlum)
      \,d\Mlum\,/\FsunX}  .
\end{equation}
For a bulge MF as in \S~\ref{sec.massfunc} we get a characteristic flux
$<\Flum_R>=0.20 \FsunR$, yielding a $(M/L)$ in the $R$ band
of $(M/L)_{R}=2.67$ and a $(B-V)=1.14\E{mag}$. For a disk MF as in
\S~\ref{sec.massfunc} we get a characteristic flux $<\Flum_R>=0.67
\FsunR$, yielding a $(M/L)$ in the $R$ band of $(M/L)_{R}=0.61$
and a $(B-V)=0.88\E{mag}$.

\begin{figure}
  \epsscale{1.0}\plottwo{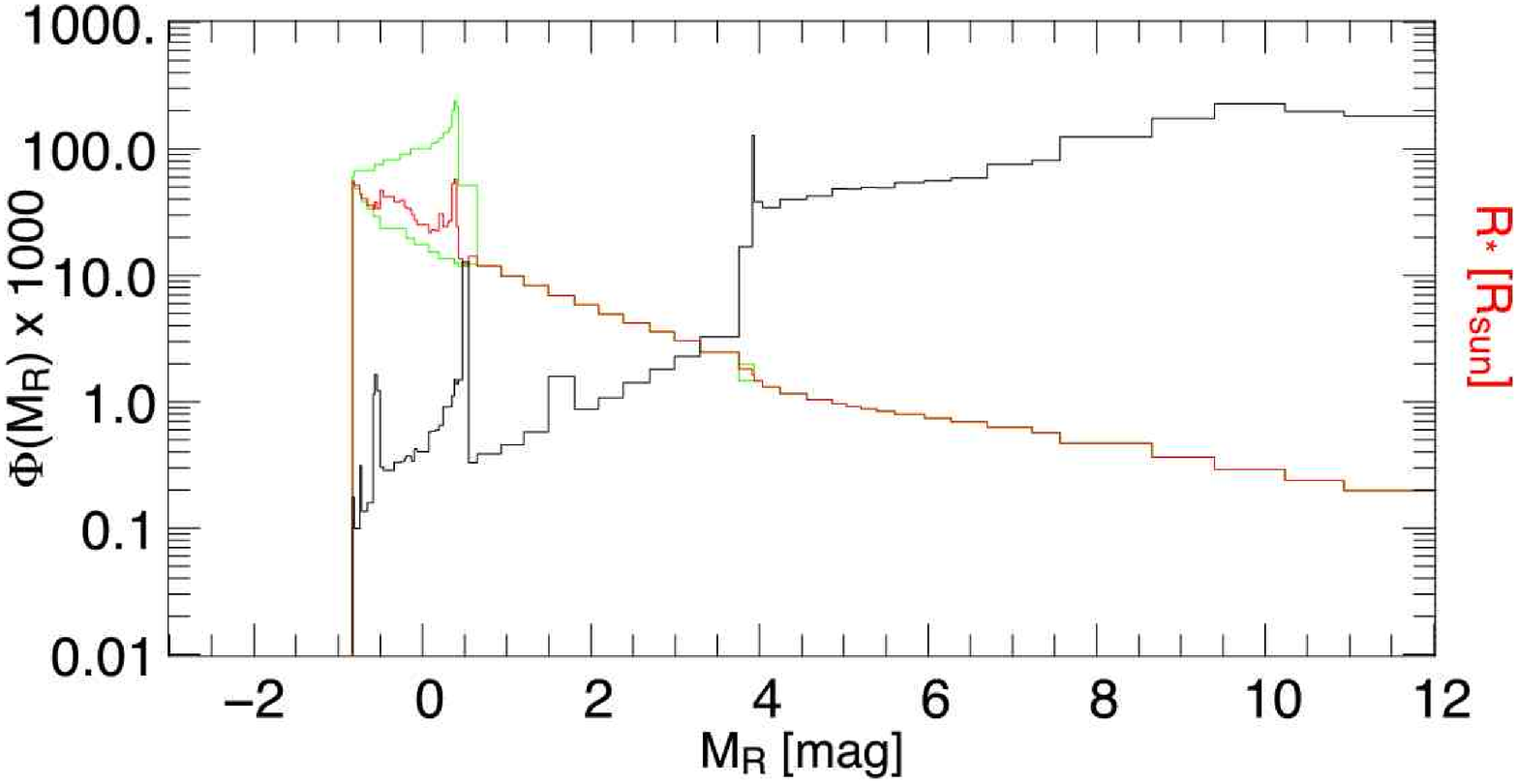}{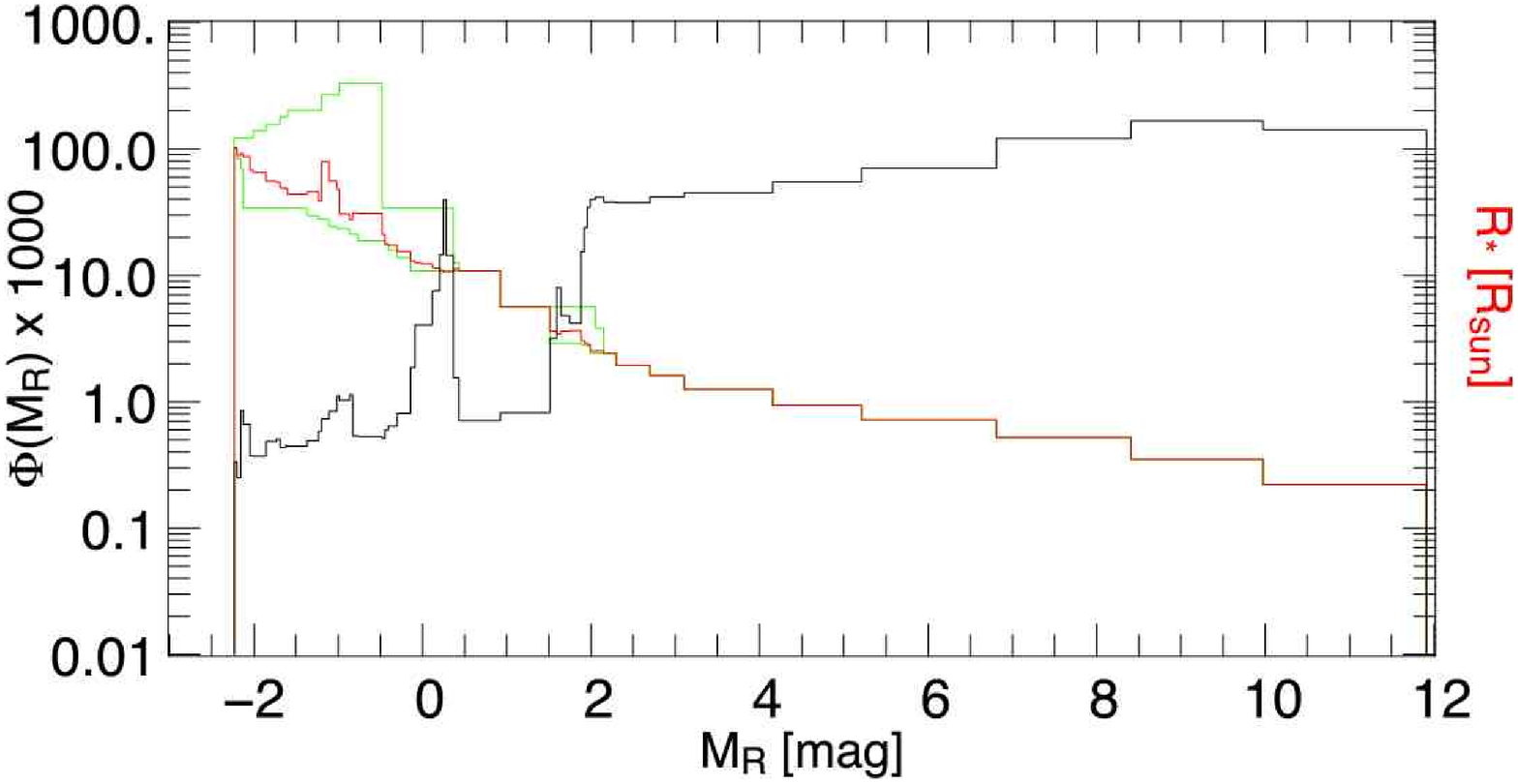}
  \caption{Theoretical LF in the $R$ band $\LF_R(\Mlum)$.  {\it Left}:
    bulge for a 12 Gyr old SSP of 2 $Z_\odot$ metalicity.  {\it
      Right}: disk for a 2 Gyr old SSP of 1 $Z_\odot$ metalicity.  In
    {red} and {green} we show the values of the stellar radii obtained
    with \S~\ref{sec.radius_flux} and the theoretical luminosities
    for the stars of the model SSP. The {red line} shows the average
    radius $\overline{R}_{\ast}$ according to equation
    (\ref{eq.Raver}). In {green} we give the minimal and maximal radii
    of stars (reflecting the different values in color space) in the
    particular magnitude range. The LF was scaled by a factor of 1000
    to show the two different histograms with the same scaling. The
    unit of the LF is number of stars per magnitude, the radii
    distribution is given in solar radii.}
   \label{fig.LF}
\end{figure}

Note that other values of $\ms_\Min$ and $\ms_\Max$ give different
mass-to-light ratios, as the decrease of $\ms_\Min$ increases only the
mass of the population, but not its luminosity. We show the LF for the
bulge population in Figure~\ref{fig.LF}, along with the stellar radii
data (see \S~\ref{sec.radius_flux}).  Note that the faint cutoff of
$\Phi(\Mlum)$ affects the characteristic luminosity $<\Flum>$ but at
the same time the normalization of $\LF(\Mlum)=\Phi(\Mlum) / \int
\Phi(\Mlum) d\Mlum$.  Therefore, the number of bright stars,
$\Flum_\M{tot}<\Flum>^{-1} \int_\M{bright} \LF(\Mlum) \, d\Mlum$, is
nearly not affected by changing the faint cutoff.

\begin{figure}[b]
  \epsscale{0.9}
  \plottwo{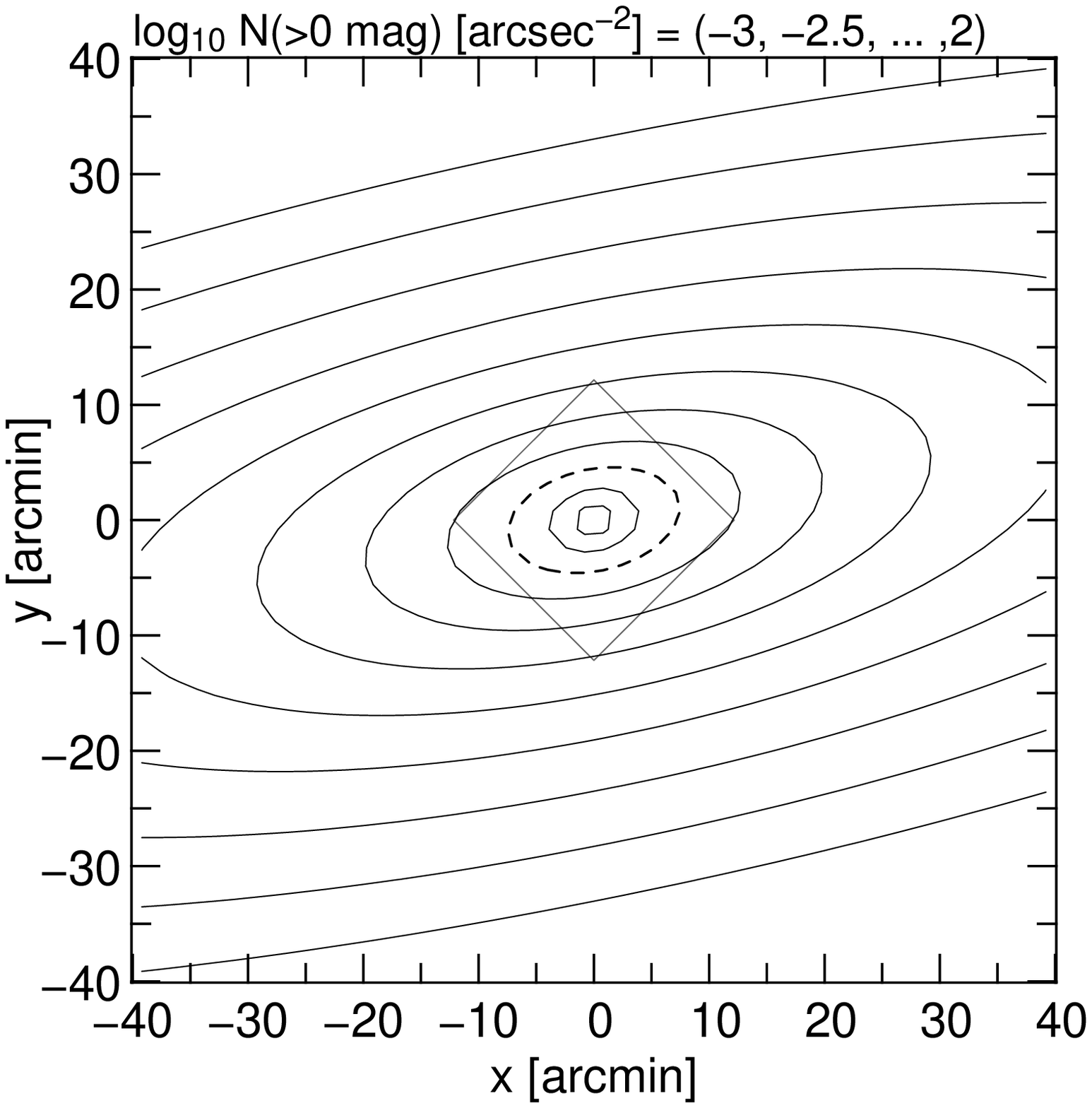}{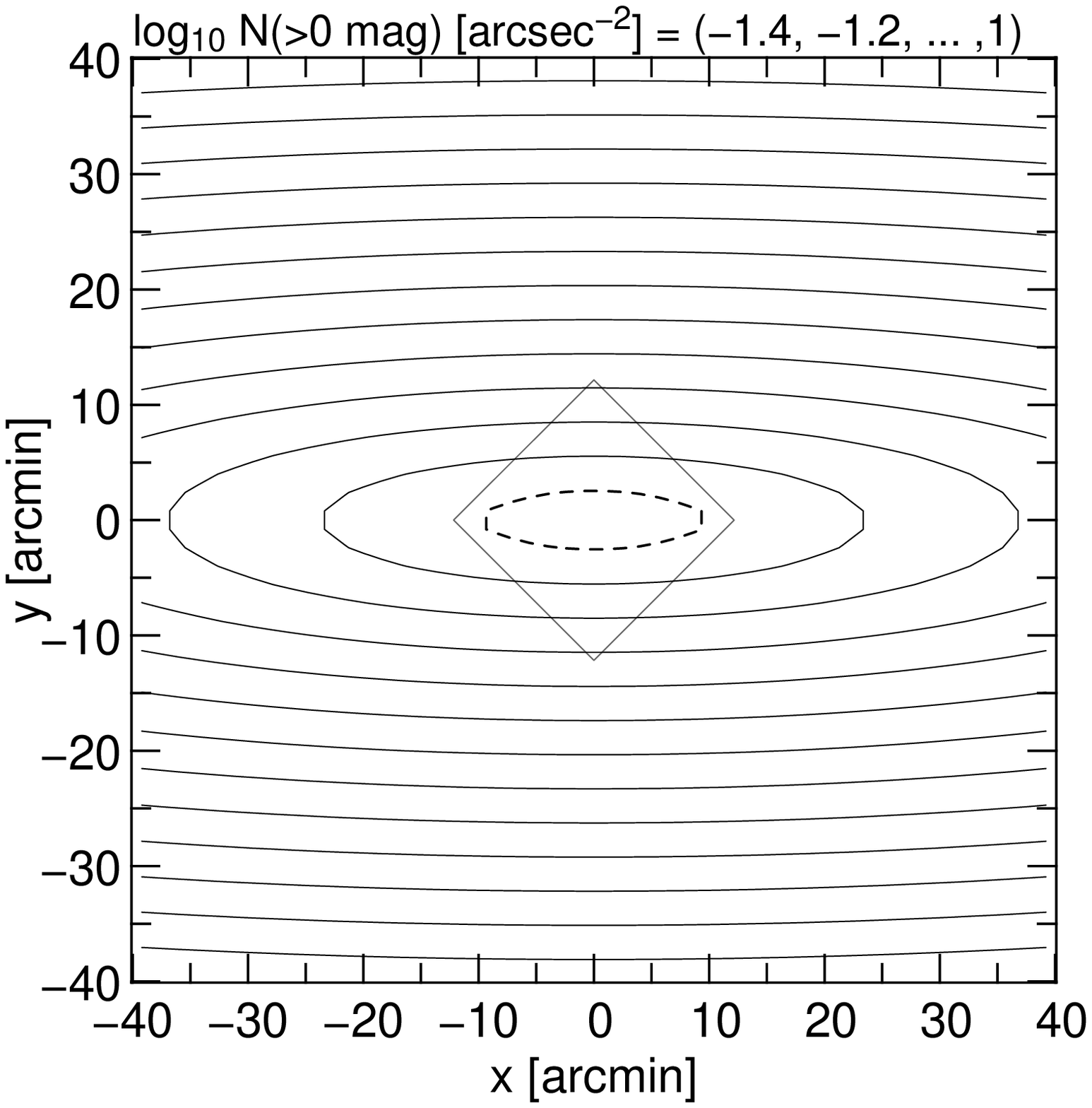}
  \caption{Number density of bulge stars ({\it left}) and disk stars
    ({\it right}) brighter than $\Mlum_R \le 0\E{mag}$ in units of
    stars arcsec$^{-2}$. The contours show the values of
    $\left.d^2N/(dx \,dy)\right|_{\Mlum_R \le 0} = \int_{-\infty}^{0}
    \int_0^\infty \LF(\Mlum_R) \,n_\iss(x,y,\Dos) \,d\Dos \,d\Mlum_R$ and
    were obtained from the number density and luminosity functions of
    the bulge and disk component of M31.  The WeCAPP field, a square
    of $17.2\arcmin$, is shown as a box.  The dashed contour outline a
    density of the $\Mlum_R \le 0$ stars of 10 stars/arcsec$^2$ and
    demonstrate that one cannot resolve even giants in the central M31
    field for the majority of ground-based data.  The coordinates are
    that of the intrinsic M31 system (see
    Figure~\ref{fig.tau_wecapp}).}
  \label{fig.N0mag}
\end{figure}

Using equation (\ref{eq.nsourcedis}) we calculate the projected
densities of bulge and disk stars brighter than $\Mlum_R \le 0\E{mag}$
and show the results in Figure~\ref{fig.N0mag}; basically at any
position monitored by WeCAPP there is more than one bright star per
square arcsec each from bulge and disk. This demonstrates that
crowding in the central bulge is very severe even for the brightest
stars with $\Mlum_R \le 0\E{mag}$ and even if image PSFs are small.

\subsection{Radius-Brightness Relations for Stars}
\label{sec.radius_flux}

For the inclusion of finite source effects one needs the
radius-brightness relation of stars.  The radius can easily be
correlated to the brightness (and to the luminosity function) using
$\lgt(L_i)$ and $\lgt(T_{\M{eff},i})$ given in the theoretical stellar
isochrones (see \S~\ref{sec.LFmodel})
\begin{equation}
  R_\ast(\Mlum_i,\Col_i) = \frac{10^{\left[\lgt L_i+\lgt L_{i+1}\right]/4}}
{\sqrt{4 \pi \sigma_\M{B}} \,10^{\lgt T_{\M{eff},i}+\lgt T_{\M{eff},i+1}}} 
\hs \Mlum_i \le \Mlum \le \Mlum_{i+1}\,, \Col_i \le \Col \le \Col_{i+1} .
\end{equation}
If we want to account for finite source effects without having any
color information, e.g., equation (\ref{eq.n_dgam_dt12_dMlum}), we use
a color-averaged source radius $\overline{R}_{\ast}$,
\begin{equation}
  \overline{R}_{\ast}(\Mlum) = \int \pcmd(\Mlum,\Col) \,\Rstar(\Mlum,\Col) \,d\Col
   ,
  \label{eq.Raver}
\end{equation}
and replace $R_{\ast}(\Mlum,\Col)$ with $\overline{R}_{\ast}(\Mlum)$
in Eqs.~\ref{eq.dgam_dt12_dFlum_nfs} and \ref{eq.dgam_dt12_dFlum_wfs} 
(see Figure~\ref{fig.LF}).

\subsection{The Velocity Distributions  for the M31 Components}
The random velocity components of bulge, disk, and halo are assumed to
be of Gaussian shape with dispersions taken from
\cite{2001MNRAS.323...13K}:
\begin{equation}
   \begin{array}{ll}
    \sigma_\M{bulge}= 100\E{km\ s}^{-1}  , \quad &
    \sigma_\M{disk} =  30\E{km\ s}^{-1}  , \\
    \sigma_\M{halo} = 166\E{km\ s}^{-1}  , \quad &
    \sigma_\M{MW-halo} = 156\E{km\ s}^{-1}  .
    \end{array}
\end{equation}
In addition, we account for rotation in bulge and disk of
$v_\M{rot,bulge}=30\E{km\ s}^{-1}$ and $v_\M{rot,disk}=235\E{km\ s}^{-1}$
\citep{2001MNRAS.323...13K}. In a previous work
\cite{1996ApJ...473..230H} used $\sigma_\M{halo} = 170\E{km\ s}^{-1}$ for
the halo, but a value of $\sigma_\M{bulge}=156\E{km\ s}^{-1}$ for the bulge
and disk (based on \cite{1983ApJ...273..562L}).

In the following two sections we derive the relative source-lens
velocity $v_0$ taking into account rotation of the source and lens
objects and the observers motion. The combination of all contributions
results in one movement depending on
\begin{equation}
  v_0(\Dos,\Dol,v_\M{rot,l},v_\M{rot,s},v_{\odot-\M{M31}}).
\end{equation}

\subsubsection{Additional Rotation for Lenses and Sources}
\label{ssec.rotation}
The additional rotation of the lens system $v_\M{rot,l}$ (for bulge
and disk lenses) and/or of the source system $v_\M{rot,s}$ changes the
relative velocity $v_0$. For the calculation of the effect we first
have to transform the positional components of a lens located at
($x,y,z:=\Dol-d_\M{m31}$) along the line-of-sight to the components
($x_0,y_0,z_0$) in the M31 system. In the internal system the position
is given by
\begin{equation}
       x_0=x  , \quad y_0=y \cos{i} - z \sin{i}  ,\quad z_0=y \sin{i} +
     z \cos{i}  ,
  \end{equation}
with inclination angle $i=77^\circ$ and the distance to M31
$d_\M{m31}=770$~kpc. Projecting on the base $\rho=(x_0^2+y_0^2)^{1/2}$,
the rotation angle can be expressed as
$\omega=\arccos(x_0/\rho)=\arcsin(y_0/\rho)$.

Reprojecting the components of the rotation velocity $v_x$ and
$v_{yz}$ (calculated for a clockwise rotation)
\begin{equation}
       v_x=v_\M{rot} \sin{\omega}=-\frac{y_0}{\sqrt{x_0^2+y_0^2}}
     \,v_\M{rot}  , \quad
     v_{yz}=\sqrt{v_\M{rot}^2-v_x^2}=\frac{x_0}{\sqrt{x_0^2+y_0^2}}
     \,v_\M{rot} ,
  \end{equation}
to the $y$- and $z$- plane yields\footnote{The relations are valid for
the first quadrant, else the sign has to change.}
\begin{equation}
   \begin{array}{l}
v_y= v_{yz} \cos{i}  ,\quad v_z= v_{yz} \sin{i}  ,
   \end{array}
\end{equation}
which depends on the position along the line-of-sight ($x,y,z$).  To
combine this velocity vector $(v_x,v_y,v_z)$ with all other velocities
(see \S~\ref{ssec.obs_motion}) it has to be projected to the lens
plane.

\subsubsection{Observer's Motion}
\label{ssec.obs_motion}

\begin{figure}   
  \epsscale{0.28}\plotone{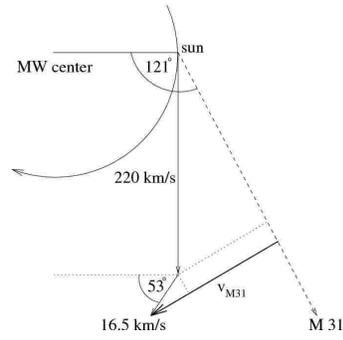}
  \caption{Geometry of the Galaxy-M31 system. A star at solar distance
    is assumed to move on a circular orbit with a rotational velocity
    of $220\E{km\ s}^{-1}$ (local standard of rest, LSR). M31 is located at
    Galactic coordinates $l_\M{M31}=121.2^\circ$ and
    $b_\M{M31}=-21.6^\circ$. The Sun has a velocity of $16.5\E{km\ s}^{-1}$
    relative to the LSR. The transversal velocity of M31 is shown as
    $v_\M{M31}$.}
  \label{sun}
\end{figure}

Finally, we have to account for the transversal velocity of M31
$v_\M{M31}$ arising from the observers motion against M31.  A
hypothetical star on a circular orbit at solar distance (local
standard of rest, LSR) has velocity $v_l(R_\odot)=220 \pm 15
\E{km\ s}^{-1}$. The Sun is moving with $v_\odot=16.5 \E{km\ s}^{-1}$ relative to
the LSR toward the directions $l=53^\circ$, $b=25^\circ$
\citep{1987gady.book.....B}.  For simplicity we neglect the
contributions to the Galactic height (see Figure~\ref{sun}) and
calculate the transversal velocity of M31 as
\begin{equation}
  \darray
\begin{array}{ll}
  v_{\odot-\M{M31}} \approx & (220\E{km\ s}^{-1}) \,\sin(l_\M{M31}-90^\circ) +
    16.5\E{km\ s}^{-1} \,\sin(121^\circ-l_\M{LSR}) = 129 \E{km\ s}^{-1}  ,
\end{array}
\end{equation}
with the Galactic coordinates of M31 $l_\M{M31}=121.2^\circ$ and
$b_\M{M31}=-21.6^\circ$. The relative velocity between the velocity
distribution of the lenses and the sources is calculated by projecting
$v^p_{\odot-\M{M31}}$ to the lens plane
\begin{equation}
  v^p_{\odot-\M{M31}} \approx \frac{\Dos-\Dol}{\Dos} 129\E{km\ s}^{-1}.
\end{equation}
For lenses residing in M31 this motion is negligible compared to the
rotation described in \S~\ref{ssec.rotation}.

\end{appendix}

\end{document}